\pdfoutput=1

\documentclass[ amsmath, amssymb, aps, prd, twocolumn, lengthcheck, sort&compress, showpacs, nofootinbib, letterpaper, eqsecnum ]{revtex4}

\usepackage{amsmath}
\usepackage{amssymb}
\usepackage{graphicx}
\usepackage{dsfont}
\usepackage{dcolumn}
\usepackage{units}
\usepackage{bm}
\usepackage{wasysym}
\usepackage{multirow}
\usepackage{slashed}
\usepackage[usenames]{color}
\usepackage[colorlinks=true,linkcolor=blue,citecolor=blue,urlcolor=blue]{hyperref}

      \newcommand{\conjg}[1]{\ensuremath{\hspace{1pt}\overline{\hspace{-1pt}#1\hspace{-1pt}}}\hspace{1pt}}
      \newcommand{\vect}[1]{\bm{#1}}

      \def\p{\partial}

      \newcommand{\be}{\begin{equation}}
      \newcommand{\ee}{\end{equation}}

      \newcommand{\Slash}[1]{\slashed{#1}}

      \def\longlongrightarrow{
      \relbar\joinrel\relbar\joinrel\relbar\joinrel\relbar\joinrel\rightarrow}

      \definecolor{violet}{RGB}{111,0,255}
      \definecolor{webgreen}{rgb}{0,0.75,0}
      \definecolor{webred}{rgb}{0.75,0,0}
      \definecolor{webblue}{rgb}{0,0,0.75}
      \definecolor{darkblue}{rgb}{0,0,0.6}
      \definecolor{darkgreen}{rgb}{0,0.5,0.5}
      \definecolor{darkpurple}{rgb}{0.5,0,0.5}
      \definecolor{darkorange}{rgb}{1,0.5,0}
      \definecolor{darkgrey}{rgb}{0.4,0.4,0.4}
      \definecolor{lgray}{rgb}{0.95,0.95,0.95}
      \definecolor{lgreen}{rgb}{0.95,1.00,0.90}
      \definecolor{lred}{rgb}{1.00,0.90,0.80}
      \definecolor{lblue}{rgb}{0.2,0.35,1.00}
      \definecolor{shadecolor}{rgb}{1.00,0.92,0.82}

\begin{document}

         \title{Nucleon resonances in Compton scattering  }

       \author{Gernot~Eichmann$^{1}$}
       \author{G.~Ramalho$^2$}

\affiliation{$^1$CFTP, Instituto Superior T\'ecnico, Universidade de Lisboa, 1049-001 Lisboa, Portugal  \\
             $^2$Laborat\'orio de F\'{i}sica Te\'orica e Computacional -- LFTC, Universidade Cruzeiro do Sul,
01506-000, S\~ao Paulo, SP, Brazil}

         \date{\today}

         \begin{abstract}
         We calculate the nucleon resonance contributions to nucleon Compton scattering,
         including all states with $J^P=1/2^\pm$ and $J^P=3/2^\pm$ where experimental data for their
         electromagnetic transition form factors exist.
         To this end, we construct a tensor basis for the Compton scattering amplitude
         based on electromagnetic gauge invariance, crossing symmetry and analyticity.
         The corresponding Compton form factors provide a Lorentz-invariant description of the process in general kinematics,
         which reduces to the static and generalized polarizabilities in the appropriate kinematic limits.
         We derive the general forms of the offshell nucleon-to-resonance transition vertices that implement
         electromagnetic and spin-$3/2$ gauge invariance, which automatically also defines onshell transition form factors
         that are free of kinematic constraints. We provide simple fits for those form factors,
         which we use to analyze the resulting Compton form factors and extract their contributions to the nucleon's polarizabilities.
         Apart from the $\Delta(1232)$, the resonance contributions to the scalar and spin polarizabilites are very small,
         although the $N(1520)$ could play a role for the proton's magnetic polarizability.
         \end{abstract}


         \maketitle


     \begin{figure*}[t!]
     \center{
     \includegraphics[scale=0.085]{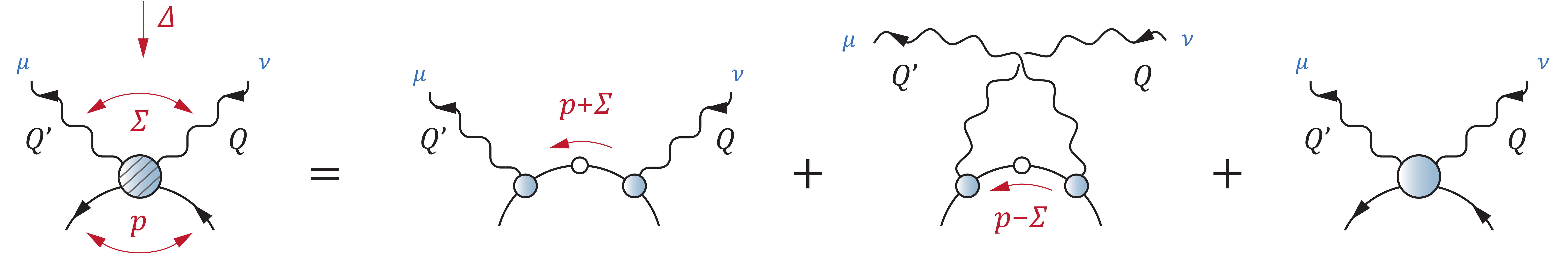}}
        \caption{Separation of the nucleon Compton amplitude into Born terms and a 1PI structure part. }
        \label{fig:qcv-born}
     \end{figure*}

        \section{Introduction} \label{sec:introduction}

       Compton scattering on the nucleon encodes a multitude of interesting physical applications.
       It is the process $\gamma^\ast N \to \gamma^\ast N$, sketched in Fig.~\ref{fig:qcv-born},
       where either of the photons can be real or virtual.
       Compton scattering probes the electromagnetic structure of the nucleon and therefore the quarks inside.
       On the one hand, it encodes the nucleon's polarizabilities which test its response to an external electromagnetic field.
       Ongoing efforts with chiral effective field theory, dispersion relations and other approaches
       aim to determine the proton's and neutron's scalar and spin polarizabilities~\cite{Drechsel:2002ar,Schumacher:2005an,Griesshammer:2012we,Holstein:2013kia,Hagelstein:2015egb}.
       On the other hand, virtual Compton scattering (VCS), where one photon is virtual and the other is real,
       provides access to the nucleon's generalized polarizabilities~\cite{Guichon:1995pu,Drechsel:1998zm,Drechsel:2002ar,Downie:2011mm,Lensky:2016nui}.
       Deeply virtual Compton scattering (DVCS) is the primary tool to extract the nucleon's
       generalized parton distributions (GPDs)~\cite{Belitsky:2001ns,Belitsky:2005qn,Guidal:2013rya,Kumericki:2016ehc};  
       and the forward limit, where the momentum transfer vanishes, is experimentally accessible
       in deep inelastic scattering and relates the Compton amplitude with the nucleon structure functions and PDFs.

       Also the integrated Compton amplitude is of interest.
       The diagram where the two photons couple to a lepton encodes the two-photon exchange (TPE) corrections
       to electromagnetic form factors. These are believed to be responsible for the difference in the proton's
       $G_E/G_M$ measurements, because the Rosenbluth separation method is sensitive to TPE effects whereas the
       polarization transfer experiments are not~\cite{Guichon:2003qm,Carlson:2007sp,Arrington:2011dn}.
       However, at present it still remains to be clarified which parts of the Compton amplitude cause the difference.
       TPE contributions also enter in the proton radius puzzle
       although so far the effect appears to be too small by an order of magnitude to explain the discrepancy~\cite{Pohl:2010zza,Birse:2012eb,Antognini:1900ns,Antognini:2013jkc,Pohl:2013yb,Carlson:2015jba}.

       At the hadronic level, the Compton amplitude can be split into `elastic' Born terms and an `inelastic' structure part as in Fig.~\ref{fig:qcv-born}.
       In principle the Born terms are determined by the nucleon electromagnetic form factors,
       whereas the one-particle irreducible (1PI) structure part encodes the structure information such as polarizabilities.
       The latter probes the spectrum of hadrons:
       in terms of its singularity structure, it contains intermediate nucleon resonances
       in the $s$ and $u$ channels such as the $\Delta(1232)$ resonance, 
       meson exchanges in the $t$ channel, and vector-meson poles for timelike photon virtualities.
       These are accompanied by multiparticle cuts in the various channels,
       which come from $N\pi$, $N\pi\pi$, $\pi\pi \dots$ loops and are directly accessible in effective field theory approaches.
       For example, it is well known that the $\Delta$ resonance provides a large contribution to the
       magnetic polarizability which is counteracted by pion loops,
       thus leading to the picture of a  `paramagnetic quark core' that is cancelled by its `diamagnetic pion cloud'~\cite{Bernard:1993bg,Hemmert:1996rw,Lensky:2009uv,Hagelstein:2015egb}.

       On the other hand,
       handbag dominance in DVCS attributes the dynamics in Compton scattering
       to an interaction of the photons with the perturbative quarks inside the nucleon.
       It is then understood that the hadronic description should be applied at low energies
       whereas the microscopic approach is appropriate when $Q^2$ is large. Still, it is
       desirable to connect these two regimes by a common underlying formulation.
       Such an approach in terms of nonperturbative quark and gluon degrees of freedom has been formulated~\cite{Eichmann:2011ec,Eichmann:2012mp,Eichmann:2016yit}
       but it is not the topic of the present work. Here we aim for a more modest goal,
       namely to establish a  connection  in terms of common amplitudes to describe the process   
       in arbitrary kinematics.

       In principle Compton scattering is completely specified by 18 Lorentz-invariant functions~\cite{Tarrach:1975tu},
       which are probed in different kinematic limits by the experimental processes mentioned above.
       The purpose of this paper is to make a step towards connecting these limits
       by providing a tensor basis based
       on electromagnetic gauge invariance, crossing symmetry and analyticity.
       This leads to a set of 18 Compton form factors (CFFs) which depend on four Lorentz-invariant variables and which are free of kinematic constraints.
       In the limit where all variables vanish these are related to the nucleon's polarizabilities,
       in VCS they are connected to the generalized polarizabilities and in the forward limit to the nucleon's structure functions.
       Each CFF has certain characteristics: the nucleon Born terms contribute to only a few of them, as well as the $t$-channel meson poles;
       and only certain subsets of them survive in the forward limit, in RCS or in VCS.

       Following the approach by Bardeen, Tung and Tarrach~\cite{Bardeen:1969aw,Tarrach:1975tu}, similar tensor bases have been employed 
       for specific applications
       such as low-energy VCS~\cite{Drechsel:1997xv,Gorchtein:2009wz} or scalar Compton scattering~\cite{Drechsel:1996ag,Bakker:2016vin}.
       Here we provide the detailed basis construction for a spin-$1/2$ target in general kinematics, using
       a procedure that  differs from  Refs.~\cite{Bardeen:1969aw,Tarrach:1975tu} and allows one to better track the occurrence or absence of kinematic singularities.
       It is still true that kinematic singularities cannot be avoided in certain limits~\cite{deCalan:1969zz},
       but this does not affect the 18 CFFs in general or the limits of RCS, VCS and the forward limit where
       direct measurements are possible.

       As a practical application we work out the CFF contributions from intermediate $s$- and $u$-channel nucleon resonances,
       which enter in the process through their electromagnetic transition form factors.
       The $\Delta(1232)$ contribution to the nucleon's polarizabilities is known~\cite{Pascalutsa:2002pi,Pascalutsa:2003zk,Lensky:2016nui},
       but in view of a precision determination of polarizabilities it is still desirable to understand the impact of higher resonances,
       which can also play a role in TPE~\cite{Kondratyuk:2005kk,Kondratyuk:2007hc}.
       In the last decade significant progress has been made in measuring the electrocouplings
       of nucleon resonances through meson electroproduction in a wide $Q^2$ range~\cite{Aznauryan:2011qj,Tiator11a}.
       In addition to the $\Delta(1232)$ resonance, the electromagnetic transitions are now relatively well known
       also for the Roper resonance $N(1440)$, the nucleon's tentative parity partner $N(1535)$, and the $N(1520)$ resonance.
       First data for higher-lying resonances have been accumulated in two-pion production~\cite{Mokeev:2015moa,CLAS2016,Mokeev-NSTAR}
       and more results are underway with the Jefferson Lab 12 GeV program.

       The fact that the resonances in Compton scattering are offshell creates additional complications.
       Electromagnetic gauge invariance and spin-$3/2$ gauge invariance for Rarita-Schwinger particles~\cite{Weinberg:1980kq,Pascalutsa:1999zz},
       which ensures the absence of the spin-$1/2$ background in matrix elements,
       induce further constraints on the offshell transition vertices.
       Here we derive the most general structure for the $J^P=1/2^\pm$ and $J^P=3/2^\pm$ transition amplitudes
       that are compatible with these constraints.
       As a result, their implementation in Compton scattering
       automatically ensures the absence of spurious contributions.

       Moreover, these expressions also determine the most general forms of the onshell
       nucleon-to-resonance electromagnetic transition currents which are free of kinematic constraints.
       One obtains two form factors $F_{1,2}(Q^2)$ in the $J=1/2$ case and
       three form factors $F_{1,2,3}(Q^2)$ for $J=3/2$ and higher spin, which are kinematically independent
       so that all their singularities and momentum dependencies are of dynamical origin.
       The experimental data are usually discussed in terms of helicity amplitudes
       or multipole form factors~\cite{Jones:1972ky,Devenish:1975jd,Aznauryan:2011qj} which are neither free of kinematics nor satisfy the offshell constraints.
       Here we provide simple fits to the experimental data for all available resonances in terms of the constraint-free form factors $F_i(Q^2)$.
       Those parametrizations we finally implement in the Compton amplitude to calculate the CFFs in the entire kinematic domain.

       The paper is organized as follows. In Sec.~\ref{sec:compton-amplitude} we establish our notation,
       discuss the kinematic regions in terms of four Lorentz-invariant variables,
       provide the tensor basis for the Compton amplitude and investigate different kinematic limits.
       In Sec.~\ref{sec:scalar-csa} we illustrate the situation for scalar Compton scattering. In Sec.~\ref{sec:spin-1/2} we discuss
       the nucleon Born terms, together with the offshell nucleon-photon vertex that enters there, in some detail and work out the corresponding CFFs.
       In Secs.~\ref{sec:spin-1/2r} and~\ref{sec:spin-3/2} we apply the same procedure to derive the $J=1/2$ and $J=3/2$ resonance contributions, respectively,
       and in Sec.~\ref{sec:ffs} we provide our fits for their transition form factors.
       The resulting CFFs and polarizabilities are discussed in Sec.~\ref{sec:cffs}. We summarize in Sec.~\ref{sec:outlook}.

       Several appendices serve the purpose of making our calculations as transparent as possible for practitioners.
       We use a Euclidean metric for practical convenience but with the formulas in App.~\ref{sec:euclidean} 
       the transcription between Euclidean and Minkowski conventions should be straightforward. 
       In App.~\ref{sec:tensor-basis} we explain the tensor basis derivation for the Compton amplitude in detail;
       we provide relations between our CFFs and the amplitude conventions in some kinematic limits used in the literature;
       and we investigate the consequences of breaking gauge invariance for the nucleon Born term.
       App.~\ref{sec:point-tfs} gives some details on spin-$3/2$ Lagrangians, and in App.~\ref{sec:ffs-onshell} we collect the relations
       between the resonance transition form factors and helicity amplitudes employed in the literature.

             \begin{figure*}[t]
                    \begin{center}
                    \includegraphics[width=1\textwidth]{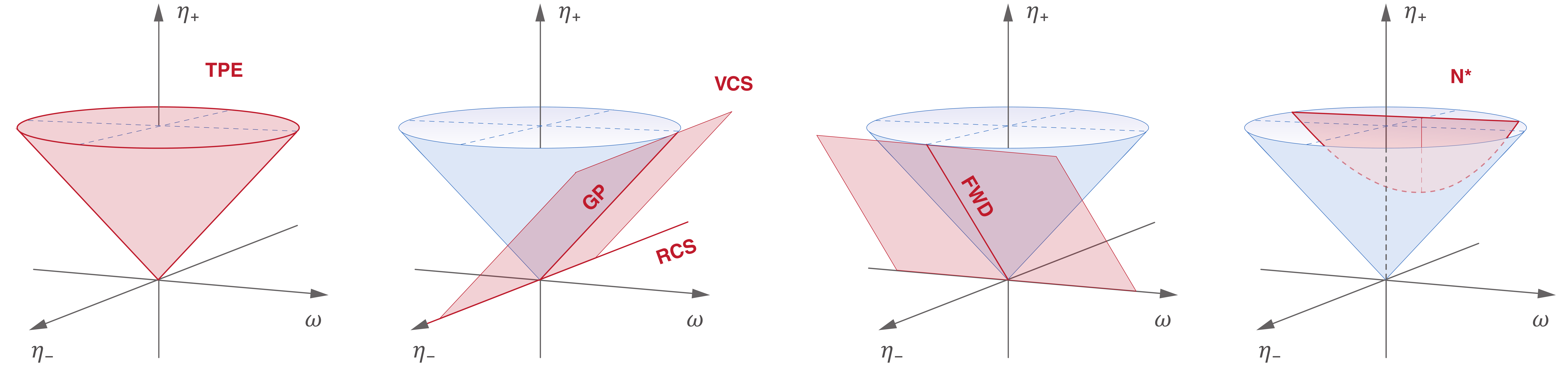}
                    \caption{Compton scattering in the variables $\eta_+$, $\eta_-$ and $\omega$.
                             The interior of the cone contributes to two-photon exchange (TPE).
                             Real Compton scattering (RCS) lives on the $\eta_-$ axis and virtual Compton scattering (VCS) on the plane $\eta_+=\omega$.
                             The boundary of the cone contains the doubly-virtual forward limit (FWD) at $t=0$ ($\eta_+=\eta_-$)   and the
                             VCS limit where the generalized polarizabilities are defined (GP, $\eta_+=\omega$ and $\eta_-=0$).
                             Inside the cone, nucleon resonances appear at $\eta_-=-\delta$.
                             }\label{fig:phasespace}
                    \end{center}
            \end{figure*}

       \pagebreak

    \renewcommand{\arraystretch}{1.3}

   \section{Compton amplitude} \label{sec:compton-amplitude}

        \subsection{Kinematics} \label{sec:kinematics}

        The onshell nucleon Compton amplitude with virtual photons has the form
        \begin{equation}
           \mathcal{M}^{\mu\nu}(p,Q',Q) = \frac{e^2}{m} \,\conjg{u}(p_f)\,\Gamma^{\mu\nu}(p,Q',Q)\, u(p_i)\,,
        \end{equation}
        where $e^2=4\pi\alpha_\text{em}$,
        $m$ is the nucleon mass,
        $Q$ and $Q'$ are the incoming and outgoing photon four-momenta, $p_i$ and $p_f$ are the initial and final onshell nucleon momenta ($p_i^2=p_f^2=-m^2$), and
        $p=(p_i+p_f)/2$ is the average nucleon momentum (see Fig.~\ref{fig:qcv-born}). $u(p_i)$ and $\conjg{u}(p_f)$ are nucleon spinors satisfying the Dirac equation;
        they are eigenspinors of the positive-energy projectors
        \begin{equation}\label{pos-energy-proj}
           \Lambda_+(p_f)  = \frac{-i\slashed{p}_f+ m}{2m}\,, \qquad
           \Lambda_+(p_i)  = \frac{-i\slashed{p}_i+ m}{2m}
        \end{equation}
        with $\Lambda_+(p_i)\,u(p_i) = u(p_i)$ and $\conjg{u}(p_f)\,\Lambda(p_f) = \conjg{u}(p_f)$.
        It is then more convenient to work with the Dirac matrix-valued Compton amplitude
        \begin{equation}\label{csa-decomp-01}
            \Gamma^{\mu\nu}(p,Q',Q) = \Lambda_+(p_f) \left[ \sum_{i=1}^{18} c_i\,X_i^{\mu\nu} \right] \Lambda_+(p_i)\,,
        \end{equation}
        where the spinors are replaced with the projectors.
        The Compton amplitude is constructed from 18 dimensionless Compton form factors (CFFs) $c_i$ which depend on four kinematical invariants,
        together with 18 Lorentz-covariant basis tensors $X_i^{\mu\nu}(p,Q',Q)$.

             We will alternatively use two sets of four-vectors, $\{p,\, Q,\, Q'\}$ and $\{ p,\, \Sigma,\, \Delta\}$,
             with the relations
             \begin{equation}\label{kinematics-1}
                 \begin{array}{rl}
                     p &= \frac{1}{2} (p_i+p_f)\,, \\[1mm]
                     \Sigma &= \frac{1}{2}(Q+Q')\,,
                 \end{array}\quad
                 \Delta=Q-Q'=p_f-p_i
             \end{equation}
             and
             \begin{equation}
                 \begin{array}{rl}
                     p_i &= p-\frac{\Delta}{2}\,, \\
                     p_f &= p+\frac{\Delta}{2}\,,
                 \end{array}\qquad
                 \begin{array}{rl}
                     Q &= \Sigma+\frac{\Delta}{2}\,, \\
                     Q' &= \Sigma-\frac{\Delta}{2}\,.
                 \end{array}
             \end{equation}
             With the constraints $p_i^2=p_f^2=-m^2$, the process is characterized
             by four Lorentz invariants.
            We work with the dimensionless variables\footnote{Introducing new symbols for these variables
            provides a compact notation but also has the following advantage: we
            use Euclidean conventions throughout this work, but since Lorentz-invariant scalar products differ from their
            Minkowski counterparts only by minus signs these variables are the same in Minkowski space if one defines them as
           \begin{equation*}  \renewcommand{\arraystretch}{1.2}
            \begin{split}
               &\eta_+ = -\frac{q^2+{q'}^2}{2m^2}, \quad
               \eta_- = -\frac{q\cdot q'}{m^2}, \quad
               \omega = -\frac{q^2-{q'}^2}{2m^2}, \\
               & \qquad\qquad\quad \lambda = \frac{\tilde{p}\cdot q}{m^2} = \frac{\tilde{p}\cdot q'}{m^2}\,,
            \end{split}
            \end{equation*}
            where $\tilde{p}$, $q$ and $q'$ are the Minkowski momenta corresponding to $p$, $Q$ and $Q'$.
            In that way all relations between Lorentz-invariant (but also Lorentz-covariant) quantities,
            such as the CFFs given in Tables~\ref{tab:nucleon-born}, \ref{tab:j=1/2+born} and \ref{tab:j=3/2+born},
            are identical in Euclidean and Minkowski conventions; see App.~\ref{sec:euclidean} for more details.}
            \begin{equation}\label{li-1} \renewcommand{\arraystretch}{1.2}
            \begin{split}
               &\eta_+ = \frac{Q^2+{Q'}^2}{2m^2}, \quad
               \eta_- = \frac{Q\cdot Q'}{m^2}, \quad
               \omega = \frac{Q^2-{Q'}^2}{2m^2}, \\
               & \qquad\qquad\quad \lambda = -\frac{p\cdot\Sigma}{m^2}= -\frac{p\cdot Q}{m^2} = -\frac{p\cdot Q'}{m^2}
            \end{split}
            \end{equation}
            and vice versa
            \begin{equation}\label{kinematics-1} \renewcommand{\arraystretch}{1.2}
            \begin{split}
                \left\{ \begin{array}{c} Q^2 \\ {Q'}^2 \end{array}\right\} &=  \Sigma^2+\frac{\Delta^2}{4} \pm \Sigma\cdot \Delta = m^2\, (\eta_+ \pm \omega), \\
                Q\cdot Q' &= \Sigma^2-\frac{\Delta^2}{4} = m^2 \,\eta_-, \\
            \end{split}
            \end{equation}
            so that the CFFs in Eq.~\eqref{csa-decomp-01} are dimensionless functions $c_i(\eta_+, \eta_-, \omega, \lambda)$.
            The variables $\eta_+$ and $\eta_-$ are even under
            photon crossing and charge conjugation, whereas $\omega$ and $\lambda$ switch signs (see Eq.~\eqref{bose+cc} below).
            Below we employ a tensor basis that is invariant under both operations,
            so that the CFFs can depend on $\omega$ and $\lambda$ only quadratically.

        The variables $\eta_+$, $\eta_-$ and $\omega$ admit a simple geometrical understanding of the phase space.
        Defining the momentum transfer as
        \begin{equation} \label{t-def}
            t = \frac{\Delta^2}{4m^2} =  \frac{\eta_+-\eta_-}{2}
        \end{equation}
        (which differs by $-4m^2$ from the usual definition),
        then for $t>0$ the region that one must integrate over in order to calculate two-photon exchange (TPE) contributions to observables 
        forms a cone  around the $\eta_+$ direction, which is shown in the leftmost panel in Fig.~\ref{fig:phasespace}.
        This is so because $\Sigma^\mu$ is the integration momentum and the integration region
        is subject to the constraints  
        \begin{equation}\label{spacelike}
            \sigma>0, \quad -1 < Z < 1, \quad -1 < Y < 1
        \end{equation}
        where $\sigma$, $Z$ and $Y$ are the hyperspherical variables from Eq.~\eqref{simple-frame} with the Lorentz-invariant definition
        \begin{equation}\label{li-3}
            \sigma = \frac{\Sigma^2}{m^2}\,, \qquad Z = \widehat{\Sigma}\cdot\widehat{\Delta}\,, \qquad Y = \widehat{p}\cdot\widehat{\Sigma_\perp} \,.
        \end{equation}
        Here, a hat denotes a normalized four-momentum (e.g., $\widehat{\Sigma} = \Sigma/\sqrt{\Sigma^2}$)
        and the subscript $\perp$ stands for a transverse projection with respect to the total momentum transfer $\Delta$.
        These variables are related to the ones in Eq.~\eqref{li-1} via
        \begin{align}
             \sigma &= \frac{\eta_++\eta_-}{2}\,, \qquad
           Z = \frac{\omega}{\sqrt{\eta_+^2-\eta_-^2}}, \label{li-relations} \\
           \lambda &= \frac{Y}{2}\,\sqrt{\omega^2 +\eta_-^2-\eta_+^2}\,\sqrt{1+\frac{2}{\eta_+-\eta_-}} \,. \label{li-relations-Y}
        \end{align}
        With $t>0$ the first two constraints in~\eqref{spacelike} entail
        \begin{equation}\label{cone-relations}
            -\eta_+ < \eta_- < \eta_+, \qquad \omega^2 +\eta_-^2 < \eta_+^2
        \end{equation}
        which defines the cone in Fig.~\ref{fig:phasespace}.
        Because of
        \begin{equation} \renewcommand{\arraystretch}{1.0}
           \eta_+\pm\omega = \frac{1}{m^2}\left\{ \begin{array}{c} Q^2 \\ {Q'}^2 \end{array}\right\}, \quad
           \frac{\eta_+\pm\eta_-}{2} = \left\{ \begin{array}{c} \sigma \\ t \end{array}\right\}
        \end{equation}
        the opposite sides of the cone in the $\{\omega,\eta_+\}$ plane define the axes of $Q^2$ and ${Q'}^2$, whereas
        in the $\{\eta_-,\eta_+\}$ plane the cone is bounded by the axes for $\sigma$ and $t$. 
        Because $Y\in(-1,1)$ is real inside the cone, the crossing variable $\lambda$ must become imaginary
        due to Eqs.~\eqref{li-relations-Y} and \eqref{cone-relations}.

        In Fig.~\ref{fig:phasespace} we show the various kinematic limits:
        \begin{itemize}
  \setlength{\itemsep}{5pt}
  \setlength{\parskip}{0pt}
  \setlength{\parsep}{0pt}
        \item \textbf{Real Compton scattering (RCS):} \\ $Q^2={Q'}^2=0$ $\Rightarrow$ $\eta_+=\omega=0$.
        \item \textbf{Virtual Compton scattering (VCS):} \\ ${Q'}^2=0$ $\Rightarrow$ $\eta_+=\omega$.
        \item \textbf{Generalized polarizabilities:} \\ ${Q'}^\mu=0$ $\Rightarrow $ $\eta_+=\omega$, $\eta_-=\lambda=0$.
        \item \textbf{Doubly-virtual forward limit:}\\ $\Delta^\mu=0$ $\Rightarrow $ $\eta_+=\eta_-$, $\omega=0$.
        \item \textbf{Static polarizabilities:} \\ $\eta_+=\eta_-=\omega=\lambda=0$.
        \end{itemize}
        In the 3D plots the static polarizabilities are defined at the origin of the coordinate system;
        the forward amplitudes and generalized polarizabilities live on the boundary of the cone
        where $t=0$ or ${Q'}^2=0$, respectively;
        the RCS limit is defined along the $\eta_-$ axis outside of the cone;
        and the VCS limit defines the plane ${Q'}^2=0$.

        The nucleon resonance poles at $s=m_R^2$ and $u=m_R^2$,
        where $s$ and $u$ are the Mandelstam variables and $m_R$ is the mass of the resonance,
        are more difficult to visualize because they also depend on the crossing variable $\lambda$:
        \begin{equation}\label{var-su}     \renewcommand{\arraystretch}{1.0}
            \left\{ \begin{array}{c} s \\ u \end{array}\right\} = -(p\pm \Sigma)^2 = m^2\,\big[ 1 - (\eta_- \mp 2\lambda) \big] = m_R^2\,,
        \end{equation}
        with $\lambda = (s-u)/(4m^2)$.
        Taking also into account the resonance width, $m_R^2 \rightarrow m_R^2 - im_R\,\Gamma_R$, and defining
        \begin{equation}\label{delta-def}
            \delta=\frac{m_R^2-m^2}{m^2} \,, \qquad \gamma = \frac{m_R\, \Gamma_R}{2m^2}\,,
        \end{equation}
        the condition for a pole becomes
        \begin{equation}\label{pole-condition}
            \eta_- = \pm 2\lambda -\delta + 2i\gamma \,.
        \end{equation}

            \begin{figure}[t]
                    \begin{center}
     \includegraphics[scale=0.14]{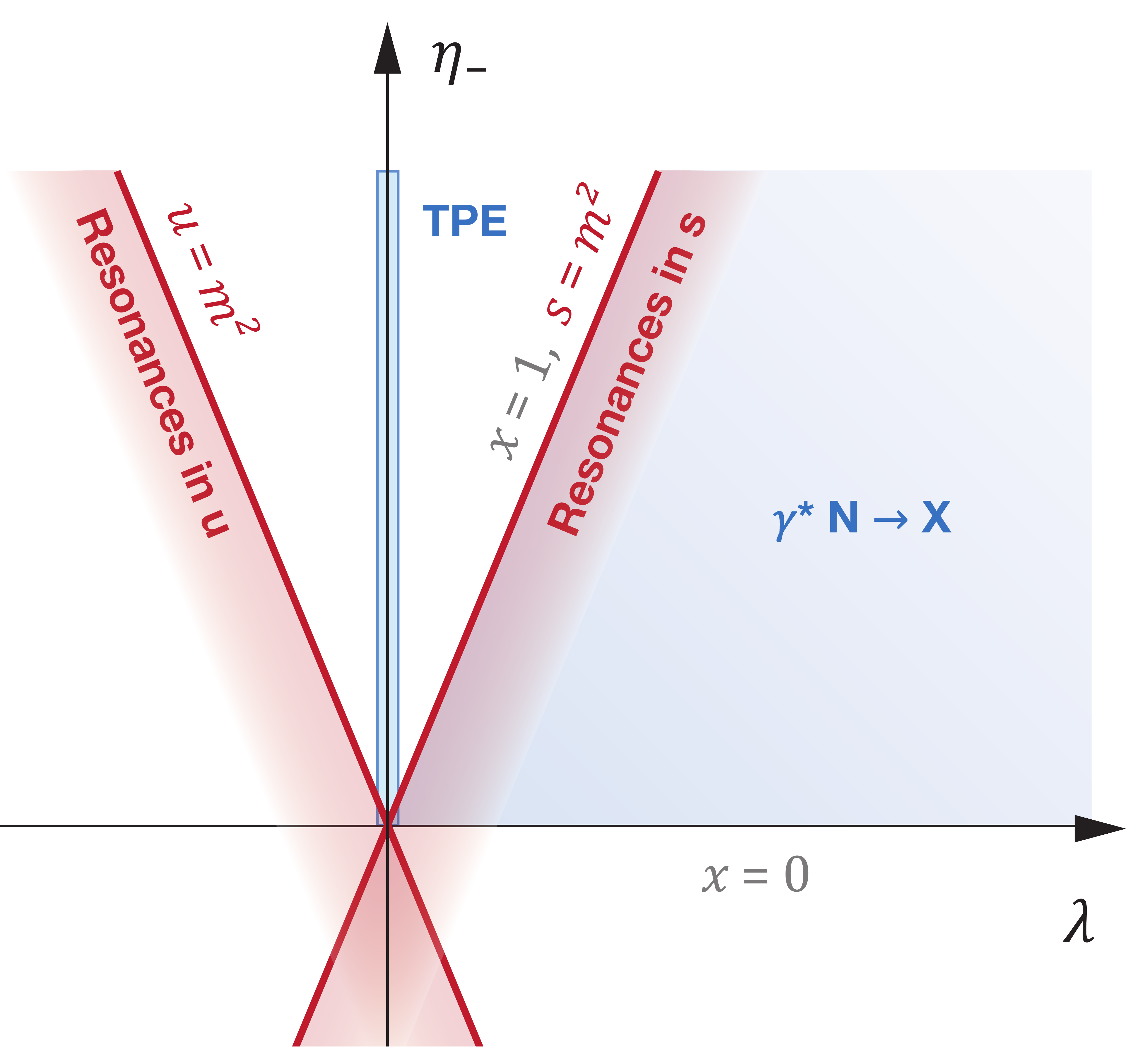}
        \caption{Kinematics in forward Compton scattering in the variables $\eta_-$ and $\lambda$.}
        \label{fig:fwd-phasespace}
        \end{center}
        \end{figure}

        Fig.~\ref{fig:fwd-phasespace} illustrates the situation in the forward limit, where the two remaining variables
        $\eta_- = \eta_+ = Q^2/m^2$ and $\lambda$ define the Mandelstam plane.
       The forward CS amplitude is of special interest because
       the optical theorem relates its imaginary part to the total photoabsorption cross section $\gamma^\ast N\to X$
       and thus to the nucleon's structure functions. 
       The physical region of that process is where
       the Bjorken variable $x=\eta_-/(2\lambda)$ takes values $0 \leq x \leq 1$.
       The nucleon resonances appear at fixed $s$ and $u$,
       starting with the nucleon poles at $s=m^2$ and $u=m^2$ (corresponding to $\eta_- = \pm 2\lambda$).
       The resonance regions are indicated by the red shaded areas in the plot, where
       at larger $s$ and $u$ the resonances are eventually washed out.
       In addition, one has branch cuts from multiparticle production:
       the right-hand cut at $s \geq (m+m_\pi)^2$, which starts at the
       pion production threshold  and extends to infinity, the left-hand cut at $u \geq (m+m_\pi)^2$,
       plus further cuts in the timelike region where $\eta_-$ is negative.

       Except for the nucleon Born poles and branch cuts, the CFFs are analytic functions in the physical sheet,
       given that they are defined through an appropriate tensor basis which does not introduce additional kinematic singularities.
       Since their imaginary parts along the cuts are known from the cross section data,
       one can exploit Cauchy's formula to determine the CFFs everywhere in the complex $\lambda$ plane via (subtracted) dispersion relations.
       Except for the subtraction functions,  
       which can be determined in chiral effective field theory (see e.g. the reviews~\cite{Griesshammer:2012we,Hagelstein:2015egb}),
       the forward CS amplitude is then in principle fully determined by experimental data.

       The TPE region is the interior of the cone where  
       the crossing variable $\lambda$ is imaginary.
       In the forward limit Eq.~\eqref{li-relations-Y} becomes $\lambda = i Y\sqrt{\eta_-}$, so that
       the remnant of the cone is the domain $\text{Re}\,\lambda=0$ and $|\text{Im}\,\lambda| \leq \sqrt{\eta_-}$ along the imaginary $\lambda$ axis,
       as indicated in Fig.~\ref{fig:fwd-phasespace}.
       On the other hand,
       for small values of $\lambda$ the CFFs can be expanded in powers of $\lambda^2$.
       The $Q^2-$dependent forward polarizabilities are accordingly defined as the coefficients in a low-energy expansion:
       \begin{equation}\label{lex}
          c_i(\eta_-,\lambda) = c_i^\text{Born}(\eta_-,\lambda) + c_i^\text{Pol.}(\eta_-,\lambda=0) +  \mathcal{O}(\lambda^2)\,,
       \end{equation}
       where the Born contributions are singular for $\lambda = \pm \eta_-/2$
       and the remaining pieces absorb all structure effects.

       Because the nucleon resonance locations only depend on $\eta_-$ and $\lambda$,
       the Mandelstam plane has the same form
       as in Fig.~\ref{fig:fwd-phasespace} also in general kinematics, such as for example in RCS and VCS,
       although the respective physical regions are different.
       The interior of the cone always corresponds to imaginary $\lambda$, so that
        the condition~\eqref{pole-condition} becomes $\eta_-=-\delta$ and $\text{Im}\,\lambda = \mp\gamma$.
        Thus, for negative values of $\eta_-$ the resonance poles can appear in the TPE integration region and must be properly taken care of.
        This is illustrated by the vertical plane in the rightmost panel of Fig.~\ref{fig:phasespace} for an exemplary resonance.
        The poles of the nucleon itself ($\delta=\gamma=0$) intersect with the cone in the limit $\eta_-=\lambda=0$.
        In the case of VCS ($\eta_+=\omega$) this is just the limit where the generalized polarizabilities are defined (second panel in Fig.~\ref{fig:phasespace}),
        so that an extraction of polarizabilities requires a sensible subtraction of the nucleon poles contained in the nucleon Born terms.

        \subsection{Tensor basis} \label{sec:cs-tensor-basis}

        The extraction of CFFs requires a suitable tensor basis.
        While in principle the tensor decomposition is arbitrary, the choice of basis matters in practice.
        Compton scattering is characterized by 18 CFFs $c_i(\eta_+,\eta_-,\omega,\lambda)$, cf.~Eq.~\eqref{csa-decomp-01}, and thus
        it is desirable to construct a basis where these functions become as simple as possible:

        (i) Gauge invariance must be properly implemented, which reduces the number of CFFs from 32 to 18.
            Below we write down a basis where transversality is manifest.

        (ii) The Compton amplitude is invariant under photon crossing and charge conjugation:
             \begin{equation}\label{bose+cc}
             \begin{split}
                 \Gamma^{\mu\nu}(p,Q',Q) &\stackrel{!}{=} \Gamma^{\nu\mu}(p,-Q,-Q')\,, \\  
                 \Gamma^{\mu\nu}(p,Q',Q) &\stackrel{!}{=} C\,\Gamma^{\nu\mu}(-p,-Q,-Q')^T C^T\,,
             \end{split}
             \end{equation}
             where $C=\gamma^4 \gamma^2$ is the charge-conjugation matrix and the superscript $T$ denotes a matrix transpose.
             Implementing these properties already at the level of the basis elements
             simplifies the discussion because the resulting CFFs
             can depend on the variables $\omega$ and $\lambda$ only quadratically: $c_i(\eta_+,\eta_-,\omega^2,\lambda^2)$.

        (iii) To make the CFFs dimensionless we
              divide the basis tensors by powers of the nucleon mass $m$.

        (iv)  The CFFs should be free of kinematic singularities; analyticity then implies that their only singularities are physical poles and cuts.
              In Fig.~\ref{fig:phasespace} the dominant poles are: the $s$- and $u$-channel nucleon and nucleon resonance poles;
              the lowest vector-meson poles in $Q^2$ and ${Q'}^2$, which live on planes parallel to the VCS plane (and their mirror planes with $\omega\to -\omega$);
              and the lowest $t$-channel meson poles on planes parallel to the forward plane outside of the cone.
              Being free of kinematic effects has several practical advantages;
              it can simplify the momentum dependence of the CFFs, as
              their dependence on the four variables $\eta_+$, $\eta_-$, $\lambda$ and $\omega$ effectively often
              collapses into a one-dimensional dependence on $\eta_+$.  
              The absence of kinematic dependencies in the CFFs is tied to using a `minimal' basis,
              which is characterized by having no kinematic singularities and featuring the lowest possible powers in the photon momenta.
              Such bases have been frequently used in the literature following the works by Bardeen, Tung and Tarrach~\cite{Bardeen:1969aw,Tarrach:1975tu}.

        Without reference to the separation into Born and 1PI terms in Fig.~\ref{fig:qcv-born},
        one can generally decompose the onshell Compton amplitude into three contributions:
        \begin{equation}\label{generic-decomposition}
            \Gamma^{\mu\nu} =  \Gamma^{\mu\nu}_\text{G} + \Gamma^{\mu\nu}_\perp + \Gamma^{\mu\nu}_{\perp\perp} \,,
        \end{equation}
        which are distinguished by their transversality properties.
        $\Gamma_{\perp\perp}^{\mu\nu}$ is transverse with respect to both photon momenta,
        \begin{equation}\label{gi-01}
            {Q'}^\mu\,\Gamma_{\perp\perp}^{\mu\nu} = 0\,, \qquad
            \Gamma_{\perp\perp}^{\mu\nu} Q^\nu = 0\,,
        \end{equation}
        whereas $\Gamma_\perp^{\mu\nu}$ is subject to the weaker constraint  
        \begin{equation}
            {Q'}^\mu\,\Gamma_\perp^{\mu\nu} Q^\nu = 0
        \end{equation}
        and the remaining `gauge part' $\Gamma_\text{G}^{\mu\nu}$ is not transverse.

        The physical Compton amplitude is gauge invariant, so it must satisfy~\eqref{gi-01} and thus only
        $\Gamma_{\perp\perp}^{\mu\nu}$ survives.
        The full amplitude depends on 32 independent Lorentz-Dirac tensors,
        18 of which belong to $\Gamma_{\perp\perp}^{\mu\nu}$, 12 to $\Gamma_{\perp}^{\mu\nu}$ and two to $\Gamma_\text{G}^{\mu\nu}$.  
        In the main text we will only consider the physical, transverse part $\Gamma_{\perp\perp}^{\mu\nu}$ which depends on 18 tensors.
        However, to quantify a potential loss of gauge invariance it is also useful to work out the remaining non-transverse terms
        $\Gamma_\text{G}^{\mu\nu}$ and $\Gamma_\perp^{\mu\nu}$, which is done in App.~\ref{sec:non-transverse-basis}.

             \begin{table}[t]

             \begin{equation*}
             \begin{array}{ l @{\quad} |  @{\quad} l   @{\;\;}  }

                 n & \text{Basis element} \\[1mm] \hline \rule{-0.0mm}{0.5cm}

                 2 & X_1   =\,         \mathsf{F}_{1,6} \\
                 2 & X_2   =\, \frac{1}{2} \,\mathsf{F}_{1,1}  \\
                 4 & X_3   =\,         \,\mathsf{F}_{1,2} \\
                 4 & X_4   =\,         \,\mathsf{F}_{2,6} \\
                 4 & X_5   =\,         \,\mathsf{F}_{1,9} \\[2mm]

                 2 & X_6  =\, \frac{1}{4} \,\mathsf{G}_{1,1} \\
                 3 & X_7  =\, \frac{1}{\lambda\omega} \,\mathsf{G}_{1,24}  \\
                 5 & X_8  =\, \frac{\omega}{\lambda}   \,\mathsf{G}_{1,23}  \\
                 5 & X_9  =\, \frac{\omega}{\lambda}  \,\mathsf{G}_{1,25}

             \end{array}
             \begin{array}{ | @{\quad} l @{\quad} |  @{\quad} l   }

                 n & \text{Basis element} \\[1mm] \hline \rule{-0.0mm}{0.5cm}

                 3 & X_{10}  =\,   \mathsf{F}_{1,21} - \frac{1}{4}\,\mathsf{F}_{1,34} + 2\mathsf{F}_{1,6} \\
                 3 & X_{11}  =\,   \mathsf{F}_{6,33} + \frac{1}{4}\,\mathsf{F}_{2,33} \\
                 3 & X_{12}  =\,   \mathsf{F}_{1,33} \\
                 5 & X_{13}  =\,   \mathsf{F}_{2,33} \\
                 5 & X_{14}  =\,   \mathsf{F}_{1,27}+2\mathsf{F}_{1,22} \\[2mm]

                 3 & X_{15}  =\,  \frac{1}{\lambda^2}\,\mathsf{F}_{9,33}\\
                 5 & X_{16}  =\,  \frac{1}{\lambda^2}\,\mathsf{F}_{10,33}\\

                 4 & X_{17}  =\,  \mathsf{F}_{1,23}  \\
                 6 & X_{18}  =\,  \mathsf{F}_{1,24}

             \end{array}
             \end{equation*}

               \caption{Transverse basis for the nucleon Compton amplitude.
                        $\mathsf{F}_{i,j}$ and $\mathsf{G}_{i,j}$ are defined in Eq.~\eqref{FG-Def}
                        and the explicit expressions for the first few tensors are given in Eq.~\eqref{basis-explicit}.}
               \label{transverse-basis-0}

             \end{table}

        The derivation of the 18 transverse tensors is straightforward and sketched in App.~\ref{sec:tensor-basis}. One starts from a set of 32 linearly independent elementary tensors,
        the $K_i^{\mu\nu}$ in Table~\ref{basis-type-I}, and applies the constraints~\eqref{gi-01} such that no kinematic singularities are introduced.
        In practice this means eliminating 14 CFFs without any division by kinematic factors, i.e., without introducing denominators that depend on
        $\eta_+$, $\eta_-$, $\lambda^2$, $\omega^2$, etc. Fortunately, in the case of Compton scattering this is possible and thus the procedure automatically generates a minimal basis.

        The resulting 18 transverse basis elements $X_i^{\mu\nu}$ are lengthy combinations of the $K_i^{\mu\nu}$ and given in Table~\ref{X-basis-in-terms-of-K}
        in the appendix,
        but they can be written in a compact way using the definitions
             \begin{equation}
             \begin{split}
                 t_A^{\mu\alpha\beta} &= \delta^{\mu\beta} A^\alpha-\delta^{\mu\alpha} A^\beta\,,\\
                 \varepsilon_A^{\mu\alpha\beta} &= \gamma_5 \,\varepsilon^{\mu\alpha\beta\lambda} A^\lambda\,.
             \end{split}
             \end{equation}
        These are the lowest-dimensional Lorentz tensors that are linear in the momenta and transverse without introducing kinematic singularities.
        $t_A^{\mu\alpha\beta}$ is transverse to the momentum $A^\mu$, whereas $\varepsilon_A^{\mu\alpha\beta}$ is transverse in all Lorentz indices:
        $A^\mu\,t_A^{\mu\alpha\beta}=0$, $A^\mu\,\varepsilon_A^{\mu\alpha\beta}=0$, etc.
        With their help we define Compton basis tensors of the form
             \begin{align}
                \mathsf{F}_{i,j}^{\mu\nu} &= \frac{1}{2m^2}\,t^{\mu\alpha\rho}_{Q'}\,t^{\nu\beta\sigma}_Q \left\{ K_i^{\alpha\beta},\,K_j^{\rho\sigma} \right\}, \label{FG-Def} \\
                \mathsf{G}_{i,j}^{\mu\nu} &= \frac{1}{2m^2}\,( t^{\mu\alpha\rho}_{Q'}\,\varepsilon^{\nu\beta\sigma}_Q
                + \varepsilon^{\mu\alpha\rho}_{Q'}\,t^{\nu\beta\sigma}_Q) \left\{ K_i^{\alpha\beta},\,K_j^{\rho\sigma} \right\}  \nonumber
             \end{align}
        which are dimensionless and manifestly transverse with respect to ${Q'}^\mu$ and $Q^\nu$.
        They define our transverse basis in Table~\ref{transverse-basis-0},
        with the $K_i^{\mu\nu}$ given in Table~\ref{basis-type-I}.

        To arrive at more explicit expressions, we further define
             \begin{equation} \label{new-transverse-projectors-2}
             \begin{split}
                   t_{AB}^{\mu\nu} &= t^{\mu\alpha\nu}_A  B^\alpha = A\cdot B\,\delta^{\mu\nu} - B^\mu A^\nu\,,  \\
                   \varepsilon^{\mu\nu}_{AB} &= \varepsilon^{\mu\alpha\nu}_A B^\alpha = \gamma_5\,\varepsilon^{\mu\nu\alpha\beta}A^\alpha B^\beta
             \end{split}
             \end{equation}
       where $A^\mu$, $B^\mu$ stand for the four-vectors $p^\mu$, $Q^\mu$ and ${Q'}^\mu$.
       These expressions are quadratic in the momenta~and also manifestly transverse: $t^{\mu\nu}_{AB} = t^{\nu\mu}_{BA}$ is transverse to $A^\mu$ and~$B^\nu$
       whereas $\varepsilon^{\mu\nu}_{AB}=\varepsilon^{\nu\mu}_{BA}$ is transverse to $A$ and $B$ in both Lorentz indices.
       With their help the Compton tensors in Table~\ref{transverse-basis-0} take the form
             \begin{equation}\label{basis-explicit}
             \begin{split}
                 X_1^{\mu\nu} &= \frac{1}{m^4}\,t^{\mu\alpha}_{Q'p} \, t^{\alpha\nu}_{pQ}\,,  \\
                 X_2^{\mu\nu} &= \frac{1}{m^2}\,t^{\mu\nu}_{Q'Q}\,, \\
                 X_3^{\mu\nu} &= \frac{1}{m^4}\,t^{\mu\alpha}_{Q'Q'} \, t^{\alpha\nu}_{QQ}\,,  \\
                 X_4^{\mu\nu} &= \frac{1}{m^6}\,t^{\mu\alpha}_{Q'Q'} \,p^\alpha p^\beta \, t^{\beta\nu}_{QQ}\,,  \\
                 X_5^{\mu\nu} &= \frac{\lambda}{m^4}\left( t^{\mu\alpha}_{Q'Q'} \, t^{\alpha\nu}_{pQ} + t^{\mu\alpha}_{Q'p} \, t^{\alpha\nu}_{QQ} \right), \\
                 X_6^{\mu\nu} &= \frac{1}{m^2}\,\varepsilon^{\mu\nu}_{Q'Q}\,, \\
                 X_7^{\mu\nu} &= \frac{1}{im^3} \left(t^{\mu\alpha}_{Q'Q'}\,\varepsilon^{\alpha\nu}_{\gamma Q} - \varepsilon^{\mu\alpha}_{Q'\gamma}\,t^{\alpha\nu}_{QQ}  \right), \\
                 X_8^{\mu\nu} &= \frac{\omega}{im^3} \left( t^{\mu\alpha}_{Q'Q'}\,\varepsilon^{\alpha\nu}_{\gamma Q} + \varepsilon^{\mu\alpha}_{Q'\gamma}\,t^{\alpha\nu}_{QQ}  \right), \\
             \end{split}
             \end{equation}
       etc. For $X_7^{\mu\nu}$ and $X_8^{\mu\nu}$ we have extended the definition~\eqref{new-transverse-projectors-2} to also include $\gamma-$matrices (see Eq.~\eqref{triple-commutator} for the definition of the triple commutator):
             \begin{equation*}
                   \varepsilon^{\mu\nu}_{\gamma A} = \gamma_5\,\varepsilon^{\mu\nu\alpha\beta}\gamma^\alpha A^\beta
                       = \tfrac{1}{6}\,[\gamma^\mu,\gamma^\nu,\slashed{A}] = \tfrac{1}{4}\big\{ \, [\gamma^\mu, \gamma^\nu], \,\slashed{A}\,\big\}\,.
             \end{equation*}
       Note that the denominators of $X_{7,8,9,15,16}^{\mu\nu}$ in Table~\ref{transverse-basis-0} do not lead to kinematic singularities because they
       are matched by corresponding factors from the $K_i^{\mu\nu}$ which enter in Eq.~\eqref{FG-Def}.

       By construction, all basis elements $X_i^{\mu\nu}$ and $K_i^{\mu\nu}$ are even under photon crossing and charge conjugation, i.e., they satisfy the requirements of Eq.~\eqref{bose+cc}:
             \begin{equation}\label{bose+cc2}
             \begin{split}
                 X_i^{\mu\nu}(p,Q',Q) &\stackrel{!}{=} X_i^{\nu\mu}(p,-Q,-Q')\,, \\  
                 X_i^{\mu\nu}(p,Q',Q) &\stackrel{!}{=} C\,X_i^{\nu\mu}(-p,-Q,-Q')^T C^T \,.
             \end{split}
             \end{equation}
       The systematic (anti-)\,symmetrization and use of commutators ensure that all tensors are either even or odd under these operations,
       and with appropriate prefactors $\lambda$, $\omega$ and $\lambda\omega$
       they become symmetric.
       Because these are also the symmetries of the Compton amplitude, the resulting CFFs are even in $\lambda$ and $\omega$
       so that they can depend on these variables only quadratically. Bose symmetry and charge conjugation amount to a permutation-group symmetry $S_2 \times S_2$
       and therefore the CFFs corresponding to Table~\ref{transverse-basis-0} are permutation-group singlets.

       For a given tensor $X_i^{\mu\nu}$ in Table~\ref{transverse-basis-0}, the number $n$  counts the powers in the photon momenta.
       It can be read off from the definitions~\eqref{FG-Def} and the $K_i^{\mu\nu}$ in Table~\ref{basis-type-I}:
       each four-momentum ${Q'}^\mu$, $Q^\mu$ as well as
       the Lorentz invariant $\lambda$ contribute $n=1$, whereas $\omega$, $\eta_+$ and $\eta_-$ contribute $n=2$.
       In principle this is useful for the construction of minimal bases
       characterized by the lowest overall photon momentum powers~\cite{Eichmann:2015nra}: collect all linearly independent tensors with $n=2$, then proceed to $n=3$, etc.

       For example, for Compton scattering on a scalar particle, which only involves the tensors $X_{1\dots 5}^{\mu\nu}$,
       the minimality is tied to the alignment $n=\{2,2,4,4,4\}$. On the one hand, it is not possible
       to find more than two tensors with $n=2$ unless one divides by kinematic variables, which leads to kinematic singularities in the basis elements.
       On the other hand, replacing tensors in the set by others with higher $n$ introduces kinematic singularities
       in the CFFs, because those higher momentum powers must be matched by respective denominators in the CFFs.
       For example, in Tarrach's original basis~\cite{Tarrach:1975tu} the following tensor with $n=6$ appears:
       \begin{equation}
          \mathsf{F}_{1,10}^{\mu\nu} = \frac{\lambda\omega}{m^4}\left( t^{\mu\alpha}_{Q'Q'} \, t^{\alpha\nu}_{pQ} - t^{\mu\alpha}_{Q'p} \, t^{\alpha\nu}_{QQ}  \right).
       \end{equation}
       Noting that the resulting basis is not minimal, it was
       subsequently exchanged with $X_4^{\mu\nu}=\mathsf{F}_{2,6}^{\mu\nu}$  which is still linearly independent but only has $n=4$.
       (In Tarrach's notation $X_4^{\mu\nu} \propto \tau_{19}^{\mu\nu}$ and the bracket above is identical
       to $-\tau_5^{\mu\nu}$, cf.~Table~\ref{tab:tarrach-1} in App.~\ref{appendix-TransversePart}.)
       Thus, only those transverse bases that are free of kinematic singularities
       \textit{and} satisfy $n=\{2,2,4,4,4\}$ are minimal and guarantee the absence of kinematic dependencies in the CFFs.
       (As a caveat, see the discussion below Eq.~\eqref{vvcs}.)

       Unfortunately, for the $X_{6\dots 18}^{\mu\nu}$ the counting is obscured by the contraction with the onshell projectors in~\eqref{csa-decomp-01}.
       The resulting Gordon identities can raise the photon momentum powers so that the definition of $n$ is no longer meaningful.
       Scalar Compton scattering is an exception because the first five tensors do not involve $\gamma-$matrices
       and can be pulled out from $\Lambda_+(p_f) \dots \Lambda_+(p_i)$.

       In any case, the $X_i$ basis in Table~\ref{transverse-basis-0} is minimal because no division is necessary in its derivation
       (see App.~\ref{sec:tensor-basis}).
       This is signalled by the fact that all CFFs in Tables~\ref{tab:nucleon-born}, \ref{tab:j=1/2+born} and~\ref{tab:j=3/2+born} below
       are free of kinematic singularities and no kinematic factors appear in their denominators.
       Any basis transformation
       whose determinant is a constant preserves this property, i.e.
       \begin{equation}
          X'_i = U_{ij}\,X_j\,, \qquad \det U = const.,
       \end{equation}
       because otherwise the transformation would become singular at specific kinematic points.
       The standard example of a minimal basis is Tarrach's (modified) basis~\cite{Tarrach:1975tu} which is given in Table~\ref{tab:tarrach-1}.

       We constructed the $X_i$ in Table~\ref{transverse-basis-0} to facilitate the physical interpretation:
       \begin{itemize}
       \item $X_1$ and $X_2$ are the Compton tensors that survive for a pointlike scalar particle (cf.~Sec.~\ref{sec:scalar-csa});
       \item $X_1$ and $X_{10}$ are the tensors for a pointlike fermion, such as the electron in tree-level QED (see Table~\ref{tab:nucleon-born} and Sec.~\ref{sec:nucleon-born});
       \item $X_2$ and $X_3$ are the tensors for a scalar $t-$channel exchange, i.e., the CFFs $c_2$ and $c_3$ have scalar poles (cf. Sec.~II in Ref.~\cite{Eichmann:2015nra});
       \item $X_6$ is the tensor for pseudoscalar $t-$channel exchange and therefore $c_6$ contains the pion pole.
       \end{itemize}

        \subsection{Kinematic limits} \label{sec:kinematic-limits}

        We conclude this section with a discussion of the various kinematic limits.
        As is well known~\cite{Bardeen:1969aw,Tarrach:1975tu,Drechsel:2002ar}, the 18 CFFs in general kinematics collapse into four CFFs in the forward limit,
        six CFFs in RCS and 12 CFFs in VCS. With the notation in Table~\ref{transverse-basis-0} and Eq.~\eqref{basis-explicit}
        these properties are comparatively easy to derive.

                \smallskip
               In the \underline{\textbf{RCS limit}} both photons are real ($\eta_+=\omega=0$).
               In that case all instances of $t^{\mu\alpha}_{Q'Q'}$ and $t^{\alpha\nu}_{QQ}$, which up to factors ${Q'}^2$ and $Q^2$ are the transverse projectors, vanish after
               contraction with the transverse polarization vectors:
               \begin{equation}\label{RCS-1}
               \begin{split}
                  {\varepsilon^\ast}^\mu(Q')\,t^{\mu\alpha}_{Q'Q'} = {Q'}^2  {\varepsilon^\ast}^\alpha(Q') &\stackrel{{Q'}^2=0}{\longlongrightarrow} 0\,, \\
                  t^{\alpha\nu}_{QQ}\,\varepsilon^\nu(Q) = Q^2  \varepsilon^\alpha(Q) &\stackrel{Q^2=0}{\longlongrightarrow} 0\,.
               \end{split}
               \end{equation}
               For example, one can see from Eq.~\eqref{basis-explicit} that the tensors $X_3$, $X_4$, $X_5$, $X_7$ and $X_8$ vanish in RCS. In total only six tensors are non-zero, namely
               $X_1$, $X_2$, $X_6$, $X_{10}$, $X_{11}$ and $X_{12}$, and thus the RCS amplitude is described by the corresponding six CFFs which depend on $\eta_-$ and $\lambda^2$.
               Their relations with the RCS amplitudes $A_i$ defined in Refs.~\cite{Lvov:1980wp,Lvov:1996rmi}
               can be found in Table~\ref{tab:lvov} in App.~\ref{sec:tensor-basis}.
               In the limit $\eta_-\to 0$ and $\lambda\to 0$ they are related with the nucleon's static polarizabilities:
               the electric and magnetic polarizabilities $\alpha$ and $\beta$,
               \begin{equation} \label{polarizability-1}\renewcommand{\arraystretch}{1.1}
                 \left[ \begin{array}{c} \alpha +\beta \\ \beta \end{array}\right] = -\frac{\alpha_\text{em}}{m^3}
                 \left[ \begin{array}{c} c_1 \\ c_2 \end{array}\right],
               \end{equation}
               and the four spin polarizabilities
               \begin{equation} \renewcommand{\arraystretch}{1.1}
                 \left[ \begin{array}{c} \gamma_{E1E1} \\ \gamma_{M1M1} \\ \gamma_{E1M2} \\ \gamma_{M1E2} \end{array}\right] = \frac{\alpha_\text{em}}{2m^4}
                 \left[ \begin{array}{c} c_6 + 4c_{11} - 4c_{12} \\ - c_6 -2c_{10}+4c_{12} \\  c_6 + 2c_{10} \\ -c_6 \end{array}\right].
               \end{equation}
               The forward polarizability $\gamma_0$ and so-called pion polarizability $\gamma_\pi$ are their linear
               combinations
               \begin{equation} \label{polarizability-3} \renewcommand{\arraystretch}{1.1}
                 \left[ \begin{array}{c} \gamma_0 \\ \gamma_\pi \end{array}\right] = -\frac{2\alpha_\text{em}}{m^4}
                 \left[ \begin{array}{c} c_{11} \\ c_6 +c_{10}+c_{11}-2c_{12} \end{array}\right].
               \end{equation}

               The magnitudes of the CFFs in this limit can be
               reconstructed from the experimental results for the polarizabilities
               as well as from ChPT and dispersion relations
               (see \textit{e.g.} Table~8 in Ref.~\cite{Drechsel:2002ar} and Table 4.2 in~\cite{Hagelstein:2015egb} for compilations).
               For example, the $\mathcal{O}(p^3)$ heavy-baryon ChPT calculations for the polarizabilities yield~\cite{Bernard:1991rq,Bernard:1995dp}
               \begin{equation}\label{spin-pol-chpt} \renewcommand{\arraystretch}{1.1}
               \begin{split}
                 \left[ \begin{array}{c} c_1 \\ c_2 \end{array}\right] &= -\mathcal{C} \,\frac{\pi m g_A }{4m_\pi}
                 \left[ \begin{array}{c} 11 \\ 1 \end{array}\right], \\[1mm]
                 \left[ \begin{array}{c} c_6 \\ c_{10} \\ c_{11} \\ c_{12} \end{array}\right] &= \mathcal{C}\,\frac{m^2}{m_\pi^2} 
                 \left[ \begin{array}{c} 12-g_A \\ g_A \\ -g_A \\ 0 \end{array}\right],
               \end{split}
               \end{equation}
               where the first term in $c_6$ is due to the $t-$channel pion pole.
               Here, $m_\pi$ and $f_\pi$ are the pion mass and decay constant, $g_A$ is the nucleon's axial charge and
               we abbreviated
               \begin{equation}
                  \mathcal{C} = \frac{g_A}{3}\left( \frac{m}{4\pi f_\pi}\right)^2.
               \end{equation}
               Note that the CFFs diverge with powers of $1/m_\pi$ in the chiral limit.

               \smallskip
               In the \underline{\textbf{VCS limit}} ($\eta_+=\omega$) one has ${Q'}^2=0$ and thus only the outgoing photon is real.
               Only instances of $t^{\mu\alpha}_{Q'Q'}$ vanish upon contracting with polarization vectors,
               such as $X_3$ and $X_4$ in Eq.~\eqref{basis-explicit}, whereas others such as $X_7$ and $X_8$ become linearly dependent.
               One arrives at six relations
               \begin{equation}\label{VCS-1}
                   X_3 = X_4 = X_{13} = 0\,, \qquad
                   \begin{array}{rl}
                   X_8 &= -\eta_+ X_7\,, \\
                   X_{16} &= \eta_+ X_{15}\,, \\
                   X_{18} &= -\eta_+ X_{17}
                   \end{array}
               \end{equation}
               which leaves 12 independent CFFs in VCS:
               \begin{equation}\label{CFFs-VCS}
               \begin{array}{l}
                 c_1, \\
                 c_2, \\
                 c_5,
               \end{array}\qquad
               \begin{array}{l}
                 c_6, \\
                 c_7-\eta_+ c_8, \\
                 c_9,
               \end{array}\quad
               \begin{array}{l}
                 c_{10}, \\
                 c_{11}, \\
                 c_{12},
               \end{array}\qquad
               \begin{array}{l}
                 c_{14}, \\
                 c_{15}+\eta_+ c_{16}, \\
                 c_{17}-\eta_+ c_{18}\,.
               \end{array}
               \end{equation}
               They are functions of $\eta_+$, $\eta_-$ and $\lambda^2$. In the limit $\eta_- \to 0$ and $\lambda \to 0$ they
               are related to the nucleon's generalized polarizabilities~\cite{Guichon:1995pu,Drechsel:1998zm,Drechsel:2002ar,Downie:2011mm,Lensky:2016nui},
               which can be reconstructed with the help of Table~\ref{tab:drechsel}
               in the appendix.

               \smallskip
               In the doubly-virtual \underline{\textbf{forward limit}}, which is defined by $\eta_+=\eta_-=Q^2/m^2$ and $\omega=0$,
               both photons are virtual but because of $Q^\mu = {Q'}^\mu$ many basis tensors vanish or become linearly dependent.
               In the simpler cases this can  be read off directly from Eq.~\eqref{basis-explicit}, for example
               \begin{equation}
               \begin{split}
                   X_3^{\mu\nu} &= \frac{1}{m^4}\,t^{\mu\alpha}_{QQ} \, t^{\alpha\nu}_{QQ} = \frac{Q^2}{m^4}\,t^{\mu\nu}_{QQ} = \eta_+\,X_2^{\mu\nu}\,, \\
                   X_6^{\mu\nu} &= \frac{1}{m^2}\,\varepsilon^{\mu\nu}_{QQ} = 0\,,
               \end{split}
               \end{equation}
               etc. Note also that
               \begin{equation*}
                    X_7^{\mu\nu} = \frac{1}{im^3} \left(Q^2 \,\varepsilon^{\mu\nu}_{\gamma Q} - Q^2 \, \varepsilon^{\mu\nu}_{Q\gamma} \right) = \frac{2\eta_+}{m}\,i\varepsilon^{\mu\nu}_{Q\gamma}\,,
               \end{equation*}
               so we can drop the factor $\eta_+$ and thereby define a new tensor with a lower power $n=1$.
               In total only four independent tensors survive in the forward limit:
             \begin{equation}\label{fwd-tensors-1} \renewcommand{\arraystretch}{1.4}
             \begin{array}{rl}
                 X_1^{\mu\nu} &= \displaystyle \frac{1}{m^4}\,t^{\mu\alpha}_{Qp} \, t^{\alpha\nu}_{pQ}\,,  \\[3mm]
                 \widetilde{X}_7^{\mu\nu} &= \displaystyle\frac{1}{m}\,i\varepsilon^{\mu\nu}_{Q\gamma},
             \end{array}\;\;
             \begin{array}{rl}
                 X_2^{\mu\nu} &= \displaystyle\frac{1}{m^2}\,t^{\mu\nu}_{QQ}\,, \\[3mm]
                 X_{12}^{\mu\nu} &= \displaystyle\frac{\lambda}{m^2}\left[ t^{\mu\alpha}_{Q\gamma},\,t^{\alpha\nu}_{\gamma Q}\right] ,
             \end{array}
             \end{equation}
             whereas $X_6=X_8=X_9=X_{14}=X_{16}=X_{18}=0$ and the remaining ones can be related to them:
             \begin{equation*}     \renewcommand{\arraystretch}{1.4}
                \begin{array}{rl}
                   X_3 &= \eta_+X_2\,, \\
                   X_4 &= \eta_+X_1 - \lambda^2 X_2\,, \\
                   X_5 &= -2\lambda^2 X_2\,, \\
                   X_{10} &=  - \eta_+ \widetilde{X}_7\,, \\
                \end{array}\quad
                \begin{array}{rl}
                   X_{11} &= 2\lambda^2 \widetilde{X}_7 + \frac{1}{4}\,\eta_+ X_{12} \\
                   X_{13} &= \eta_+ X_{12}\,, \\
                   X_{15} &= - 2\eta_+\widetilde{X}_7 - X_{12} \,, \\
                   X_{17} &= 2\lambda^2 X_2\,.
                \end{array}
             \end{equation*}
             The forward Compton amplitude -- more precisely, the bracket in Eq.~\eqref{csa-decomp-01} -- then becomes
             \begin{equation}
                [ \dots ] = \overline c_1\,X_1 + \overline c_2\,X_2 + \overline c_3\,\widetilde{X}_7 + \overline c_4\,X_{12}\,,
             \end{equation}
             where the four CFFs depend on 
             $\eta_+=\eta_-$ and $\lambda^2$:
             \begin{equation} \label{CFFs-FWD}
             \begin{split}
                \overline c_1 &= c_1 + \eta_+ c_4\,, \\
                \overline c_2 &= c_2 + \eta_+ c_3 - \lambda^2(c_4+2c_5-2c_{17})\,,\\
                \overline c_3 &= \eta_+(2c_7-c_{10}-2c_{15})+2\lambda^2 c_{11}\,,\\
                \overline c_4 &= c_{12}-c_{15} + \eta_+ (c_{13}+\tfrac{1}{4}\,c_{11})\,.
             \end{split}
             \end{equation}
             Their relations to the forward amplitudes $T_{1,2}$ and $S_{1,2}$ defined as in~\cite{Hagelstein:2015egb},
             whose imaginary parts are proportional to the nucleon structure functions, are given by
             \begin{equation}\label{fwd-amplitudes} \renewcommand{\arraystretch}{1.1}
                \left[ \begin{array}{c} T_1 \\ T_2 \\ S_1 \\ S_2 \end{array}\right] = -\frac{4\pi\alpha_\text{em}}{m}
                \left[ \begin{array}{c} \lambda^2 \overline c_1 + \eta_+ \overline c_2 \\ \eta_+ \overline c_1 \\ \overline c_3 \\ -2\lambda \overline c_4 \end{array}\right].
             \end{equation}
             From their expansion around $\eta_\pm  = \lambda = \omega = 0$ one can extract several further relations
              such as the one for the longitudinal-transverse polarizability $\delta_{LT}$:
             \begin{equation}
               \delta_{LT} = -\frac{2\alpha_\text{em}}{m^4}\left( c_{11} - c_{12} + c_{15} \right).
             \end{equation}

               \smallskip
               Another example is the doubly-virtual but off-forward \underline{\textbf{VVCS limit}},
               where $Q^2={Q'}^2$ and therefore \mbox{$\omega=0$}
               but $\eta_+ \neq \eta_-$.
               Also here the tensor basis becomes redundant, however in a way where kinematic
               singularities cannot be avoided.
               The characteristics already appear in scalar Compton scattering defined by the tensors $X_{1 \dots 5}$,
               cf.~Refs.~\cite{Tarrach:1975tu,Drechsel:1996ag,Birse:2012eb,Bakker:2016vin}.
               In terms of the $K_i$ from Table~\ref{basis-type-I} one can see that $K_5=K_8=K_{10}=0$ for $\omega=0$.
               At first sight this does not seem to affect the $X_i$ because Table~\ref{X-basis-in-terms-of-K} still implies
               \begin{equation}\label{vvcs}
               \begin{split}
                 X_1    &=   \lambda^2 \,K_1 + \eta_- K_6 + K_7\,, \\
                 X_2    &=  \eta_- K_1 - K_3\,, \\
                 X_3    &=  \eta_+^2 K_1 + \eta_- K_2 - \eta_+ K_4 \,, \\
                 X_4    &=  \lambda^2 K_2 + \eta_+^2 K_6 + \eta_+ K_9 \,, \\
                 X_5    &=  -\lambda^2\, (2\eta_+ K_1 - K_4) - \eta_+ K_7  + \eta_- K_9 \,.
               \end{split}
               \end{equation}
               However, the combination\footnote{This is the tensor $\lambda\omega \tau_5/m^4$ in Tarrach's basis, cf.~Table~\ref{tab:tarrach-1},
               which vanishes for $\omega=0$.}
               \begin{equation}
                 \eta_+^2 X_1 + \lambda^2 X_3 - \eta_- X_4 + \eta_+ X_5 = 0
               \end{equation}
               vanishes, as one can verify, and thus one has a non-trivial relation between the $X_i$
               which cannot be solved without introducing kinematic singularities.
               In the limit $\omega=0$ one then needs a redundant basis to avoid them.
               In general kinematics there is no problem: the 18 CFFs and corresponding tensors are regular in the limit $\omega\to 0$;
               only when they collapse into fewer independent functions those functions can acquire kinematic singularities.

   \newpage

        \section{Scalar Compton amplitude} \label{sec:scalar-csa}

        To illustrate the procedure of working out the resonance contributions,
        we start with the tree-level Compton amplitude for a pointlike scalar particle
        as a template; see also Refs.~\cite{Drechsel:1996ag,Fearing:1996gs,Bakker:2016vin}.
        With the momentum definitions in Eq.~\eqref{kinematics-1}
        the Born terms in Fig.~\ref{fig:qcv-born} read
             \begin{equation*}
             \begin{split}
                 \!\!\Gamma_\text{B}^{\mu\nu}(p,Q,Q')
                                  &= \overline{\Gamma}^\mu(p_i^+,Q')\,D(p+\Sigma)\,\Gamma^\nu(p_f^+,Q) \\
                                  & +\overline{\Gamma}^\nu(p_f^-,-Q)\,D(p-\Sigma)\,\Gamma^\mu(p_i^-,-Q').
             \end{split}
             \end{equation*}
        The scalar propagators depend on the $s$- and $u$-channel momenta $p\pm \Sigma$;
        at tree level they are given by
        \begin{equation}
           D(p \pm \Sigma) = \frac{1}{(p\pm\Sigma)^2+m^2} = \frac{1}{m^2}\,\frac{1}{\eta_-\mp 2\lambda}\,.
        \end{equation}
        The arguments of the scalar-photon vertices
        are the photon momenta $Q$, $Q'$ and the average momenta of the scalar particle:
             \begin{equation}\label{born-kinematics}
                 p_f^\pm = p \pm \frac{Q'}{2}\,, \qquad
                 p_i^\pm = p \pm \frac{Q}{2}\,.
             \end{equation}
        The tree-level vertex is $\Gamma^\mu(k,Q) = 2k^\mu$ and its charge conjugate is defined as $\overline{\Gamma}^\mu(k,Q) = \Gamma^\mu(-k,-Q) = -2k^\mu$.

        The Born contribution thus becomes
             \begin{equation}
             \begin{split}
                 \Gamma_\text{B}^{\mu\nu}  &= -\frac{1}{m^2}\left[ \frac{ p_i^{+\mu}  p_f^{+\nu}} {\eta_--2\lambda} + \frac{ p_i^{-\mu} p_f^{-\nu} }{\eta_-+2\lambda} \right] \\
                                  &= -\frac{8}{\eta_-^2-4\lambda^2}\left[ \eta_-\left( \frac{K_3^{\mu\nu}}{4} + K_6^{\mu\nu} \right) + K_7^{\mu\nu} \right],
             \end{split}
             \end{equation}
        where the $K_i^{\mu\nu}$ are the tensors from Table~\ref{basis-type-I}:  
        \begin{equation}
        \begin{split}
             K_3^{\mu\nu} &= \frac{Q^\mu {Q'}^\nu}{m^2}, \\
             K_6^{\mu\nu} &= \frac{p^\mu {p}^\nu}{m^2}, \\
             K_7^{\mu\nu} &= \frac{\lambda}{m^2}\,(p^\mu\,{Q'}^\nu + Q^\mu\,p^\nu)\,.
        \end{split}
        \end{equation}
        Note that the $s$ and $u$-channel poles at $\eta_-=\pm 2\lambda$ enter the denominator in combination
        and thereby ensure crossing symmetry.

        Comparing this with Table~\ref{X-basis-in-terms-of-K}, we can recast the result in terms of the transverse tensors $X_i^{\mu\nu}$ from Table~\ref{transverse-basis-0},
        \begin{equation}\label{scalar-result-1}
           \Gamma_\text{B}^{\mu\nu} = \frac{1}{\eta_-^2-4\lambda^2}\left( \widetilde{c}_1\, X_1^{\mu\nu} + \widetilde{c}_2\, X_2^{\mu\nu} \right) - 2K_1^{\mu\nu}\,,
        \end{equation}
        and read off the resulting CFF residues:
        \begin{equation}
           \widetilde{c}_1 = -8, \qquad
           \widetilde{c}_2 =  2\eta_-\,.
        \end{equation}

        The Born term is not gauge invariant due to the remainder proportional to $K_1^{\mu\nu}=\delta^{\mu\nu}$,
        but this is only so because the scalar theory has a pointlike seagull interaction similar to the rightmost diagram in Fig.~\ref{fig:qcv-born}:
        \begin{equation}\label{scalar-result-2}
             \Gamma^{\mu\nu}_\text{1PI}(p,Q,Q') = 2\delta^{\mu\nu}.
        \end{equation}
        Adding it cancels the gauge part and ensures that the full Compton amplitude $\Gamma^{\mu\nu}_\text{B} + \Gamma^{\mu\nu}_\text{1PI}$ is transverse.
        As a result, it is completely specified by $\widetilde{c}_1$ and $\widetilde{c}_2$.

             \renewcommand{\arraystretch}{1.4}

       One could generalize the discussion by calculating corrections to the propagator, the vertex, and the 1PI structure part, for example in an effective field theory.
       As long as the theory respects electromagnetic gauge invariance, the resulting Compton amplitude is fully transverse.
       The most general form of the offshell vertex allowed by gauge invariance, which is free of kinematic singularities, is
       \begin{equation}\label{scalar-offshell-vertex}
          \Gamma^\mu(k,Q) = 2f_1\,k^\mu + f_2\,t^{\mu\nu}_{QQ} k^\nu\,.
       \end{equation}
       $t^{\mu\nu}_{QQ}$ is defined in~\eqref{new-transverse-projectors-2} and $f_1$, $f_2$ are functions of $k^2$, $w=k\cdot Q$ and $Q^2$.
       The form factor $f_1$ is determined by the Ward-Takahashi identity (WTI)
       \begin{equation}\label{scalar-wti}
       \begin{split}
           & Q^\mu \Gamma^\mu(k,Q) = D(k_+)^{-1} - D(k_-)^{-1}  \\
           & \Rightarrow \;  f_1 = \frac{D(k_+)^{-1} - D(k_-)^{-1}}{k_+^2-k_-^2}\,,
       \end{split}
       \end{equation}
       with $k_\pm = k \pm Q/2$ and thus $k_+^2-k_-^2 = 2k\cdot Q$, so that only $f_2$ carries dynamical information.

       The recipe for deriving Eq.~\eqref{scalar-offshell-vertex} is the same as for the more complicated cases in the
       following sections, such as the nucleon-photon vertex in Sec.~\ref{sec:spin-1/2}, the nucleon-to-resonance transition vertices in Secs.~\ref{sec:spin-1/2r} and~\ref{sec:spin-3/2},
       and finally the Compton amplitude in App.~\ref{sec:tensor-basis}. We start with the general decomposition
       \begin{equation}
          \Gamma^\mu(k,Q) = a_1\,k^\mu + a_2\,w\,Q^\mu\,,
       \end{equation}
       where $w=k\cdot Q$ ensures the correct charge-conjugation parity: $\overline\Gamma^\mu(k,Q) = -\Gamma^\mu(k,-Q)$.
       As a consequence, $a_1$ and $a_2$ are even in $w$ and only depend on $w^2$.
       Next, we derive the transverse part of the vertex by solving
       \begin{equation}
          Q^\mu \Gamma^\mu(k,Q) = w \,(a_1 + a_2\,Q^2) = 0\,.
       \end{equation}
       This must be done without introducing kinematic singularities, i.e., we must solve for $a_1$ (and not $a_2$) which leads to
       the transverse part $\propto t^{\mu\nu}_{QQ} \,k^\nu$. Relaxing again the transversality constraint,
       we then add the term $\propto k^\mu$ that we eliminated (and not $Q^\mu$), which constitutes the gauge part and leads to the result~\eqref{scalar-offshell-vertex}.
       Finally, solving the WTI in Eq.~\eqref{scalar-wti} determines the coefficient $f_1$.

       The same procedure is carried out in App.~\ref{sec:tensor-basis} to derive the tensor basis for the Compton amplitude itself, although in that case
       the gauge parts must vanish because the amplitude is transverse.
       In general neither the Born terms nor the structure part alone are gauge invariant,
       but one can project them onto a complete basis  
       where the sum of the gauge parts must cancel in the end like in the simple case~(\ref{scalar-result-1}--\ref{scalar-result-2}).

       In the following we are interested in the nucleon Born terms
       and nucleon resonance contributions to Compton scattering.
       In those cases one can enforce gauge invariance from the beginning by imposing
       appropriate constraints on the vertices (which is also possible because there is no seagull term for fermions).


    \newpage

    \renewcommand{\arraystretch}{1.4}

        \section{Nucleon Born term} \label{sec:spin-1/2}

         Returning to nucleon Compton scattering,
         the Born term for the nucleon has the form
             \begin{equation}\label{qcv-born}
             \begin{split}
                 \Gamma_\text{B}^{\mu\nu}  &= \Lambda_+(p_f) \,\Big[ \overline{\Gamma}^\mu(p_i^+,Q')\,S(p+\Sigma)\,\Gamma^\nu(p_f^+,Q) \\
                                                    & +\overline{\Gamma}^\nu(p_f^-,-Q)\,S(p-\Sigma)\,\Gamma^\mu(p_i^-,-Q') \Big] \,\Lambda_+(p_i).
             \end{split}
             \end{equation}
         Here, $\Gamma^\mu(k,Q)$ is the dressed offshell nucleon-photon vertex that depends on the average nucleon momentum $k$ and
         the total photon momentum $Q$. $p_f^\pm$ and $p_i^\pm$ were defined in~\eqref{born-kinematics}.
         The charge-conjugate vertex is
            \begin{equation}\label{cc-vertex}
                \overline{\Gamma}^\mu(k,Q) = C\,\Gamma^\mu(-k,-Q)^T C^T\,,
            \end{equation}
         where $C=\gamma^4 \gamma^2$ is the charge-conjugation matrix that satisfies $C^T=C^{-1}=-C$.
         The charge-conjugation symmetry of the nucleon-photon vertex amounts to
            \begin{equation}\label{cc-invariance}
                \overline{\Gamma}^\mu(k,Q) = - \Gamma^\mu(k,-Q)\,.
            \end{equation}
         The Born term~\eqref{qcv-born} shares the symmetries of the full Compton amplitude, namely
         Bose (photon-crossing) and charge-conjugation invariance as in Eq.~\eqref{bose+cc}.

         The nucleon propagator and its inverse are given by
         \begin{equation}\label{spin-1/2-propagator}
         \begin{split}
             S(k) &= \frac{1}{A(k^2)}\,\frac{-i\slashed{k} + M(k^2)}{k^2+M(k^2)^2}, \\
             S(k)^{-1} &= A(k^2)\,\left(i\slashed{k} + M(k^2)\right),
         \end{split}
         \end{equation}
         where $M(k^2)$ and $A(k^2)$ are momentum-dependent functions.
         In practice we treat the nucleon
         as a constituent-like particle and set $M(k^2)=m$ and $A(k^2)=1$, which holds on the mass shell $k^2=-m^2$,
         but to keep the discussion general we will retain the momentum dependence in the following two subsections.

     \begin{figure}[t]
     \center{
     \includegraphics[scale=0.07]{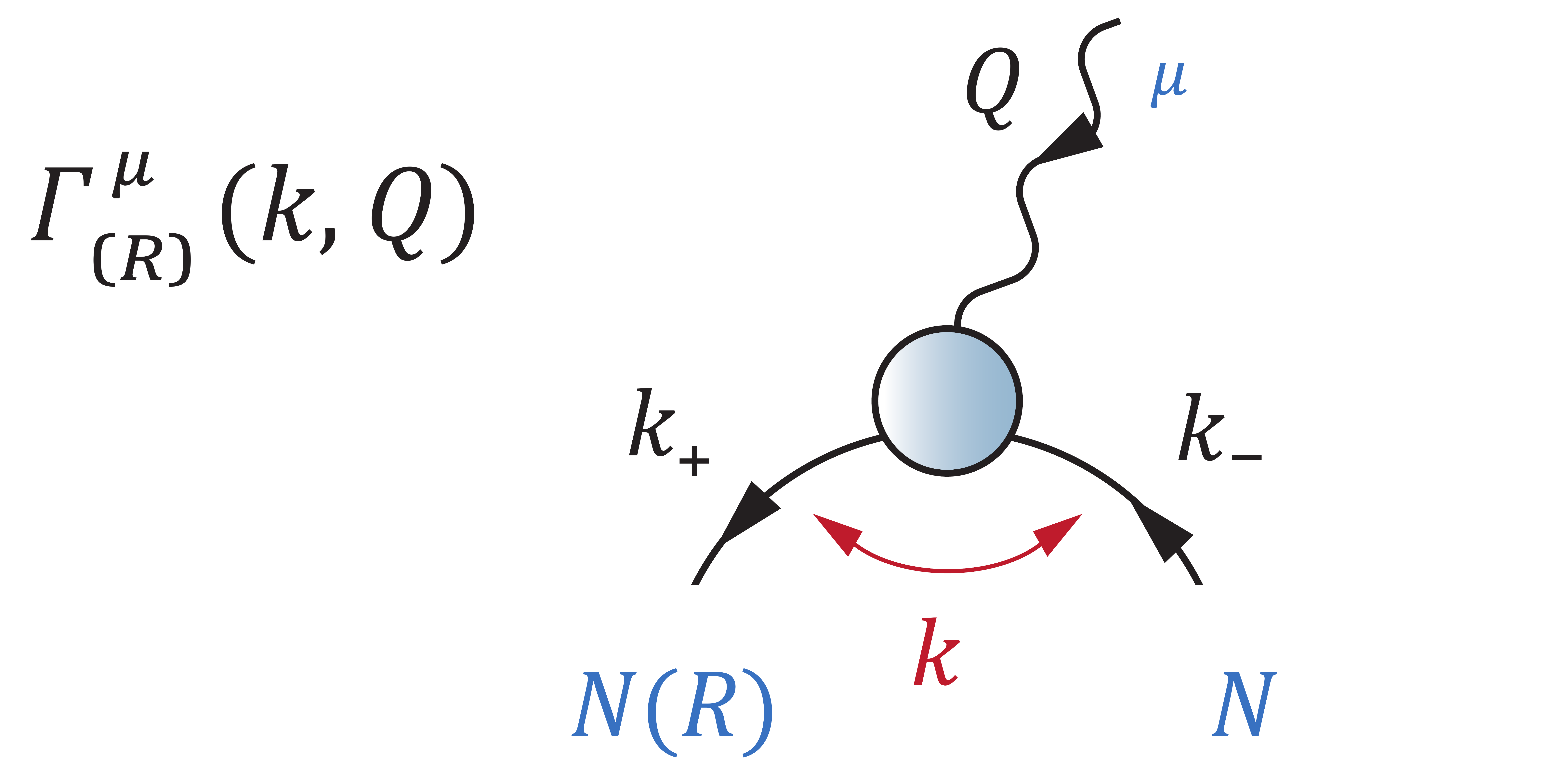}}
        \caption{Kinematics in the nucleon-photon and $N\to \tfrac{1}{2}^\pm$ transition vertex.}
        \label{fig:npv-kinematics}
     \end{figure}

        \subsection{Offshell nucleon-photon vertex}

         First we derive the general form of the offshell nucleon-photon vertex.
         The discussion is based on the quark-photon vertex~\cite{Kizilersu:1995iz,Skullerud:2002ge,Eichmann:2012mp} but it can be equally applied
         to nucleon resonances.
         The kinematics are sketched in Fig.~\ref{fig:npv-kinematics}; $k_\pm = k \pm Q/2$ are the incoming and outgoing nucleon momenta.
         Electromagnetic gauge invariance leads to a Ward-Takahashi identity (WTI) for the vertex,
                  \begin{equation}\label{qpv-wti}
                      Q^\mu \,\Gamma^\mu (k,Q) = \mathcal{Z} \left[ S(k_+)^{-1}-S(k_-)^{-1} \right],
                  \end{equation}
         with $\mathcal{Z}=1$ ($\mathcal{Z}=0$) for the proton (neutron).
         It can thus be written as the sum of a `gauge part' and a transverse part, where the former is constrained by the WTI:
         \begin{equation}\label{npv-decomp-1}
            \Gamma^\mu(k,Q) = \Gamma^\mu_\text{G}(k,Q) + \Gamma^\mu_\perp(k,Q)\,.
         \end{equation}
         In the case of a nucleon resonance the r.h.s. of Eq.~\eqref{qpv-wti} is zero and the vertex is purely transverse.

         To derive both contributions, we start from
         the general offshell fermion-photon vertex for a spin-1/2 particle:
         \begin{equation}\label{full-npv-1}
            \Gamma^\mu(k,Q) = \sum_{n=1}^{12} h_n(k^2, w, Q^2)\,i\tau_n^\mu(k,Q)\,.
         \end{equation}
         The $h_i$ are Lorentz-invariant functions, with $w=k\cdot Q$, and the $\tau_i^\mu$ are the 12 possible tensors permitted by Lorentz covariance and parity invariance:
            \begin{equation}\label{qpv-general-basis}
            \begin{array}{r @{\!\;} l}
                      & \gamma^\mu    \\
                   iw & \left[ \gamma^\mu, \slashed{k} \right]   \\
                   i  & \left[ \gamma^\mu, \slashed{Q} \right] \\
                      & \left[ \gamma^\mu, \slashed{k}, \slashed{Q} \right]
            \end{array}\qquad
            \begin{array}{r @{\!\;} l}
                    i & k^\mu \\
                      & k^\mu \slashed{k}\\
                    w & k^\mu \slashed{Q}\\
                    i & k^\mu [ \slashed{k},\slashed{Q}]
            \end{array}\qquad
            \begin{array}{r @{\!\;} l}
                   iw & Q^\mu\\
                    w & Q^\mu \slashed{k}  \\
                      & Q^\mu \slashed{Q}  \\
                   iw & Q^\mu [ \slashed{k},\slashed{Q}] .
            \end{array}
            \end{equation}
            We took commutators and attached factors of $w$ to ensure that they all
            share the charge-conjugation symmetry~\eqref{cc-invariance} of the full vertex,
        so that the $h_i$ are even in $w$ and only depend on $w^2$.
        We label the tensors column-wise:
        $\tau_{1\dots 4}^\mu$, $\tau_{5\dots 8}^\mu$ and $\tau_{9\dots 12}^\mu$ are the elements in the first, second and third column,
        respectively.

         To derive the transverse part of the vertex we work out the condition $Q^\mu \,\Gamma^\mu = 0$.
         The contraction produces four independent tensors $\sim\mathds{1}$, $\slashed{Q}$, $\slashed{k}$, $[\slashed{k},\slashed{Q}]$
         and thus four relations between the dressing functions, which must be solved so that no kinematic singularities are introduced in the process.
         The result
         \begin{equation} \renewcommand{\arraystretch}{1.2}
         \begin{array}{rl}
             h_1 &= -w^2 \,h_7 - Q^2\,h_{11}\,, \\
             h_2 &= h_8 + Q^2\,h_{12}\,,
         \end{array}\qquad
         \begin{array}{rl}
             h_5 &= -Q^2\,h_9\,, \\
             h_6 &= -Q^2\,h_{10}
         \end{array}
         \end{equation}
         is \textit{almost} unique: without dividing by factors of $Q^2$ or $w^2$, our only freedom is to solve for either $h_2$ or $h_8$.
         Substitution into~\eqref{full-npv-1} yields the transverse vertex
         \begin{equation}\label{npv-transverse}
            \Gamma^\mu_\perp(k,Q) = \sum_{n=1}^8 f_n(k^2,w, Q^2)\,i T_n^\mu(k,Q)\,,
         \end{equation}
         where the $f_n$ are the remaining independent functions and
          $T_n^\mu$ the transverse tensors in Table~\ref{n-offshell}.
        This defines a minimal basis where transversality and analyticity are manifest:
        the $T_n^\mu$ are transverse and regular for $Q^\mu \rightarrow 0$ and the
         $f_n$ are free of kinematic singularities at $Q^2\rightarrow 0$ and kinematically independent.

        \begin{table}[t]

             \begin{equation*}  \renewcommand{\arraystretch}{1.6}
             \begin{array}{r@{\!\;\,}l   }

                 m^2\,T_1^\mu &= t^{\mu\nu}_{QQ}\,\gamma^\nu  \\
                 m^5\,T_2^\mu &= t^{\mu\nu}_{QQ}\,w  \tfrac{i}{2} [\gamma^\nu,\Slash{k}]  \\
                 m \,T_3^\mu &= \tfrac{i}{2}\,[\gamma^\mu,\slashed{Q}]  \\
                 m^2\,T_4^\mu &= \tfrac{1}{6}\,[\gamma^\mu, \slashed{k}, \slashed{Q}]

             \end{array} \qquad\quad
             \begin{array}{r@{\!\;\,}l   }

                  m^3\,T_5^\mu &= t^{\mu\nu}_{QQ}\,ik^\nu  \\
                  m^4\,T_6^\mu &= t^{\mu\nu}_{QQ}\,k^\nu \Slash{k}  \\
                  m^4\,T_7^\mu &= t^{\mu\nu}_{Qk}\,w \gamma^\nu  \\
                  m^3\,T_8^\mu &= t^{\mu\nu}_{Qk}\,\tfrac{i}{2}\,[\gamma^\nu,\Slash{k}]

             \end{array}
             \end{equation*}
             \begin{equation*}  \renewcommand{\arraystretch}{1.4}
             \begin{array}{r@{\!\;\,}l   }

                 G_1^\mu &= \gamma^\mu  \\
                 m^2\,G_2^\mu &= k^\mu \slashed{k}

             \end{array} \qquad\quad
             \begin{array}{r@{\!\;\,}l   }

                 m\,G_3^\mu &= ik^\mu \\
                 m^3\,G_4^\mu &= w \tfrac{i}{2} [\gamma^\mu,\slashed{k}]

             \end{array}
             \end{equation*}

               \caption{\textit{Top:} Eight tensors $T_i^\mu$ constituting the transverse part $\Gamma_\perp^\mu(k,Q)$ of the offshell nucleon-photon vertex
                        without introducing kinematic singularities.
                        $t_{AB}^{\mu\nu}$ and the triple commutator are defined in Eqs.~\eqref{new-transverse-projectors-2}
                        and~\eqref{triple-commutator}.
                        \textit{Bottom:} Four tensors $G_i^\mu$ of the gauge part $\Gamma_\text{G}^\mu(k,Q)$.
                        We attached powers of the nucleon mass $m$ to make all tensors dimensionless.}
               \label{n-offshell}

             \end{table}

    \renewcommand{\arraystretch}{1.3}

         The remaining gauge part in~\eqref{npv-decomp-1} can only depend on the tensors for $h_1$, $h_2$, $h_5$, and $h_6$
         that we eliminated under the assumption that the vertex was transverse; these are the $G_i^\mu$ in Table~\ref{n-offshell}.
         Putting them back into the WTI~\eqref{qpv-wti} together with the nucleon propagator~\eqref{spin-1/2-propagator}
         determines their coefficients and leads to the Ball-Chiu vertex~\cite{Ball:1980ay}:
                  \begin{equation*}\label{vertex:BC}
                      \Gamma_\text{G}^\mu(k,Q) =   i\mathcal{Z} \left[ \Sigma_A \, G_1^\mu + 2m^2 \Delta_A \, G_2^\mu  -2m \Delta_B \, G_3^\mu \right].
                  \end{equation*}
      It is fully specified by the nucleon propagator and depends on sums and difference quotients of the propagator dressing functions
      $A(k^2)$ and $M(k^2)$:
      \begin{equation}
          \Sigma_F = \frac{F(k_+^2) + F(k_-^2)}{2}, \;\;
          \Delta_F = \frac{F(k_+^2) - F(k_-^2)}{k_+^2-k_-^2},
      \end{equation}
      where $F \in \{A,B\}$ and $B(k^2)=A(k^2)M(k^2)$.
      Note that $G_4^\mu$ drops out as a consequence
      of electromagnetic gauge invariance.
      For a tree-level nucleon propagator the gauge part reduces to $\Gamma_\text{G}^\mu(k,Q) =  \mathcal{Z}\, i\gamma^\mu$.

      We should emphasize that the gauge part is \textit{not} longitudinal. One could equally split the vertex into longitudinal and transverse parts,
      where the longitudinal tensors are proportional to $Q^\mu$  and defined by the rightmost column in~\eqref{qpv-general-basis}.
      In that case the WTI would still only affect the longitudinal part,
      but because the transverse projector has a kinematic singularity at $Q^2=0$ the longitudinal and transverse dressing functions
      would become kinematically related at the origin and/or show kinematic zeros.
      Thus, in analogy to Eq.~\eqref{scalar-offshell-vertex} for the scalar vertex,
      only the separation into $\Gamma^\mu_\text{G}$ and $\Gamma^\mu_\perp$ ensures that the resulting dressing functions are truly
      kinematically independent.

        \subsection{Onshell nucleon-photon current}

        The onshell current follows from sandwiching the vertex between nucleon spinors (or positive-energy projectors)
        and taking the nucleon momenta onshell:
        \begin{equation}\label{onshell-nucleon}
           J^\mu(k,Q) = \Lambda_+(k_+)\,\Gamma^\mu(k,Q)\,\Lambda_+(k_-)\big|_\text{onshell} \,.
        \end{equation}
        The limit $k_+^2=k_-^2=-m^2$ entails $k^2=-m^2-Q^2/4$ and $w=0$, so the only remaining independent variable is $Q^2$.
        The 12 offshell tensors
        collapse into two, $G_1^\mu$ and $T_3^\mu$,
        by means of the Gordon identities in Table~\ref{n-onshell-gordon}. The current takes the standard Dirac form:
             \begin{equation}\label{current-onshell-standard}
                 J^\mu(k,Q) =  i\Lambda_+(k_+)\left[ F_1\,\gamma^\mu + \frac{iF_2}{4m}\,[\gamma^\mu,\slashed{Q}]\right]\Lambda_+(k_-)\,,
             \end{equation}
        where $F_1(Q^2)$ and $F_2(Q^2)$ are the onshell Dirac and Pauli form factors.

        \begin{table}[t]

             \begin{equation*} \renewcommand{\arraystretch}{1.4}
             \begin{array}{l   }

                 T_1 - 4\tau\,G_1  \\
                 T_2 \\
                 T_4 + 2\tau\,G_1 + T_3 \\
                 T_5 + 4\tau \left( G_1 - \tfrac{1}{2}\,T_3 \right) \\
                 T_6 + 4\tau \left( G_1 - \tfrac{1}{2}\,T_3 \right) \\

             \end{array} \qquad\quad
             \begin{array}{l   }

                  T_7 \\
                  T_8 - 2\tau \left( G_1 - \tfrac{1}{2}\,T_3 \right) \\
                  G_2 + G_1 - \tfrac{1}{2}\,T_3  \\
                  G_3 + G_1 - \tfrac{1}{2}\,T_3  \\
                  G_4

             \end{array}
             \end{equation*}

               \caption{Combinations of tensors that vanish
                        in the onshell projection~\eqref{onshell-nucleon}. We abbreviated $\tau=Q^2/(4m^2)$.}
               \label{n-onshell-gordon}

             \end{table}

        Even though the offshell current has a gauge part, it becomes `accidentally' transverse in the onshell projection~\eqref{onshell-nucleon}:
        \begin{equation}
            Q^\mu\,\Lambda_+(k_+) \,\gamma^\mu \Lambda_+(k_-) \big|_\text{onshell} = 0\,,
        \end{equation}
        and the same is true for the remaining tensors $G_{2,3,4}^\mu$ because $w=k\cdot Q=0$ on the mass shell.
        It follows from Table~\ref{n-onshell-gordon} that the gauge part contributes to $F_1$ and $F_2$:
             \begin{equation}\label{qpv-onshell-dressing-functions-gauge}
             \begin{split}
                 F_1(Q^2) &= \mathcal{Z} \left[ \Sigma_A + 2m\,(\Delta_B-m \Delta_A) \right] + \dots , \\
                 F_2(Q^2) &= -2m\,\mathcal{Z}\,(\Delta_B-m \Delta_A) + \dots ,
             \end{split}
             \end{equation}
        where the dots refer to the transverse pieces.
        On the mass shell, however, the nucleon propagator is that of a free particle
        and therefore $\Sigma_A = 1$ and $\Delta_A = \Delta_B = 0$.
        As a result, the Dirac and Pauli form factors are related with the offshell dressing functions via  
             \begin{equation}\label{qpv-onshell-dressing-functions}
             \begin{split}
                 F_1(Q^2) &= \mathcal{Z} + \frac{Q^2}{m^2} \left[ f_1 - \frac{f_4}{2} - \left(f_5 + f_6 - \frac{f_8}{2}\right) \right] , \\
                 F_2(Q^2) &= 2\,(f_3- f_4) + \frac{Q^2}{m^2} \left( f_5 + f_6 - \frac{f_8}{2}\right) ,
             \end{split}
             \end{equation}
        with the $f_n$ evaluated at the onshell point.

        \subsection{Compton form factors} \label{sec:nucleon-born}

         With the offshell nucleon-photon vertex at hand,
         we proceed to work out the Compton form factors for the nucleon Born term~\eqref{qcv-born}.
         Although the Born term does not contribute to nucleon polarizabilities,
         it is still relevant for two-photon exchange effects to form factors. 

         We restrict ourselves to the tree-level propagator
         \begin{equation}\label{ }
             S(k) = \frac{-i\slashed{k} + m}{k^2+m^2} \,,
         \end{equation}
        and instead of the full vertex in Eq.~\eqref{npv-decomp-1} we employ the `Dirac form' for the offshell nucleon-photon vertex:
        \begin{equation}\label{nucleon-off-curr-1}
            \Gamma^\mu(k,Q) = i\left( F_1 \,G_1^\mu + F_2 \,\frac{T_3^\mu}{2}  \right).
        \end{equation}
         With the definition of charge conjugation in Eq.~\eqref{cc-vertex},
         the charge-conjugate vertex $\overline{\Gamma}^\mu(k,Q)$ differs from the above only by a minus sign in front of the $F_1$ term.

        While the Dirac and Pauli form factors $F_i(k^2, w, Q^2)$ in these expressions are offshell,
        we will identify them with their onshell expressions $F_i(Q^2)$
        since this is the only information we can gather from experiments.
        Employing the Dirac form is also the minimal requirement for keeping the Born term gauge invariant~\cite{Scherer:1996ux}.  
        We would lose transversality if we
        \begin{itemize}
        \item equipped $F_{1,2}(Q^2)$ with a $k^2$ or $w$ dependence,
        \item added other tensors $G_{2,3,4}^\mu$ from Table~\ref{n-offshell},
        \item but also other tensors $T_i^\mu$ (except for $T_1^{\mu}$)
              because they lead to interference terms with $G_1^\mu$ from the second vertex,
        \item or if we implemented momentum-dependent dressing functions in the nucleon propagator with ramifications for the gauge part of the vertex.  
        \end{itemize}
        This is all due to the gauge part in the vertex and does not happen for the nucleon resonances which we consider later.
        It is also not a serious conceptual problem because the two-photon WTI allows one to construct a gauge-invariant completion of the Born term
        for a general offshell nucleon-photon vertex and nucleon propagator, which can be found in Ref.~\cite{Eichmann:2012mp}.  
        A simpler alternative is to project the (non gauge-invariant) Born term
        onto a full basis and  afterwards retain only the transverse part, since
        all non-transverse pieces must cancel when they are calculated from some consistent underlying theory.
        We will not further pursue this here and instead provide examples in App.~\ref{sec:gi-broken}.

        \begin{table}[t]
             \begin{equation*}
             \renewcommand{\arraystretch}{1.5}
             \begin{array}{rl   }

                 \widetilde c_1 &= - (4H_1 + \eta_- H_2)  \\
                 \widetilde c_2 &= -\eta_-(H_2+2H_3)+\lambda^2 H_2  \\
                 \widetilde c_6 &=  \eta_-(H_3-\tfrac{1}{4}\,\eta_+ H_2) \\
                 \widetilde c_{10} &=  2\,(H_1+H_3) -\tfrac{1}{2}\, (\eta_+-\eta_-)H_2 \\
                 \widetilde c_{11} &=  -H_2 \\
                 \widetilde c_{12} &=  -(H_2+H_3) - \tfrac{1}{4} \,(\eta_+-\eta_-)H_2  \\
                 \widetilde c_{14} &=  \tfrac{1}{4} H_2 + H_4 \\
                 \widetilde c_{15} &= -\tfrac{1}{4} \eta_- H_2

             \end{array}
             \end{equation*}
             \caption{Compton form factor residues for the nucleon Born term.
                      The $H_i$ are defined in Eq.~\eqref{Hi}.
                      The remaining $\widetilde c_i$ are zero.}
             \label{tab:nucleon-born}
        \end{table}

        Inserting the above expressions into Eq.~\eqref{qcv-born}
        yields four mixed terms $F_i({Q'}^2)\,F_j(Q^2) \equiv F_i' F_j$.
        We take their symmetric combinations $H_1 = F_1'\, F_1$, $H_2 = F_2'\, F_2$ and
        \begin{equation}\label{Hi}
           H_3 = \frac{F_1' \,F_2 + F_2' \,F_1}{2}\,, \quad
           H_4 = \frac{F_1' \,F_2 - F_2' \,F_1}{2\omega}\,,
        \end{equation}
        with $\omega$ defined in~\eqref{li-1}.
        Note that $H_4(\omega\to 0)$ is regular.
        The nucleon Born term then takes the form
        \begin{equation}\label{nucleon-born-final}
           \Gamma^{\mu\nu}_\text{B} = \frac{1}{\eta_-^2-4\lambda^2}\,\sum_{i=1}^{18} \widetilde c_i\, \Big[ \Lambda_+(p_f) \,X_i^{\mu\nu} \Lambda_+(p_i)\Big]\,,
        \end{equation}
        where the resulting CFF residues $\widetilde c_i$
        are collected in Table~\ref{tab:nucleon-born}.
        For a pointlike fermion ($F_1=1$ and $F_2=0$) only $\widetilde c_{1}=-4$ and $\widetilde c_{10}=2$ survive, i.e.,
        $X_1^{\mu\nu}$ and $X_{10}^{\mu\nu}$
        defined in Table~\ref{transverse-basis-0} are the Compton tensors of a structureless fermion
        such as the electron in tree-level QED.

        \pagebreak

        Because $Q^2$ and ${Q'}^2$ are linear combinations of $\eta_+$ and $\omega$, the $H_i$ can only depend on $\eta_+$ and $\omega^2$.
        In addition, the CFF residues in Table~\ref{tab:nucleon-born} depend on the variable $\lambda$ at most quadratically (which is also true for the resonance
        terms in Tables~\ref{tab:j=1/2+born} and~\ref{tab:j=3/2+born} below). It is then customary to rearrange $a-4\lambda^2 b = (\eta_-^2 - 4\lambda^2)\,b + (a- \eta_-^2\, b)$
        and split the CFFs into non-resonant and resonant terms:
        \begin{equation}\label{nucleon-born-split}
           c_i = \frac{\widetilde c_i}{\eta_-^2-4\lambda^2} = c_i^{(0)} + \frac{c_i^{(1)}}{\eta_-^2-4\lambda^2}\,,
        \end{equation}
        where $c_i^{(0)}$ and $c_i^{(1)}$ no longer depend on $\lambda^2$.

\begin{figure}[t]
                    \begin{center}
     \includegraphics[width=1.0\columnwidth]{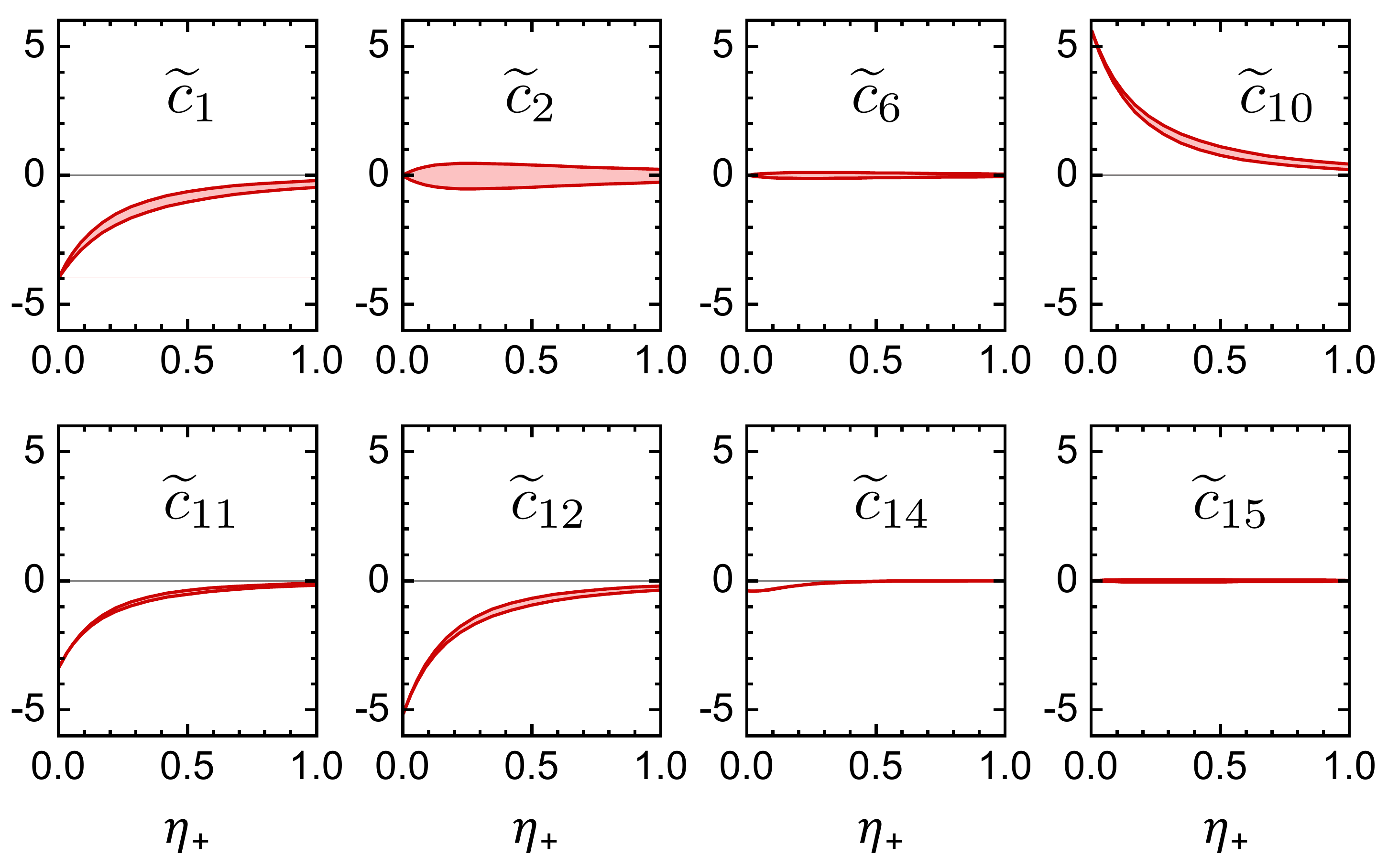}
        \caption{Compton form factor residues from the nucleon Born term inside the TPE cone and plotted over the variable $\eta_+$.}
        \label{fig:nucleon-born}
        \end{center}
        \end{figure}

        In Fig.~\ref{fig:nucleon-born} we plot the $\widetilde c_i$ from Table~\ref{tab:nucleon-born}
        inside the TPE cone shown in Fig.~\ref{fig:phasespace},
        using simple multipole parametrizations for the proton's Dirac and Pauli form factors~\cite{Friedrich:2003iz}.
        The bands show the variation with $\eta_-$, $\omega$ and $\lambda$, which turns out to be almost negligible. Therefore,
        the dependence on the four variables effectively collapses into
        a one-dimensional dependence on $\eta_+$.
        This is the typical behavior also for the resonance Born terms in the following sections,
        which happens in different systems as well~\cite{Eichmann:2014xya,Eichmann:2014qva,Eichmann:2015nra}: when implementing Lorentz invariance, permutation-group symmetries
        and minimal tensor bases, the potentially complicated momentum dependencies of three- and four-point amplitudes
        often collapse into a simple one-dimensional dependence on the symmetric variable, which in our case is $\eta_+$.

        In passing we can also verify the low-energy theorem by Low~\cite{Low:1954kd}, Gell-Mann and Goldberger~\cite{GellMann:1954kc}.
        In the forward limit where $\eta_+ = \eta_- = Q^2/m^2={Q'}^2/m^2$,
        the contribution from the nucleon Born term to the forward amplitudes $A_i = \{ T_1, T_2, 2S_1, -S_2/\lambda \}$
        in Eq.~\eqref{fwd-amplitudes}  is
        \begin{equation}
           A_i = \frac{4\pi\alpha_\text{em}}{m} \left( -A_i^{(0)} + \frac{A_i^{(1)}}{\eta_-^2 - 4\lambda^2}\right),
        \end{equation}
        where
        \begin{equation} \renewcommand{\arraystretch}{1.2}
            A_i^{(0)} =  \left[\begin{array}{c} F_1^2 \\ 0 \\  F_2^2 \\ 0 \end{array}\right], \quad
            A_i^{(1)} =  \left[\begin{array}{c} \eta_-^2  G_M^2 \\ \eta_-(4F_1^2 + \eta_- F_2^2) \\ 4\eta_- F_1 \,G_M \\ 2F_2 \,G_M \end{array}\right]
        \end{equation}
        and $G_M = F_1+F_2$ is the proton's magnetic Sachs form factor.
        In the static limit ($\eta_- = \lambda=0$) the amplitude $T_1$ vanishes according to Eq.~\eqref{fwd-amplitudes},
        except when the CFFs are singular.
        The only singularities in that limit come from the nucleon Born terms,
        as illustrated in Fig.~\ref{fig:fwd-phasespace}, which produces the Thomson term in $T_1$:
        \begin{equation}
           T_1 - T_1^\text{Pole} \Big|_{\eta_- = \lambda=0} = -\frac{4\pi\alpha_\text{em}}{m} \,\mathcal{Z}^2\,.
        \end{equation}

        \section{Spin-1/2 resonances} \label{sec:spin-1/2r}

         We proceed with the discussion of $J=1/2^\pm$ resonances.
         In Sec.~\ref{sec:ff-spin-1/2} we will explicitly
         consider the Roper resonance $N(1440)$, the $N(1710)$, the nucleon's parity partner $N(1535)$ and its first excitation $N(1650)$,
         and the $\Delta(1620)$.
         However, the following considerations are valid for all $J=1/2^\pm$ states.
         In these cases the resonance `Born terms' conceptually enter in the structure part of Fig.~\ref{fig:qcv-born}.
         It has the same form as in~\eqref{qcv-born},
             \begin{equation}\label{qcv-born-res}
             \begin{split}
                 \Gamma_{1/2}^{\mu\nu}  &= \Lambda_+(p_f) \,\Big[ \overline{\Gamma}_\text{R}^\mu(p_i^+,Q')\,S_\text{R}(p+\Sigma)\,\Gamma^\nu_\text{R}(p_f^+,Q) \\
                                                    & +\overline{\Gamma}_\text{R}^\nu(p_f^-,-Q)\,S_\text{R}(p-\Sigma)\,\Gamma^\mu_\text{R}(p_i^-,-Q') \Big] \,\Lambda_+(p_i)\,,
             \end{split}
             \end{equation}
         except that $S_\text{R}(k)$ is the propagator of the resonance and $\Gamma^\mu_\text{R} (k,Q)$
         the nucleon-to-resonance transition vertex. Eqs.~\eqref{born-kinematics} and~\eqref{cc-vertex}
         remain valid, but
         the transition vertex is no longer charge-conjugation invariant because the fermion legs correspond to different particles.

         In view of a compact notation we abbreviate
        \begin{equation}\label{nr-var-2}
        \begin{split}
            r &= \frac{m_R}{m} = \sqrt{1+\delta}\,, \\
           \delta_\pm &= \frac{m_R \pm m}{2m} = \frac{r\pm 1}{2}, \\
            \lambda_\pm &= \frac{(m_R\pm m)^2+Q^2}{4m^2} =\tau +  \delta_\pm^2 \,,
        \end{split}
        \end{equation}
        with $\delta = 4\delta_+ \delta_-$ from Eq.~\eqref{delta-def} and $\tau=Q^2/(4m^2)$.

        \subsection{\texorpdfstring{$N\to \tfrac{1}{2}^\pm$}{N to 1/2+-} transition current}\label{sec:spin-1/2-transition}

        The offshell transition vertex
        requires no separate derivation because we only need to drop the gauge part from Eq.~\eqref{npv-decomp-1}.
        The WTI simplifies to the transversality condition
        \begin{equation}\label{res-wti}
           Q^\mu \,\Gamma^\mu_\text{R} (k,Q) = 0\,,
        \end{equation}
        so the vertex is purely transverse and can be expressed through the eight tensors $T_i^\mu$ in Table~\ref{n-offshell}:
         \begin{equation}\label{npv-decomp-2}\renewcommand{\arraystretch}{1.0}
            \Gamma^\mu_\text{R}(k,Q) =  \left[ \begin{array}{c}  \mathds{1} \\ \gamma_5  \end{array}\right] \sum_{n=1}^8 f_n^\text{R}(k^2,w,Q^2)\,i T_n^\mu(k,Q)\,,
         \end{equation}
         where the upper (lower) entry holds for resonances with positive (negative) parity.

        The onshell transition current is analogous to~\eqref{onshell-nucleon},
        \begin{equation}\label{J12-onshell-current}
           J^\mu_\text{R}(k,Q) = \Lambda_+(k_+)\,\Gamma_\text{R}^\mu(k,Q)\,\Lambda_+(k_-)\big|_\text{onshell}\,,
        \end{equation}
        except that `onshell'  now refers to the kinematic limit $k_-^2=-m^2$, $k_+^2=-m_R^2$. Therefore,
        \begin{equation}\label{12-k2-w}
            k^2 = -m^2 \left( 1 + \tau + \frac{\delta}{2}\right), \quad
            w = -\frac{m^2}{2}\,\delta
        \end{equation}
        and the positive-energy projectors are
        \begin{equation}
           \Lambda_+(k_+) = \frac{-i \slashed{k}_++m_R}{2m_R}\,, \quad
           \Lambda_+(k_-) = \frac{-i\slashed{k}_- +m}{2m}\,.
        \end{equation}
        Also in this case the eight tensors collapse into two structures
        on the mass shell; the corresponding identities are given in Table~\ref{n-res-onshell-gordon} and slightly differ from before.
        In combination with~\eqref{J12-onshell-current} we can write the onshell current as
             \begin{equation}\label{current-onshell-resonance}\renewcommand{\arraystretch}{1.0}
                 \Gamma_\text{R}^\mu(k,Q) =  i\left[ \begin{array}{c}  \mathds{1} \\ \gamma_5  \end{array}\right]\left( F_1 \,T_1^\mu + F_2\,\frac{T_3^\mu}{2}  \right) .
             \end{equation}
        To avoid clutter we use the same notation for the form factors as before ($F_1$ and $F_2$) but they should not be confused
        with those of the nucleon.

        The notable difference here is the appearance of $T_1^\mu$ instead of $G_1^\mu$ because the latter no longer appears
        in the offshell current.
        It is usually written as
        \begin{equation}\label{Roper-F1}
             F_1\,T_1^\mu = F_1^\ast\,\gamma^\mu_\perp \quad \Rightarrow \quad F_1^\ast = \frac{Q^2}{m^2}\,F_1\,,
        \end{equation}
        where $\gamma_\perp^\mu = \gamma^\mu - \slashed{Q}\,Q^\mu /Q^2$ is the transverse projection of the $\gamma-$matrix.
        This quantity has a kinematic singularity at $Q^2\rightarrow 0$, which must be compensated by a kinematic zero in $F_1^\ast$.
        Therefore, $F_1^\ast(Q^2\to 0)=0$ is a consequence of transversality and analyticity
        and holds for all $J=\tfrac{1}{2}$ resonance transition form factors alike.
        This (trivially) exemplifies the advantage of using constraint-free tensor bases:
        if the current is written in terms of $T_1^\mu$, the form factor $F_1$
        approaches a constant and non-zero value for $Q^2\rightarrow 0$.

        \begin{table}[t]

             \begin{equation*} \renewcommand{\arraystretch}{1.4}
             \begin{array}{l   }

                 T_2 + \tfrac{1}{2}\,\delta\delta_- T_1  \\
                 T_4 + \tfrac{1}{2}\,T_1 + \delta_+ T_3 \\
                 T_5 + \delta_+ T_1 - 2\tau\,T_3

             \end{array} \qquad\quad
             \begin{array}{l   }

                  T_6 + \delta_+ (\delta_+ T_1 - 2\tau\,T_3) \\
                  T_7 - \tfrac{1}{2}\,\delta\delta_- T_3 \\
                  T_8 - \tfrac{1}{2}\,(\delta_+ T_1-2\tau\,T_3) + \delta_-^2\,T_3

             \end{array}
             \end{equation*}

               \caption{Combinations of tensors (defined in Table~\ref{n-offshell}) for the offshell nucleon-to-resonance transition vertex that vanish
                        in the onshell projection~\eqref{J12-onshell-current} in the positive-parity case.
                        For negative parity, replace $\delta_+ \leftrightarrow -\delta_-$.
                        The variables $\delta$, $\delta_\pm$ and $\tau$ are defined in~\eqref{nr-var-2}.}
               \label{n-res-onshell-gordon}

             \end{table}

        \subsection{\texorpdfstring{$J^P= \tfrac{1}{2}^\pm$}{JP = 1/2+-} resonance Born terms}\label{sec:spin-1/2-Born}

        The offshell transition vertex
        does not have a gauge part, so there is also no restriction in the sense of Eq.~\eqref{nucleon-off-curr-1}
        because all eight tensors result in a transverse Born term and there are no gauge parts to interfere with.
        However, experiment only provides information about onshell form factors, and therefore we restrict ourselves again to
        tree-level propagators
         \begin{equation}\label{J12-propagator}
             S_\text{R}(k) = \frac{-i\slashed{k} + m_R}{k^2+m_R^2}
         \end{equation}
        and form factors $F_1(Q^2)$ and $F_2(Q^2)$ only:
        \begin{equation}\label{J12-transition-current}\renewcommand{\arraystretch}{1.0}
        \begin{split}
            \Gamma^\mu_\text{R}(k,Q) &= i\left[ \begin{array}{c}  \mathds{1} \\ \gamma_5  \end{array}\right]\left( F_1 \,T_1^\mu + F_2\,\frac{T_3^\mu}{2}  \right), \\
            \overline{\Gamma}^\mu_\text{R}(k,Q) &= i\left( -F_1 \,T_1^\mu + F_2\,\frac{T_3^\mu}{2}  \right)\left[ \begin{array}{c}  \mathds{1} \\ \gamma_5  \end{array}\right],
        \end{split}
        \end{equation}
         where upper (lower) entries correspond to $J^P=\tfrac{1}{2}^+$ ($\tfrac{1}{2}^-$).

        \begin{table}[t]

             \begin{equation*}
             \renewcommand{\arraystretch}{1.5}
             \begin{array}{rl   }

                 \widetilde c_1 &= -4 H_1 - (\eta_-+\delta)\, H_2 \\
                 \widetilde c_2 &= -(\eta_-+\delta)\,(\delta_+ H_2+2H_3)+\lambda^2 H_2  \\
                 \widetilde c_3 &=  4\,\delta_- (\eta_-+\delta)\,\Sigma_{11} \\
                 \widetilde c_4 &= -4\,\delta\,\Sigma_{11} \\
                 \widetilde c_5 &= 0 \\
                 \widetilde c_6 &= -\delta\,H_1 +\delta_-(\eta_-+\delta)H_2 + \eta_-(H_3-\frac{1}{4}\eta_+ H_2)    \\
                 \widetilde c_7 &= -2\,\delta_- (\eta_-+\delta)\, \Sigma_{12} -2\,\delta\,(\eta_+\Sigma_{11}+\Sigma_{12}) + \tfrac{1}{2}\,\delta H_2  \\
                 \widetilde c_8 &=  -2\,\delta_-(\eta_-+\delta)\,\Delta_{12} - 2\,\delta\,(\Sigma_{11}+\Delta_{12}) \\
                 \widetilde c_9 &= 0 \\
                 \widetilde c_{10} &=  2\,(H_1+H_3) - \tfrac{1}{2} (\eta_+-\eta_--\delta)H_2 \\
                 \widetilde c_{11} &=  -H_2\\
                 \widetilde c_{12} &=  -(\delta_+H_2+H_3) -  \tfrac{1}{4}(\eta_+-\eta_-)H_2 \\
                 \widetilde c_{13} &= -4\,\delta_-^2 \,\Sigma_{11} \\
                 \widetilde c_{14} &=  H_4+\tfrac{1}{4} H_2 \\
                 \widetilde c_{15} &= -\delta \,(\eta_+ \Sigma_{11} + \Sigma_{12}) - \tfrac{1}{4} \eta_- H_2 \\
                 \widetilde c_{16} &=  \delta\,(\Sigma_{11} + \Delta_{12})  \\
                 \widetilde c_{17} &=  -4\,\delta_- \Sigma_{12}  \\
                 \widetilde c_{18} &=  -4\,\delta_- \Delta_{12}\,.

             \end{array}     \renewcommand{\arraystretch}{1.0}
             \end{equation*}

             \caption{Compton form factor residues for a $J^P =\tfrac{1}{2}^+$ resonance.
                      The $H_i$ are defined in Eq.~\eqref{Hi'} and $\delta$, $\delta_\pm$ in~\eqref{nr-var-2}.}

             \label{tab:j=1/2+born}

          \end{table}

        In analogy to~\eqref{Hi} we employ the symmetric combinations
        \begin{equation}
           \Sigma_{ij} = \frac{F_i' \, F_j + F_j' \, F_i}{2}\,, \quad
           \Delta_{ij} = \frac{F_i' \, F_j - F_j' \, F_i}{2\omega}
        \end{equation}
        but we redefine the $H_i$ as
        \begin{equation}\label{Hi'}
        \begin{split}
            H_1 &= (\eta_+^2-\omega^2)\,\Sigma_{11}\,, \\
            H_2 &= \Sigma_{22}\,, \\
            H_3 &= \eta_+ \Sigma_{12} - \omega^2 \Delta_{12}\,, \\
            H_4 &= \eta_+ \Delta_{12} -\Sigma_{12} \,.
        \end{split}
        \end{equation}
        If we replaced $(Q^2/m^2)\,F_1 \to F_1$ they would coincide with our earlier definition~\eqref{Hi} for the nucleon.

        The Born term for an intermediate nucleon resonance then becomes
        \begin{equation}
           \Gamma^{\mu\nu}_{1/2} = \frac{1}{D}\,\sum_{i=1}^{18} \widetilde c_i\,\Big[ \Lambda_+(p_f) \,X_i^{\mu\nu} \Lambda_+(p_i)\Big]\,,
        \end{equation}
        where the resonance pole is given by (cf.~Eq.~\eqref{var-su})
        \begin{equation}
           D = \frac{(s-m_R^2)(u-m_R^2)}{m^4}  =  (\eta_-+\delta)^2 - 4\lambda^2 \,.
        \end{equation}
        The CFF residues $\widetilde c_i$ for $J^P=1/2^+$ are given in Table~\ref{tab:j=1/2+born}.
        For $m_R=m$ and with the replacement $(Q^2/m^2)\,F_1 \to F_1$ they coincide with Table~\ref{tab:nucleon-born} as they should.

        The case of negative-parity resonances requires no separate discussion:
        the vertices only differ by $\gamma_5$ factors,
        so that in the Born term~\eqref{qcv-born-res} we must replace
        \begin{equation}\label{prop-sign}
           S(p\pm\Sigma) \rightarrow \gamma_5 \,S(p\pm\Sigma) \,\gamma_5 =  S(-(p\pm\Sigma))\,.
        \end{equation}
        With Eq.~\eqref{J12-propagator} this only amounts to a global sign switch
        together with an exchange $m_R \rightarrow -m_R$,
        because we defined the transition currents so that no  $m_R$ factors explicitly appear therein.
        The CFFs for negative-parity resonances are then obtained from Table~\ref{tab:j=1/2+born}
        simply by switching all signs $\widetilde c_i \to -\widetilde c_i$ and exchanging $\delta_+ \leftrightarrow -\delta_-$.

        It is easy to work out the various kinematic limits discussed in Sec.~\ref{sec:kinematic-limits}:

        (i) In RCS ($\eta_+=\omega=0$) $H_1=H_3=0$ and
              \begin{equation}
                 H_2 = F_2(0)^2, \quad
                 H_4 = -F_1(0)\,F_2(0)\,.
              \end{equation}
              Only the CFFs $c_1$, $c_2$, $c_6$, $c_{10}$, $c_{11}$ and $c_{12}$ survive
              because the remaining tensors drop out;
              in the static limit where $\eta_-=\lambda=0$ they are related to the polarizabilities
              through Eqs.~(\ref{polarizability-1}--\ref{polarizability-3}).

        (ii)  In VCS ($\eta_+ = \omega$) one has
              \begin{equation}
              \begin{split}
                 H_1 &= 0, \\
                 H_2 &= F_2(Q^2)\,F_2(0), \\
                 H_3 &= -\eta_+ H_4 = \eta_+ F_1(Q^2)\,F_2(0) \,.
              \end{split}
              \end{equation}
              The twelve CFFs in Eq.~\eqref{CFFs-VCS} survive and
              contribute to the generalized polarizabilities.

        (ii)  In the doubly-virtual forward limit ($\eta_+ = \eta_-$ and $\omega=0$)
              the CFFs collapse into the four amplitudes $\overline c_i$ from Eq.~\eqref{CFFs-FWD}.
              Splitting them into non-resonant and resonant terms as in~\eqref{nucleon-born-split},
              where only the latter contribute to the imaginary part,
              one obtains the resonance contributions to the nucleon's structure functions.


        \section{Spin-3/2 resonances} \label{sec:spin-3/2}

         We now turn to $J^P=3/2^\pm$ resonances.
         Although the generalization to this case seems straightforward,
         one encounters several pathologies related to offshell ambiguities
         coming from the unphysical lower-spin components in the Lorentz representations of the fields;
         see~\cite{Nath:1971wp,Hagen:1972ea,Singh:1973gq,Benmerrouche:1989uc,Pascalutsa:1998pw,Pascalutsa:1999zz}
         and references therein.
         Such problems can be resolved by imposing spin-3/2 gauge symmetry
         on the effective Lagrangian~\cite{Pascalutsa:1999zz},
         which leads to additional constraints for the offshell transition vertex. Here we will derive
         the most general offshell spin-1/2 to spin-3/2 transition vertex
         that is compatible with these constraints.

\begin{figure}[t]
     \center{
     \includegraphics[scale=0.07]{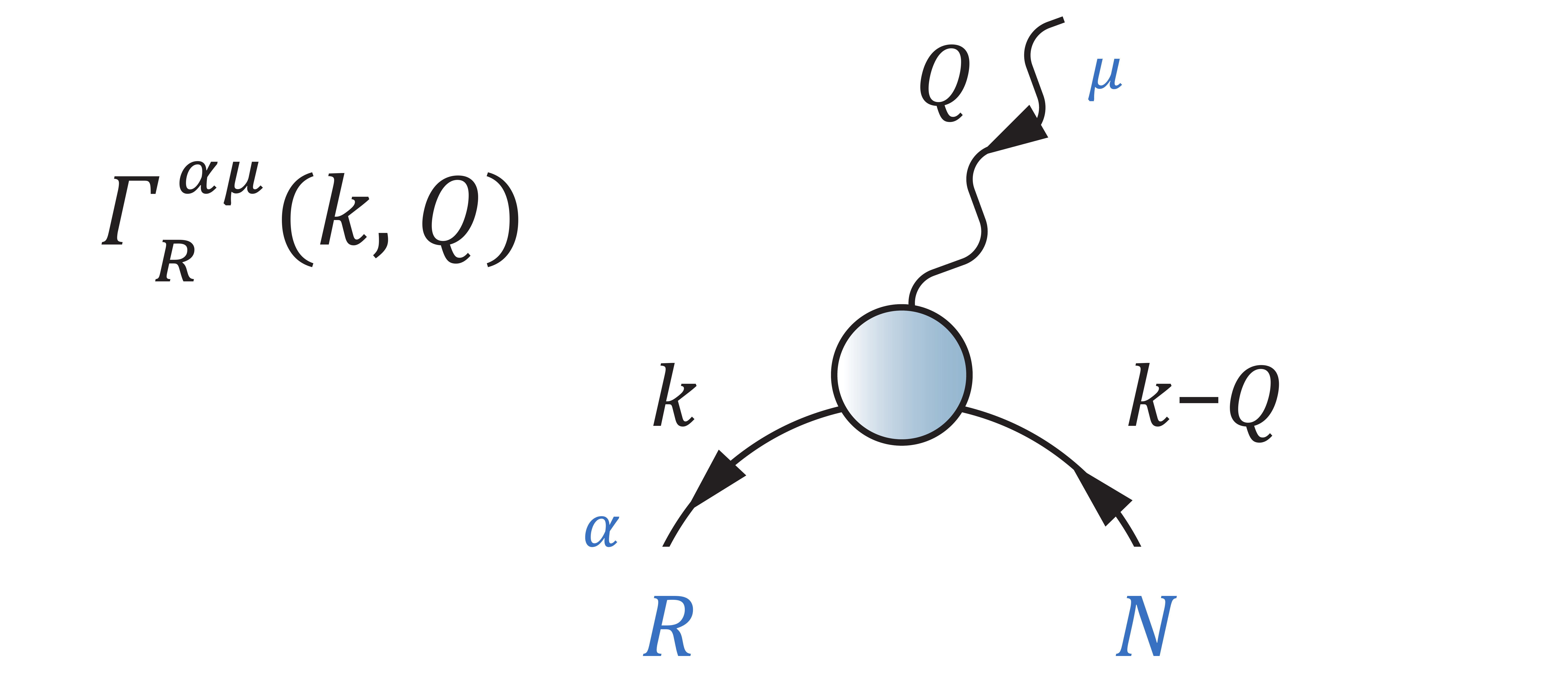}}
        \caption{Kinematics in the $N\to \tfrac{3}{2}^\pm$ transition vertex.}
        \label{fig:npv2-kinematics}
     \end{figure}

         The tree-level Compton amplitude with intermediate spin-3/2 resonances has the form
             \begin{equation}\label{qcv-born-res-3/2}
             \begin{split}
                 \Gamma_{3/2}^{\mu\nu}  &= \Lambda_+(p_f) \,\Big[ \overline{\Gamma}_\text{R}^{\mu\alpha}(p_+,Q')\,S^{\alpha\beta}_\text{R}(p_+)\,\Gamma^{\beta\nu}_\text{R}(p_+,Q) \\
                                                    & +\overline{\Gamma}_\text{R}^{\nu\alpha}(p_-,-Q)\,S^{\alpha\beta}_\text{R}(p_-)\,\Gamma^{\beta\mu}_\text{R}(p_-,-Q') \Big] \,\Lambda_+(p_i)\,,
             \end{split}
             \end{equation}
             where photon indices are denoted by $\mu$, $\nu$ and vector-spinor indices by $\alpha$, $\beta$.
             $S^{\alpha\beta}_\text{R}(k)$ is the tree-level propagator for a spin-3/2 particle and
         $\Gamma^{\alpha\mu}_\text{R} (k,Q)$ the offshell nucleon-to-resonance transition vertex.
         From now on the argument $k$ in $\Gamma^{\alpha\mu}_\text{R} (k,Q)$ denotes the momentum
         of the spin-3/2 particle and not the relative momentum, cf.~Fig.~\ref{fig:npv2-kinematics}.
         We abbreviated the resonance momenta by $p_\pm = p \pm \Sigma$.
         The charge-conjugated quantities are given by
         \begin{equation}
         \begin{split}
             \overline{\Gamma}_\text{R}^{\mu\alpha}(k,Q) &= C \,\Gamma_\text{R}^{\alpha\mu}(-k,-Q)^T C^T\,, \\
             \bar{S}^{\alpha\beta}_\text{R}(k) &= C S^{\beta\alpha}_\text{R}(-k)^T C^T
         \end{split}
         \end{equation}
         and it is straightforward to verify the Bose- and charge-conjugation invariance~\eqref{bose+cc}
         of the resonance Born terms above.

         The tree-level propagator for a spin-3/2 particle is the Rarita-Schwinger propagator
              \begin{equation}\label{RS-prop-1}
                S^{\alpha\beta}_\text{R}(k) = \frac{-i\slashed{k}+m_R}{k^2+m_R^2}\,\Delta^{\alpha\beta}\,,
              \end{equation}
         where $m_R$ is the mass of the spin-3/2 particle
         and the Rarita-Schwinger tensor is defined as
              \begin{equation}\label{RS-tensor}
                  \Delta^{\alpha\beta} = \delta^{\alpha\beta}- \frac{\gamma^\alpha \gamma^\beta}{3}  + \frac{2\,k^\alpha k^\beta}{3m_R^2} + \frac{k^\alpha \gamma^\beta - \gamma^\alpha k^\beta}{3im_R}   \,.
              \end{equation}

        It is well known that in the construction of vector-spinors from the Lorentz-group representations
        \begin{equation*}
            \big(\tfrac{1}{2},\tfrac{1}{2}\big) \otimes  \left[ \big(\tfrac{1}{2},0\big) \oplus (\dots) \right] =
             \big(\tfrac{1}{2},1\big) \oplus  \big(\tfrac{1}{2},0\big) \oplus (\dots)
        \end{equation*}
        the spin-3/2 part is contaminated by spin-1/2 contributions
        from the $\big(\tfrac{1}{2},1)$ and $\big(\tfrac{1}{2},0\big)$ subspaces. The standard way to isolate them is to define the projectors~\cite{VanNieuwenhuizen:1981ae}
        \begin{align}
        \bullet \; &\text{spin-}\tfrac{3}{2} \,\text{in}\, \big(\tfrac{1}{2},1): \quad  \mathds{P}_{3/2}^{\alpha\beta} = P^{\alpha\beta}_k -\tfrac{1}{3} \gamma^\alpha_\perp \gamma^\beta_\perp\,, \label{spin-3/2-proj}\\[1mm]
        \bullet \; & \text{spin-}\tfrac{1}{2} \,\text{in}\, \big(\tfrac{1}{2},1): \quad \mathds{P}_{11}^{\alpha\beta} = \tfrac{1}{3} \gamma^\alpha_\perp \gamma^\beta_\perp\,, \\
        \bullet \; & \text{spin-}\tfrac{1}{2} \,\text{in}\, \big(\tfrac{1}{2},0): \quad  \mathds{P}_{22}^{\alpha\beta} = \frac{k^\alpha k^\beta}{k^2}\,, \label{spin-1/2-proj}
        \end{align}
        where $P^{\alpha\beta}_k = \delta^{\alpha\beta} - k^\alpha k^\beta/k^2$ and $\gamma^\alpha_\perp = P^{\alpha\beta}_k\gamma^\beta$
        denote the transverse projector with respect to the momentum $k$ and the transverse projection of the $\gamma-$matrix (with $\gamma^\alpha_\perp \,\gamma^\alpha_\perp = 3$), respectively.
        The spin-3/2 projector satisfies
        \begin{equation}\label{spin-3/2-cond-proj}
            \mathds{P}_{3/2}^{\alpha\beta} \,k^\beta = 0\,, \qquad  \mathds{P}_{3/2}^{\alpha\beta} \,\gamma^\beta = 0\,.
        \end{equation}
        The Rarita-Schwinger field can then be decomposed into
        \begin{equation}
            \psi^\alpha = ( \mathds{P}_{3/2}^{\alpha\beta} + \mathds{P}_{11}^{\alpha\beta} + \mathds{P}_{22}^{\alpha\beta} )\,\psi^\beta\,.
        \end{equation}
        If we further define (note that $\gamma^\alpha_\perp$ anticommutes with $\slashed{k}$)
        \begin{equation}  \label{spin-1/2-proj-mixed}
            \mathds{P}^{\alpha\beta}_{12} = -\frac{\gamma^\alpha_\perp\,k^\beta\,\slashed{k}}{\sqrt{3}\, k^2}\,, \qquad
            \mathds{P}^{\alpha\beta}_{21} = \frac{k^\alpha \gamma^\beta_\perp\,\slashed{k}}{\sqrt{3} \,k^2}\,,
        \end{equation}
        the Rarita-Schwinger propagator~\eqref{RS-prop-1} takes the equivalent form
        \begin{equation}\label{RS-prop-2}
        \begin{split}
          S_\text{R}^{\alpha\beta}(k) =&\, \frac{-i\slashed{k}+m_R}{k^2+m_R^2}\,\mathds{P}_{3/2}^{\alpha\beta}  +\frac{2\,(-i\slashed{k} + m_R)}{3m_R^2} \,\mathds{P}_{22}^{\alpha\beta} \\
                        &+\frac{1}{\sqrt{3} m_R}\,  (\mathds{P}_{12}+\mathds{P}_{21})^{\alpha\beta}\,.
        \end{split}
        \end{equation}

        The pole part of the propagator is proportional to $\mathds{P}_{3/2}$ and corresponds to the spin-3/2 subspace.
        The regular terms provide the offshell spin-1/2 background
        which should not contribute to matrix elements such as the Compton scattering amplitude.
        In addition, $\mathds{P}_{3/2}$ has a kinematic singularity at $k^2=0$ which cannot survive in observables either.

        Both problems can be resolved at the level of the offshell vertices that connect the Rarita-Schwinger propagators in matrix elements.
        In the case of Compton scattering this is the transition vertex $\Gamma^{\alpha\mu}_\text{R} (k,Q)$.
        Demanding spin-3/2 gauge symmetry for effective Lagrangians
        is equivalent to imposing the transversality condition $k^\alpha \Gamma^{\alpha\mu}_\text{R} (k,Q)=0$:
        if both ends of the propagator~\eqref{RS-prop-2} are contracted with a vertex that is transverse in $k^\alpha$,
        only the pole term $\sim\mathds{P}_{3/2}$ survives
        because the projectors $\mathds{P}_{22}$, $\mathds{P}_{12}$ and $\mathds{P}_{21}$ all contain instances of $k^\alpha$ or $k^\beta$.
        Hence, a vertex that satisfies $k^\alpha \Gamma^{\alpha\mu}_\text{R} (k,Q)=0$ automatically ensures the absence
        of the spin-$1/2$ background in observables.

         On the other hand, the expressions~\eqref{RS-prop-1} and~\eqref{RS-prop-2} do not yet
         represent the most general form of a spin-3/2 propagator.
         They follow from the kinetic term of the free Rarita-Schwinger Lagrangian
         $\mathcal{L} = \conjg{\psi}^\alpha \Lambda^{\alpha\beta}  \psi^\beta$, where $\psi^\alpha$ is the spin-3/2 field
         and $\Lambda^{\alpha\beta}$ the inverse tree-level propagator.
         In momentum space it takes the form
              \begin{equation}\label{Lagrangian-3/2}
                  \Lambda^{\alpha\beta} = -\frac{i}{2} \left\{ \sigma^{\alpha\beta},\,i\slashed{k}+m_R \right\} , \quad \sigma^{\alpha\beta} = -\tfrac{i}{2}\,[\gamma^\alpha,\gamma^\beta]\,.
              \end{equation}
         This is a special case of a family of Lagrangians which are related to each other by
         point transformations~\cite{Nath:1971wp,Benmerrouche:1989uc,Pascalutsa:1999zz};
         see~App.~\ref{sec:point-tfs} for details.
         The Rarita-Schwinger form corresponds to $\xi=1$,
         where $\xi$ is the respective gauge parameter.
         For $\xi \neq 1$, the general propagator is given in Eqs.~(\ref{rs-prop-11}--\ref{rs-prop-12}): the pole part remains unchanged,
        but the spin-1/2 contributions depend on $\xi$ and also on
        the remaining projectors $\mathds{P}_{11}$ and $(\mathds{P}_{12}-\mathds{P}_{21})$.
        The latter still vanishes in matrix elements if the transition vertex is transverse in $k^\alpha$,
        but in order to eliminate $\mathds{P}_{11}$ one must additionally impose $\gamma^\alpha \Gamma^{\alpha\mu}_\text{R}(k,Q)=0$,
        which at the same time ensures the invariance of the Lagrangian under point transformations.
        The transversality in both $k^\alpha$ and $\gamma^\alpha$ is therefore necessary to decouple the
        spin-1/2 background for $\xi \neq 1$.

    \renewcommand{\arraystretch}{1.6}

        \begin{table*}[t]

             \begin{equation*}
             \begin{array}{r@{\!\;\,}l   }

                 m^2 \, T_1^{\alpha\mu} &=\, \varepsilon^{\alpha\mu}_{kQ} \\
                 m^2 \, T_2^{\alpha\mu} &=\, t^{\alpha\mu}_{kQ} \\
                 m^3 \, T_3^{\alpha\mu} &=\, i\,t^{\alpha\beta}_{k\gamma}\,t^{\beta\mu}_{QQ} \\
                 m^4 \, T_4^{\alpha\mu} &=\, t^{\alpha\beta}_{kk}\,t^{\beta\mu}_{QQ} \\
                 m^3 \, T_5^{\alpha\mu} &=\, i\,t^{\alpha\mu}_{kQ}\,\slashed{k}      \\
                 m^3 \, T_6^{\alpha\mu} &=\, i\,t^{\alpha\mu}_{kQ}\,\slashed{Q}\\[2mm]

             \end{array} \qquad
             \begin{array}{r@{\!\;\,}l   }

                 m^3 \, T_7^{\alpha\mu}    &=\, i\,t^{\alpha\beta}_{kk}\,t^{\beta\mu}_{\gamma Q} \\
                 m^3 \, T_8^{\alpha\mu}    &=\, \tfrac{i}{6}\,t^{\alpha\beta}_{kQ}\,[\gamma^\beta, \slashed{Q}, \gamma^\mu] \\
                 m^4 \, T_{9}^{\alpha\mu} &=\, \tfrac{1}{2}\,t^{\alpha\beta}_{kQ}\,[\gamma^\beta,\gamma^\nu]\,t^{\nu\mu}_{kQ} \\
                 m^4 \, T_{10}^{\alpha\mu} &=\, \tfrac{1}{2}\,t^{\alpha\beta}_{kQ}\,[\gamma^\beta,\gamma^\nu]\,t^{\nu\mu}_{QQ}     \\
                 m^4 \, T_{11}^{\alpha\mu} &=\, \tfrac{1}{2}\,t^{\alpha\beta}_{kk}\,Q^\beta [\slashed{Q}, \gamma^\mu] \\
                 m^5 \, T_{12}^{\alpha\mu} &=\, i\,t^{\alpha\beta}_{kQ}\,k^\beta \gamma^\nu\,t^{\nu\mu}_{QQ}\\[2mm]

             \end{array} \qquad
             \begin{array}{r@{\!\;\,}l   }

                 m^2 \, T_{13}^{\alpha\mu} &=\, \tfrac{1}{2}\,[t^{\alpha\beta}_{k\gamma},\,t^{\beta\mu}_{\gamma Q}]\\
                 m^3 \, T_{14}^{\alpha\mu} &=\, \tfrac{i}{6}\,t^{\alpha\beta}_{kk}\,[\gamma^\beta, \slashed{Q},\gamma^\mu]\\
                 m^4 \, T_{15}^{\alpha\mu} &=\, \tfrac{1}{2}\,[\gamma^\alpha, \slashed{k}]\,k^\beta\,t^{\beta\mu}_{QQ} \\
                 m^5 \, T_{16}^{\alpha\mu} &=\, i\,t^{\alpha\beta}_{kk}\,\gamma^\beta k^\nu\,t^{\nu\mu}_{QQ}      \\[2mm]

             \end{array} \qquad
             \begin{array}{r@{\!\;\,}l   }

                 m^3 \, T_{17}^{\alpha\mu} &=\, \tfrac{i}{6}\,[\gamma^\alpha,\slashed{k},\gamma^\beta]\,t^{\beta\mu}_{kQ} \\
                 m^3 \, T_{18}^{\alpha\mu} &=\, \tfrac{i}{6}\,[\gamma^\alpha,\slashed{k},\gamma^\beta]\,t^{\beta\mu}_{QQ} \\

                 m^4 \, T_{19}^{\alpha\mu} &=\, \tfrac{1}{2}\,t^{\alpha\beta}_{kk}\,[\gamma^\beta,\gamma^\nu]\,t^{\nu\mu}_{kQ} \\
                 m^4 \, T_{20}^{\alpha\mu} &=\, \tfrac{1}{2}\,t^{\alpha\beta}_{kk}\,[\gamma^\beta,\gamma^\nu]\,t^{\nu\mu}_{QQ} \\[2mm]

             \end{array}
             \end{equation*}

               \caption{20 tensors for the $1/2^+ \to 3/2^\pm$ transition vertex $\Gamma_\text{R}^{\alpha\mu}(k,Q)$, Eq.~\eqref{32-offshell-vertex}, which implement the two constraints
                        $Q^\mu \Gamma_\text{R}^{\alpha\mu}=0$ and $k^\alpha \Gamma_\text{R}^{\alpha\mu}=0$ without introducing kinematic singularities.
                        Eight of them are redundant if $\gamma^\alpha \Gamma_\text{R}^{\alpha\mu}=0$ is imposed as well, cf.~Eq.~\eqref{spin-3/2-annihilated}.}
               \label{n-delta-gamma-offshell}

             \end{table*}

    \renewcommand{\arraystretch}{1.3}

        In summary, the resulting three constraints on the offshell vertex $\Gamma_\text{R}^{\alpha\mu}(k,Q)$ 
        are given by
              \begin{equation}\label{delta-transversality}
                  Q^\mu\,\Gamma_\text{R}^{\alpha\mu}=0, \qquad
                  k^\alpha \,\Gamma_\text{R}^{\alpha\mu}=0, \qquad
                  \gamma^\alpha \,\Gamma_\text{R}^{\alpha\mu}=0\,.
              \end{equation}
        The first incorporates electromagnetic gauge invariance;
        it ensures transversality with respect to $Q^\mu$ and therefore also onshell current conservation.
        The second and third relations are automatically satisfied for the onshell transition current due to the properties~\eqref{spin-3/2-cond-proj} of the projector $\mathds{P}_{3/2}$
        (or the Rarita-Schwinger spinors);
        however, for offshell generalizations of the vertex they yield additional constraints that must be worked out separately.
        In the `Rarita-Schwinger gauge' $\xi=1$ the first two conditions are sufficient whereas
        the third is only relevant for $\xi\neq 1$.

              Finally, these conditions
              should be solved so that no
              kinematic singularities at $k^\alpha=0$ or $Q^\mu=0$ are introduced, which entails that $\Gamma_\text{R}^{\alpha\mu}(k,Q)$ must be at least linear in $k^\alpha$ and $Q^\mu$.
              The combination of two vertices and a propagator then also cancels the kinematic $1/k^2$ singularity in $\mathds{P}_{3/2}$
              stemming from the transverse projectors,
              so that all matrix elements are free of kinematic singularities.
              Given such a vertex,
       it is sufficient to employ either
              \begin{equation}\label{RS-propagator-14}
                S_\text{R}^{\alpha\beta}(k) \simeq \frac{-i\slashed{k}+m_R}{k^2+m_R^2}\,\mathds{P}_{3/2}^{\alpha\beta}
              \end{equation}
        or the Rarita-Schwinger propagator~\eqref{RS-prop-1} because both of them produce identical matrix elements.

        \subsection{Offshell \texorpdfstring{$N\to \tfrac{3}{2}^\pm$}{N to 3/2+-} transition vertex}\label{sec:spin-3/2-transition}

        To construct the general offshell form of $\Gamma_\text{R}^{\alpha\mu}(k,Q)$, we write down
         the analogue of Eq.~\eqref{qpv-general-basis} and collect all possible 40 tensor structures that it can contain according to Lorentz covariance and parity invariance:
             \begin{equation}\label{current-basis-general}
             \left\{ \begin{array}{c} \delta^{\alpha\mu} \\  \gamma^\alpha \gamma^\mu \end{array}\;
                     \begin{array}{c} \gamma^\alpha k^\mu \\ k^\alpha \gamma^\mu  \\ \gamma^\alpha Q^\mu \\ Q^\alpha \gamma^\mu   \end{array}\;
                     \begin{array}{c} k^\alpha k^\mu \\ Q^\alpha Q^\mu \\ k^\alpha Q^\mu \\ Q^\alpha k^\mu    \end{array}
             \right\}
             \times  \left\{ \mathds{1}, \, \slashed{k},\, \slashed{Q},\, \slashed{k}\,\slashed{Q} \right\},
             \end{equation}
        with an extra factor $\gamma_5$ attached for positive-parity resonances.
        In analogy to the spin-1/2 case we take commutators whenever more than one $\gamma-$matrix appears in a tensor element.
        For example, with the definition~\eqref{new-transverse-projectors-2} and the three- and four-commutators defined in Eqs.~(\ref{triple-commutator}--\ref{eps-01}) we have
        \begin{equation}
            \gamma^\alpha \gamma^\mu \slashed{k} \,\, \slashed{Q} \; \rightarrow \;
            [\gamma^\alpha, \gamma^\mu, \slashed{k},\, \slashed{Q} ] =  -24\,\varepsilon^{\alpha\mu}_{kQ}
        \end{equation}
        which already satisfies the first two transversality constraints in Eq.~\eqref{delta-transversality}.

        In analogy to the derivation of Table~\ref{n-offshell},
        the solution of $Q^\mu\,\Gamma_\text{R}^{\alpha\mu}=0$ and $k^\alpha \,\Gamma_\text{R}^{\alpha\mu}=0$,
        where no kinematic singularities are introduced in the process,
        leads to the resulting 20 tensors in Table~\ref{n-delta-gamma-offshell}.
        Their transversality in $k^\alpha$ and $Q^\mu$ is manifest because they contain
        instances of $\varepsilon^{\alpha\mu}_{kQ}$, $t^{\alpha\cdots}_{k\cdots}$, or $t^{\cdots\mu}_{\cdots Q}$ defined in~\eqref{new-transverse-projectors-2},
        or commutators with $\slashed{k}$ or $\slashed{Q}$ that vanish upon contraction with $k^\alpha$ or $Q^\mu$.
        When inserted in the Compton amplitude, these tensors
        eliminate the projectors $\mathds{P}_{22}$ and $(\mathds{P}_{12}+\mathds{P}_{21})$ in the propagator so that only the spin-3/2 pole part
        survives.\footnote{We note that $T_1$, $T_2$ and $T_3$ in Table~\ref{n-delta-gamma-offshell} coincide with the electromagnetic couplings of the effective $N\to\Delta\gamma$ Lagrangian in Refs~\cite{Pascalutsa:1999zz,Pascalutsa:2002pi}:
        \begin{equation*} 
        \begin{split}
            g_M\,(\p^\mu \conjg{\psi}^\alpha)  \widetilde{F}^{\alpha\mu} \,\psi & \quad \simeq \quad g_M \,\conjg{\psi}^\alpha \gamma_5\,\varepsilon^{\alpha\mu}_{kQ} \,A^\mu \,\psi\,, \\
            g_E\,(\p^\mu \conjg{\psi}^\alpha) \gamma_5 \,F^{\alpha\mu} \,\psi &\quad \simeq \quad g_E \,\conjg{\psi}^\alpha \gamma_5\,t^{\alpha\mu}_{kQ} \,A^\mu \,\psi\,,
        \end{split}
        \end{equation*}
        and similarly for $g_C$.
        Here, $\psi^\alpha$, $\psi$ and $A^\mu$ are the $\Delta$, nucleon and photon fields
        and $F^{\mu\nu}$ is the electromagnetic field-strength tensor, with $\widetilde{F}^{\mu\nu}$ its dual.
        For comparison, the couplings $g_1$ and $g_2$ employed in Ref.~\cite{Kondratyuk:2001qu} correspond to $T_{13} \simeq T_1-T_2$ and $T_2$, respectively.}

        \pagebreak

        In principle one should also work out the remaining condition $\gamma^\alpha \,\Gamma_\text{R}^{\alpha\mu}=0$ in~\eqref{delta-transversality},
        which would leave 12 independent tensors. However, this is not necessary in the Rarita-Schwinger gauge $\xi=1$ because
        the projector $\mathds{P}_{3/2}$ automatically annihilates the redundant tensors: the combinations 
        \begin{equation} \label{spin-3/2-annihilated}
            \begin{array}{l}
                 T_{13}+T_2-T_1\,, \\
                 T_{14}-T_7\,,  \\
                 T_{15} \,, \\
                 T_{16} \,,
            \end{array}\qquad
            \begin{array}{l}
                 T_{17}-T_5 \,, \\
                 T_{18}-T_3 \,, \\
                 T_{19}+(k^2/m^2) \,T_2 \,, \\
                 T_{20}+T_4
            \end{array}
        \end{equation}
        vanish upon contraction with $\mathds{P}_{3/2}$, e.g.
        \begin{equation}\label{spin-3/2-annihilated-2}
            \mathds{P}_{3/2}^{\alpha\beta}(k) \left( T_{13}^{\beta\mu}+T_2^{\beta\mu}-T_1^{\beta\mu} \right)  = 0\,.
        \end{equation}
        Therefore, the first twelve elements in Table~\ref{n-delta-gamma-offshell} are sufficient when implemented in the Compton amplitude: $T_{13}$ is equivalent to $T_1-T_{2}$, etc.
        These relations hold for $J^P=3/2^\pm$ alike because $\gamma_5$ commutes with the projector.
        The offshell $N\to 3/2^\pm$ transition vertex can then be written as
        \begin{equation}\renewcommand{\arraystretch}{1.0}\label{32-offshell-vertex}
            \Gamma_\text{R}^{\alpha\mu}(k,Q) = \left[ \begin{array}{c} \gamma_5 \\ \mathds{1} \end{array}\right] \sum_{n=1}^{12} f_n^\text{R}(k^2, k\cdot Q, Q^2)\,T_n^{\alpha\mu}(k,Q) \,,
        \end{equation}
        where the upper (lower) entry holds for resonances with positive (negative) parity.

        \subsection{Onshell \texorpdfstring{$N\to \tfrac{3}{2}^\pm$}{N to 3/2+-} transition current}\label{sec:spin-3/2-transition-onshell}

        The onshell transition current follows from sandwiching the vertex $\Gamma^{\alpha\mu}_\text{R}(k,Q)$ between the respective projectors
        and taking both momenta onshell:
        \begin{equation} \label{onshell-3/2-current-0}
           J_\text{R}^{\alpha\mu} = \Lambda_+(k)\,\mathds{P}_{3/2}^{\alpha\beta}(k)\,\Gamma^{\beta\mu}_\text{R}(k,Q)\,\Lambda_+(k-Q)\big|_\text{onshell} \,.
        \end{equation}
        Again, $k$ is here the outgoing momentum of the spin-3/2 resonance and $Q$ is the incoming photon momentum;
        the incoming nucleon momentum is $k-Q$.
        `Onshell' refers to the kinematic limit $(k-Q)^2=-m^2$ and $k^2=-m_R^2$,
        which entails
        \begin{equation}
           k\cdot Q = 2m^2 \left( \tau - \frac{\delta}{4}\right).
        \end{equation}
        The positive-energy projectors are
        \begin{equation}
           \Lambda_+(k) = \frac{-i \slashed{k}+m_R}{2m_R}\,, \quad
           \Lambda_+(k-Q) = \frac{-i (\slashed{k}-\slashed{Q}) +m}{2m}\,.
        \end{equation}
        For $k^2=-m_R^2$ the two forms~\eqref{RS-prop-1} and~\eqref{RS-prop-2}  of the propagator become equivalent:
        \begin{equation}
           \Lambda_+(k)\,\mathds{P}_{3/2}^{\alpha\beta}(k) = \Lambda_+(k)\,\Delta^{\alpha\beta}(k)\,,
        \end{equation}
         so that on shell  it does not matter whether we use the projector $\mathds{P}_{3/2}^{\alpha\beta}(k)$
         or the Rarita-Schwinger tensor $\Delta^{\alpha\beta}(k)$.

        On the mass shell, the 12 structures in Table~\ref{n-delta-gamma-offshell}
        collapse into three tensors via the identities in Table~\ref{n-delta-gamma-gordon}:
        for example, $T_4 + r T_3$ vanishes in the contraction of Eq.~\eqref{onshell-3/2-current-0}.
        The onshell current then takes the form
               \begin{equation}\label{spin-3/2-offshell-1} \renewcommand{\arraystretch}{1.0}
                   \Gamma_\text{R}^{\alpha\mu}(k,Q) = \!\left[ \begin{array}{c} \gamma_5 \\ \mathds{1} \end{array}\right]
                   \left( F_1\, T_1^{\alpha\mu} - F_2\, T_2^{\alpha\mu}  - F_3\, T_3^{\alpha\mu}   \right)\!,
               \end{equation}
        which defines three dimensionless and constraint-free form factors $F_i(Q^2)$.
        The isospin factors are implicit in the form factors.

        With the help of Eq.~\eqref{spin-3/2-annihilated} and Table~\ref{n-delta-gamma-gordon}
        one can construct equivalent forms:
        for example, since either $T_1$ or $T_2$ can be traded for $T_{13}$ one could replace
        \begin{equation}\label{onshell-3/2-current-2}
            F_1\,T_1 - F_2\,T_2 \; \rightarrow \; F_1\,T_{13} +(F_1- F_2)\,T_2\,,
        \end{equation}
         which is the combination used in Ref.~\cite{Kondratyuk:2001qu}.
        The $F_i$ have simple relations with the form factors $g_E$, $g_M$ and $g_C$ of Ref.~\cite{Pascalutsa:2002pi}:
                \begin{equation}\label{spin-3/2-ff-connection} \renewcommand{\arraystretch}{1.2}
                    \left[ \begin{array}{c} g_M \\ g_E \\ g_C \end{array}\right] = \sqrt{\frac{2}{3}}\,\frac{2\lambda_+}{\delta_+}
                    \left[ \begin{array}{r} F_1 \\ F_2 \\ rF_3 \end{array}\right],
                \end{equation}
        but due to the factor $\lambda_+$ (defined in~\eqref{nr-var-2}) $g_E$, $g_M$ and $g_C$
        have a slower falloff with $Q^2$ and
        kinematic zeros at $\lambda_+=0 \, \Leftrightarrow \,Q^2 = -(m_R+m)^2$.

        \begin{table}[t]

             \begin{equation*} \renewcommand{\arraystretch}{1.5}
             \begin{array}{l   }

                 T_{4} +r\, T_3 \\
                 T_{5} - r\,T_2 \\
                 T_{6} - 2\delta_+T_2\\
                 T_{7} - r\,(T_1+T_2) \\
                 T_{8} - 2\delta_-T_1-T_3
             \end{array} \qquad
             \begin{array}{l   }
                 T_9 - 2 \,(\tau-\tfrac{1}{4}\delta) T_1 -2 r\delta_+ T_2 \\
                 T_{10} -4\tau\,T_1-2\delta_+T_3 \\
                 T_{11} -r\left( 2\delta_- T_1 - 2\delta_+ T_2 + T_3 \right) \\
                 T_{12} - 2r\left( 2\tau\,(T_1+T_2) + \delta_+ T_3\right)

             \end{array}
             \end{equation*}

               \caption{Combinations of tensors that vanish in the onshell projection of Eq.~\eqref{onshell-3/2-current-0} with $\Gamma^{\beta\mu}_\text{R}(k,Q) = \gamma_5 \,T_i^{\beta\mu}$,
                        i.e., for resonances with $J^P =3/2^+$. For negative-parity resonances with $J^P=3/2^-$, replace $r\to-r$ and $\delta_\pm \to -\delta_\mp$.
                        The variables $r$, $\delta$ and $\delta_\pm$ are defined in~\eqref{nr-var-2}.
                        Whereas the combinations in Eq.~\eqref{spin-3/2-annihilated} automatically also vanish on the mass shell,
                        the expressions above do not vanish offshell when contracted with $\mathds{P}_{3/2}$.
                        }
               \label{n-delta-gamma-gordon}

             \end{table}

        Moreover, several equivalent forms for the onshell current exist in the literature
        which are constructed from tensors different from those in Table~\ref{n-delta-gamma-offshell}.
        While they respect current conservation,
        they do not satisfy the second and third constraints in Eq.~\eqref{delta-transversality};
        in the diction of Ref.~\cite{Pascalutsa:1999zz} they correspond to `inconsistent couplings' in the effective Lagrangian.
        An example is the $J^P = 3/2^+$ current defined by the Jones-Scadron form factors $G_E^\ast$, $G_M^\ast$ and $G_C^\ast$~\cite{Jones:1972ky}:
        \begin{equation}\label{delta-current-1}
        \begin{split}
            \Gamma_\text{R}^{\alpha\mu} &= \sqrt{\frac{3}{2}}\,\frac{\delta_+}{2 m^4 \,\lambda_+ \lambda_-}\,
             \gamma_5\,\bigg[ m^2 \lambda_-\,(G_M^\ast-G_E^\ast)\,\varepsilon^{\alpha\mu}_{kQ} \\
                                                                 & \qquad\qquad -G_E^\ast\,\,\varepsilon^{\alpha\beta}_{kQ}\,\varepsilon^{\beta\mu}_{kQ} - \frac{G_C^\ast}{2}\,\,Q^\alpha  k^\beta \,t^{\beta\mu}_{QQ}\bigg]\,.
        \end{split}
        \end{equation}
        The tensor for $G_E^\ast$ is related to Table~\ref{n-delta-gamma-offshell} via
        \begin{equation}
           \frac{1}{m^4}\,\varepsilon^{\alpha\beta}_{kQ}\,\varepsilon^{\beta\mu}_{kQ} =  \frac{k\cdot Q}{m^2}\,T_2^{\alpha\mu} - T_4^{\alpha\mu} \,,
        \end{equation}
        but the one for $G_C^\ast$ has no counterpart because it violates the second condition in~\eqref{delta-transversality} and thus cannot be used offshell.
        On the mass shell the projector $\mathds{P}_{3/2}$ enforces these constraints automatically; however,
        sensible offshell generalizations must also satisfy $k^\alpha \Gamma_\text{R}^{\alpha\mu}=0$ and therefore acceptable tensors
        must be of the form given in Table~\ref{n-delta-gamma-offshell}.
        The onshell relations between the $F_i$ and the various conventions for $N\to 3/2^\pm$ transition form factors employed in the literature are collected in App.~\ref{sec:32-onshell-ff-relations}.

        \subsection{\texorpdfstring{$J^P= \tfrac{3}{2}^\pm$}{JP = 3/2+-} resonance Born terms}\label{sec:spin-3/2-Born}

        We proceed by working out the resonance Born terms and resulting CFFs for $J=3/2^\pm$ resonances
        according to Eq.~\eqref{qcv-born-res-3/2}. For the offshell vertex~\eqref{32-offshell-vertex}
        we employ again the onshell form~\eqref{spin-3/2-offshell-1}, which depends on the three form factors $F_i(Q^2)$
        that can be extracted from experiment.
        Concerning the propagator of the resonance we can employ
        either the Rarita-Schwinger form~\eqref{RS-prop-1} or Eq.~\eqref{RS-propagator-14};
        both of them produce the same results because the tensors $T_i^{\alpha\mu}$ satisfy the required offshell constraints.

        The resulting contribution to the Compton amplitude has the form
        \begin{equation}
           \Gamma^{\mu\nu}_{3/2} = \frac{1}{3D}\,\sum_{i=1}^{18} \widetilde c_i\,\Big[ \Lambda_+(p_f) \,X_i^{\mu\nu} \Lambda_+(p_i)\Big]\,,
        \end{equation}
        where the pole is given by
        \begin{equation}
           D =  \frac{(s-m_R^2)(u-m_R^2)}{m^4}  =  (\eta_-+\delta)^2 - 4\lambda^2 \,.
        \end{equation}
        The $\widetilde c_i$ are the residues of the CFFs and collected in Table~\ref{tab:j=3/2+born} for the $J^P=3/2^+$ case.
        Unfortunately the expressions become very lengthy so we only show the result for $F_3=0$.
        This form factor drops out in RCS and does not contribute to the static polarizabilities.
        In our numerical calculations we retain all three form factors.

        In analogy to Eq.~\eqref{nucleon-born-split} one could rearrange the terms proportional to $\lambda^2$ such that
        the CFFs  split into pole and non-pole pieces:
        \begin{equation}
           c_i = \frac{\widetilde c_i}{3D} = c_i^{(0)} + \frac{\delta^2 \,c_i^{(1)}}{(\eta_-+\delta)^2-4\lambda^2}\,.
        \end{equation}
        In that way $c_i^{(0)}$ and $c_i^{(1)}$ depend on $\eta_+$, $\eta_-$ and $\omega^2$ but no longer on $\lambda^2$.
        In Sec.~\ref{sec:cffs} we will see that they effectively become functions of $\eta_+$ only.  

        The various kinematic limits can be analyzed in the same way as for the $J=1/2$ case.
        The contribution from the $\Delta(1232)$ resonance was recently also worked out in the VCS limit~\cite{Lensky:2016nui}
        and the forward limit~\cite{Hagelstein:2018bdi}.

        As before, the  $J^P=3/2^-$ case requires no separate discussion. Deleting the $\gamma_5$ factor
        in the offshell vertex~\eqref{32-offshell-vertex} only changes the sign of the argument in the propagator as in Eq.~\eqref{prop-sign},
        which amounts to replacing $m_R \to -m_R$ together with a global sign change. The CFFs for negative-parity
        resonances are then obtained from Table~\ref{tab:j=3/2+born} by replacing $r\to-r$, $\delta_+ \leftrightarrow -\delta_-$
        and flipping the global sign.

        The remaining task is to convert the available experimental data for the resonance electrocouplings
        into parametrizations  for the transition form factors $F_{1,2}(Q^2)$ and $F_{1,2,3}(Q^2)$
        that enter in the various transition vertices, so that they can be implemented in Compton scattering.
        This is what we turn to next.

        \begin{table}[p]

             \begin{equation*}
             \renewcommand{\arraystretch}{1.5}
             \begin{array}{rl   }

                 \widetilde c_1 &= 4 \left[ \vartheta \,(\mathbf{Z}_2-3\mathbf{X}^-_1) + \rho\,( \mathbf{W}_{5/2} +2r\Delta_{12})
                                           - 2\lambda^2\,\mathbf{Y}^+_{-1}   \right] \\[1mm]

                 \widetilde c_2 &= -\vartheta\,\Big[ 8\delta_+ \Sigma_{11} +4\eta_+ \mathbf{X}^-_{1/2} +2r\,\omega^2 \Delta_{12}  -\eta_-\big(3\mathbf{X}^-_{-2/3}  \\
                     & \quad\quad  +\delta_- \mathbf{Y}^+_{-2} +4\mathbf{X}^+_0\big) \Big]  +4\lambda^2\big( \mathbf{Z}_1 -\delta_- \mathbf{Y}^-_{-2} -5\mathbf{X}^+_{-3/5}\big)  \\[1mm]

                 \widetilde c_3 &= -\vartheta\,(3\mathbf{X}^-_0+\delta_-\mathbf{Y}^+_{-2}) - 12\lambda^2 \,r \Delta_{12}  \\[1mm]

                 \widetilde c_4 &=  -3\vartheta\,\mathbf{W}_2 +4 \delta \,(\mathbf{W}_{5/2}  +2r\Delta_{12}) \\[1mm]

                 \widetilde c_5 &= -12\,(\mathbf{X}^-_1+\eta_+ r\Delta_{12})\\[1mm]

                 \widetilde c_6 &= \vartheta \Big[12\delta_+\Sigma_{12}-4\delta_-\Sigma_{22} +3\eta_+ \mathbf{Y}^+_2 - \eta_-(\delta_+ \mathbf{Y}^+_{10}-2\mathbf{X}^+_{1/2})\Big]  \\
                      & +2\lambda^2 \Big[ 3\eta_+\mathbf{Y}^+_2  - (r+2)\,\mathbf{Y}^+_{-10} + (1+3\delta)\,\mathbf{Y}^-_0 - 18 \Sigma_{12} \Big]  \\
                      &  -\eta_- (\eta_+ \mathbf{Z}_1 + 5 \omega^2 r\Delta_{12}) +\tfrac{1}{4} \rho\,(\delta \mathbf{Y}^+_4-3\eta_-\mathbf{Y}^+_2) \\[1mm]

                 \widetilde c_7 &= -\vartheta\left[  (1-\eta_-)\, (\Sigma_{12}+\Sigma_{22}) + r (\Sigma_{12}-\Sigma_{22}) \right] \\
                      & -\eta_+ \delta\,\mathbf{Y}^+_1  +2\eta_-\mathbf{Z}_0+2\lambda^2 \,(3\Sigma_{11}+\Sigma_{22}+4\Sigma_{12})\\[1mm]

                 \widetilde c_8 &= -\tfrac{1}{2} \vartheta\left[ \tfrac{9}{2}\,\mathbf{W}_2 - (\eta_-+2\delta_-) \Delta_{12}\right] + 2\delta \mathbf{W}_{5/2}  - 2\lambda^2 \Delta_{12} \\[1mm]

                 \widetilde c_9 &= \tfrac{3}{4} \vartheta\left[ \mathbf{W}_2+2\,(\eta_-+2\delta_-)\Delta_{12}\right] - 6\lambda^2 \Delta_{12}  \\[1mm]

                 \widetilde c_{10} &= -2\,\big( \vartheta \,\mathbf{Z}_2  + \eta_+ \mathbf{Z}_1  +\rho \mathbf{W}_{5/2} + 5\eta_+^2 r \Delta_{12}
                              - 2\lambda^2\,\mathbf{Y}^+_{-1}\big) \\[1mm]

                 \widetilde c_{11} &= -3\vartheta\,\mathbf{Y}^+_2 + 4\,(\mathbf{X}^-_0+\delta\,\mathbf{Y}^+_{1/2}) \\[1mm]

                 \widetilde c_{12} &= -\vartheta\left( \mathbf{Z}_2 + \tfrac{3}{2}\eta_+ \mathbf{Y}^+_2 + 3r^2 \Sigma_{11}- \tfrac{1}{2}\vartheta\, \mathbf{Y}^+_{-1}\right)  \\
                        & - \eta_+\big( \mathbf{Z}_1 - \mathbf{Y}^+_4 - r \mathbf{Y}^-_0 - \tfrac{3}{2}\,\delta\mathbf{Y}^+_2\big) \\
                        & - \omega^2 r \Delta_{12} + (2\delta_++\delta r)2r\Sigma_{11} + \delta \delta_+ (\mathbf{Y}^+_1+r\mathbf{Y}^-_1)  \\[1mm]

                 \widetilde c_{13} &= -\tfrac{3}{4}\,(\vartheta - 2\eta_+)\,\mathbf{Y}^+_2  + 4\delta_-^2 \mathbf{Y}^+_{5/2} + 3r \mathbf{X}^-_2 \\[1mm]

                 \widetilde c_{14} &= \tfrac{3}{4}\,(\vartheta - 2\eta_+)\,\mathbf{W}_2  - \eta_+ r\Delta_{12} - \mathbf{Z}_0\\[1mm]

                 \widetilde c_{15} &= \vartheta\,(\mathbf{X}^-_2 -\tfrac{3}{4}\,\eta_+ \mathbf{Y}^+_2) + \eta_+ \delta \mathbf{Y}^+_{5/2} - \eta_- \mathbf{Z}_1\\[1mm]

                 \widetilde c_{16} &= \tfrac{3}{4} \vartheta\,\mathbf{W}_2 - \delta \mathbf{W}_{5/2}   \\[1mm]

                 \widetilde c_{17} &= 3\,(\vartheta - \eta_+)\,\mathbf{Y}^+_2  - 12 \delta_+ r\Sigma_{12} + 4\delta_- \mathbf{X}^-_2 \\[1mm]

                 \widetilde c_{18} &= -3 \mathbf{W}_2 + 4\delta_- r  \Delta_{12}  \\[4mm]

             \end{array}
             \end{equation*}
             \begin{equation*} \renewcommand{\arraystretch}{1.5}
             \begin{array}{rl   }

                 \Sigma_{ij} &= \displaystyle\frac{F_i' \,F_j + F_j' \, F_i}{2}  \\
                 \Delta_{ij} &= \displaystyle\frac{F_i' \,F_j - F_j' \, F_i}{2\omega}

             \end{array} \qquad
             \begin{array}{rl   }

                 \vartheta &= \eta_-+\delta \\
                 \rho &= \eta_+^2 - \omega^2

             \end{array}
             \end{equation*}
             \bigskip
             \begin{equation*} \renewcommand{\arraystretch}{1.5}
             \begin{array}{rl}

                 \mathbf{X}^\pm_n  &= \delta_+ \Sigma_{11} \pm \delta_- \Sigma_{22} + n\,\Sigma_{12} \\
                 \mathbf{Y}^\pm_n  &= \Sigma_{11} \pm \Sigma_{22} + n\,\Sigma_{12} \\

                 \mathbf{W}_n &= \mathbf{Y}^+_n - 2nr\Delta_{12} \\

                 \mathbf{Z}_0 &= (\tfrac{3}{2}+\delta)\,\mathbf{Y}^+_1 + \tfrac{3}{2}\,(1+\delta)\,\mathbf{Y}^-_{-1/3} \\
                 \mathbf{Z}_1 &= \tfrac{3}{4}\,(\vartheta-2\eta_+)\,\mathbf{Y}^+_2 - \mathbf{Z}_0 \\
                 \mathbf{Z}_2 &= \mathbf{X}^-_3 - \tfrac{1}{4}\eta_- \mathbf{Y}^+_8

             \end{array}
             \end{equation*}

             \caption{CFF residues for a $J^P =3/2^+$ resonance.
                      We only display the result for $F_3=0$.
                      The necessary definitions and abbreviations are collected at the bottom. $\delta$, $\delta_\pm$ and $r$ are defined
                      in Eq.~\eqref{nr-var-2}.}

             \label{tab:j=3/2+born}

          \end{table}

          \renewcommand{\arraystretch}{1.0}


        \section{Transition form factors}  \label{sec:ffs}

        To work out the resonance contributions to the CFFs according to the formulas in Tables~\ref{tab:j=1/2+born} and~\ref{tab:j=3/2+born},
        we need to construct parametrizations for their electromagnetic transition form factors.
        The currently known $J^P=1/2^\pm$ and $J^P=3/2^\pm$ nucleon resonances listed in the PDG are collected in Table~\ref{tab:nucleondeltaresonances}.
        Experimental data for their $Q^2-$dependent electrocouplings are available for the four-star resonances (with the exception of the $\Delta(1910)$)
        and the three-star resonance $N(1710)$.
        The data are mainly from JLab and extend up to $Q^2=5 \dots 7$ GeV$^2$
        depending on the experiment~\cite{CLAS09,CLAS2012,CLAS2016,Mokeev-Database}.
        The MAID analysis~\cite{MAID2007,Tiator09a,Tiator11a} also includes data from different experiments
        where not all multipoles are measured;
        the resulting parametrizations typically show some deviations from the JLab/CLAS analyses.

        The experimental data are commonly
        presented in terms of helicity amplitudes, which are closely connected with the
        electroproduction amplitudes from where they are extracted~\cite{Aznauryan:2011qj}.
        To implement them in Compton scattering, however, it is mandatory to translate them into the
        constraint-free form factors $F_i(Q^2)$ defined by the currents~\eqref{current-onshell-resonance} and~\eqref{spin-3/2-offshell-1}.
        As explained in the previous sections, electromagnetic and spin-$3/2$ gauge invariance
        preclude using tensors others than the $T_i^\mu$ in Table~\ref{n-offshell}
        and $T_i^{\alpha\mu}$ in Table~\ref{n-delta-gamma-offshell} for the transition currents.
        For example, using the Jones-Scadron basis in Eq.~\eqref{delta-current-1}
        or the helicity basis in Eq.~\eqref{3/2-current-helicity}
        can lead to spurious singularities in the CFFs.

        Furthermore, the helicity amplitudes
        are subject to timelike kinematic constraints,
        which typically also lead to complicated momentum dependencies in the spacelike region $Q^2>0$.
        By constrast, the $F_i(Q^2)$ are kinematically independent and thus simpler:
        without kinematic effects,
        their momentum dependence is governed by physical singularities, namely
        the pion production cuts at timelike values $Q^2 < -4m_\pi^2$
        and vector-meson poles in the complex plane.
        Up to logarithmic corrections, the form factors follow a multipole behavior at large $Q^2$~\cite{Lepage:1979za,Brodsky:1981kj,Carlson86,Carlson98}.
        For decreasing $Q^2$ it is then reasonable to expect a monotonous
        increase towards the nearest $\rho-$meson pole, which is the closest non-analyticity
        relevant for the spacelike region.
        In the absence of resonance dynamics,
        the vector-meson poles would appear on the timelike real $Q^2$ axis
        (cf. Sec.~4.2 in Ref.~\cite{Eichmann:2016yit} for a discussion of the explicit mechanism).
        The cuts signal the onset of pion-cloud effects, which push the poles onto higher Riemann sheets
        and induce deviations from monotonicity at low $Q^2>0$.
        This is our guiding assumption for ground states, whereas
        for excited states some form factors will naturally have
        zero crossings for $Q^2>0$.

        A simple parametrization that is  flexible enough to accommodate
        these features is
        \begin{equation}\label{ff-fit}
          F(Q^2) = \frac{1}{1+x} \,\frac{1}{(1+y)^{n-1}}\left( H(x) \pm E(x) \right) ,
        \end{equation}
        where $x = Q^2/m_\rho^2$, $y=Q^2/m_R^2$ and
        \begin{equation}\label{ff-fit-2}
        \begin{split}
          H(x) &= \frac{a_0 + a_1\, x + a_2 \, b_2 \, x^2 }{1 + b_2 \, x^2}\,, \\
          E(x) &= e_0\,\sqrt{1+e_1\,x^2}\,.
        \end{split}
        \end{equation}
        $E(x)$ defines the error estimate.
        $F(Q^2)$ depends on two scales,
        the $\rho-$meson mass and the resonance mass $m_R$.
        While all form factors should have vector-meson poles,
        the additional poles in $m_R$ effectively implement the proper multipole falloff at large $Q^2$,
        with $n=3$ or $n=4$ depending on the form factor.
        For ground states the remainders $H(x)$ should then become roughly constant;
        they approach the constant values $a_0$ for $Q^2=0$ and $a_2$ for $Q^2\to \infty$.
        In most cases it is sufficient to set $a_1=0$.
        We assume that  $a_0$ and $a_2$ have the same sign,
        except for form factors with zero crossings,
        and we demand $b_2 > 0$ to avoid extra singularities.
        Although this form
        has no particle production cuts and only one $\rho$ pole on the real axis (which
        can be easily remedied by introducing a width),
        it does capture the spacelike properties reasonably well, in particular in the low- and intermediate
        $Q^2$ region where experimental data exist.

\begin{table}[t] \renewcommand{\arraystretch}{1.2}
\footnotesize
\begin{center}
\begin{tabular}{ c @{\qquad} c@{\qquad} c@{\qquad} c}
        $J^P=\tfrac{1}{2}^+$         & $\tfrac{3}{2}^+ $        & $\tfrac{1}{2}^-$          & $\tfrac{3}{2}^-$          \\[2mm]  \hline \rule{-0.0mm}{0.5cm}
        $\mathbf{N(940)}$            & $\mathbf{N(1720)}$       & $\mathbf{N(1535)}$        & $\mathbf{N(1520)}$         \\
        $\mathbf{N(1440)}$           & $N(1900)$                & $\mathbf{N(1650)}$        & $N(1700)$                 \\
        $N(1710)$                    &                          & $\color{darkgrey}N(1895)$ & $N(1875)$                  \\
        $\color{darkgrey}N(1880)$    &                          &                           &                             \\[2mm]  \hline \rule{-0.0mm}{0.5cm}
     $\mathbf{\Delta(1910)}$      & $\mathbf{\Delta(1232)}$  & $\mathbf{\Delta(1620)}$   & $\mathbf{\Delta(1700)}$   \\
                                  & $\Delta(1600)$           & $\color{darkgrey}\Delta(1900)$ & $\color{darkgrey}\Delta(1940)$ \\
                                  & $\Delta(1920)$           &                           &
\end{tabular}
\end{center}
 \caption{Two- to four-star nucleon and $\Delta$ resonances below 2~GeV for $J^P = \tfrac{1}{2}^\pm$ and $\tfrac{3}{2}^\pm$~\cite{PDG2016}.
          The four-star resonances are shown in bold font and the two-star resonances in gray.
          In a spectroscopic notation they are labelled by the incoming partial wave $L_{2I,2J}$ in elastic $\pi N$ scattering;
          from left to right: $P_{11}$, $P_{13}$, $S_{11}$, $D_{13}$ for the nucleon resonances with $I=\tfrac{1}{2}$ and
          $P_{31}$, $P_{33}$, $S_{31}$, $D_{33}$ for the $\Delta$ resonances with $I=\tfrac{3}{2}$.}
\label{tab:nucleondeltaresonances}
\end{table}

        In practice we convert the experimental data for the helicity amplitudes to the form factors $F_i(Q^2)$, using
        the relations in App.~\ref{sec:ffs-onshell}, and
        divide out the poles in Eq.~\eqref{ff-fit} so that the data and their error bars are given in terms of $H_x \pm \Delta H_x$.
        Those we subsequently fit by the function $H(x)$ given above.
        To arrive at the uncertainty bands shown in the plots, we fit
        \begin{equation}
           \sqrt{ \left(H_x - H(x)\right)^2 + (\Delta H_x)^2 }
        \end{equation}
        by the rather conservative ansatz $E(x)$: in that way, the error bands grow linearly at large $Q^2$ (unless $e_1=0$)
        so that the form factors can change their multipole falloff by one power of $Q^2$ within the uncertainty.
        We prefer this form because in several cases the asymptotic powers at large $Q^2$ are under dispute
        and logarithmic corrections can modify them as well.

        At the $Q^2=0$ point we use the PDG 2016 estimates
        for the helicity amplitudes from photoproduction experiments~\cite{PDG2016}.
        For the electroproduction data at $Q^2>0$ we only included data sets which measure the complete set of helicity amplitudes, because otherwise
        one cannot extract all form factors.
        Whereas for the lowest-lying resonances --- $\Delta(1232)$, $N(1440)$, $N(1520)$ and $N(1535)$ ---
        sufficient data are available, the data sets for the higher-lying resonances are scarce so that in those cases the fits are only qualitative.
        In addition, with the exception of the $\Delta(1232)$ all cases suffer from the lack of data below $Q^2 \lesssim 0.3$ GeV$^2$.
        This is unfortunate because the most important CFF contributions come from the region at low momenta,
        which in some cases are difficult to parametrize.
        This clearly motivates the need for future measurements at low $Q^2$.

        Our resulting fits for the form factors and helicity amplitudes are shown in Figs.~\ref{fig:res-N(1440)}--\ref{fig:res-N(1720)},
        where they are represented by solid lines with bands.
        The dashed (blue) lines are the MAID parametrizations~\cite{MAID2007,Tiator09a,Tiator11a} which are included for comparison.
        The fit parameters are collected in Tables~\ref{tabCoeff-Nstar1} and~\ref{tabCoeff-Nstar2}.
        For the parameters $m_R$ entering in the fits we simply used the names in Table~\ref{tab:nucleondeltaresonances},
        e.g. $m_R = 1.535$ GeV for the $N(1535)$ resonance, and we employed $m_\rho=0.77$ GeV.
        In the following we discuss the resonance transition form factors one by one.

\subsection{States with  $J^P=1/2^\pm$}\label{sec:ff-spin-1/2}

        In these cases there are two transition form factors,
        the Dirac-like $F_1(Q^2)$ and Pauli-like
        $F_2(Q^2)$ form factor. As discussed in connection with Eq.~\eqref{Roper-F1},
         our $F_1$ differs from the standard convention $F_1^\ast$
         by a factor $Q^2/m^2$ which removes its kinematic zero at $Q^2=0$.
         From the figures one can see that in most cases $F_1$ is indeed compatible
         with a monotonous rise towards $Q^2 \to 0$.

        The relations between the form factors and helicity amplitudes $A_{1/2}$ and $S_{1/2}$
        are given in Eqs.~(\ref{ffs-ha1}--\ref{ffs-ha2}).
        They imply in particular that at the pseudothreshold
        (the Siegert limit~\cite{Siegert:1937yt}) where
        \begin{equation}
           |{\bf k}|=0 \quad \Leftrightarrow \quad  Q^2=-(m_R-m)^2\,,
        \end{equation}
        with $|{\bf k}|$ denoting the virtual photon three-momentum in the resonance rest frame and defined in Eq.~\eqref{|k|},
        the helicity amplitudes behave as~\cite{Drechsel92,Tiator-Trento,N1535-Siegert}
        \begin{equation}\label{pseudothreshold}
        \begin{split}
            J^P = 1/2^+: \quad &  A_{1/2} \propto |{\bf k}|\,, \quad   S_{1/2} \propto |{\bf k}|^2\,,  \\
            J^P = 1/2^-: \quad &  S_{1/2} \propto |{\bf k}|\,.
        \end{split}
        \end{equation}
        For larger timelike momenta they become imaginary.
        Without knowledge  of the constraint-free form factors these features would not be evident,
        whereas they are automatic if one starts directly from the $F_i$.
        As a consequence, even simple monotonous ans\"atze for the $F_i$
        typically lead to complicated shapes for the helicity amplitudes, as can be seen in the figures below.

     \begin{figure*}[p]
     \center{
     \includegraphics[width=1\textwidth]{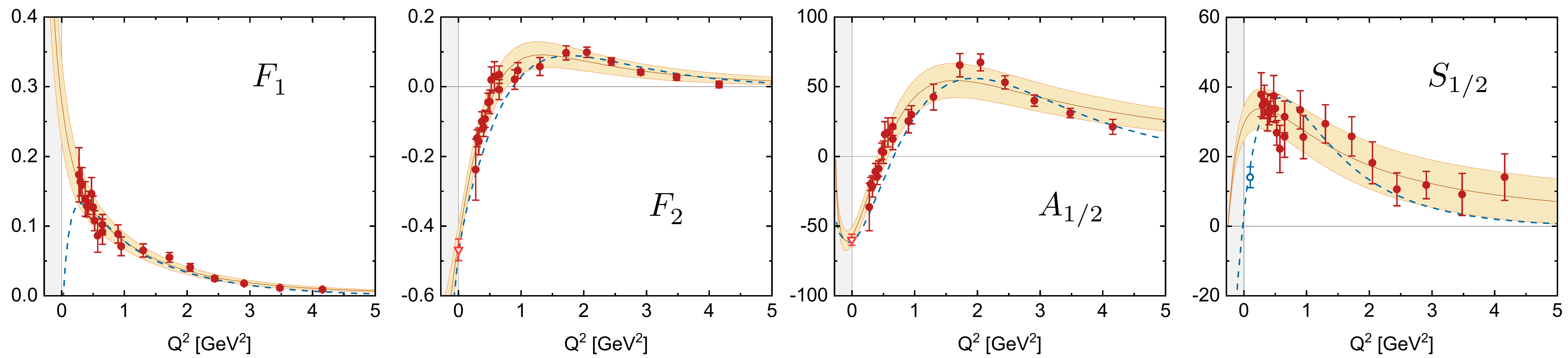}}
        \caption{Parametrization of the $\gamma^\ast N \to N(1440)$ form factors and helicity amplitudes (solid lines with bands).
                 The data points at $Q^2=0$ are from the PDG~\cite{PDG2016} and those for $Q^2>0$ from CLAS/JLab~\cite{CLAS09,CLAS2012,CLAS2016}.
                 For $S_{1/2}$ we also include the A1/MAMI point at $Q^2 \simeq 0.1$~GeV$^2$~\cite{Stajner17}.
                 The MAID parametrization (dashed, blue) is from Refs.~\cite{MAID2007,Tiator09a,Tiator11a}.
                 The form factors are dimensionless and the helicity amplitudes carry units of $10^{-3}$ GeV$^{-1/2}$.}
        \label{fig:res-N(1440)}

     \center{
     \includegraphics[width=1\textwidth]{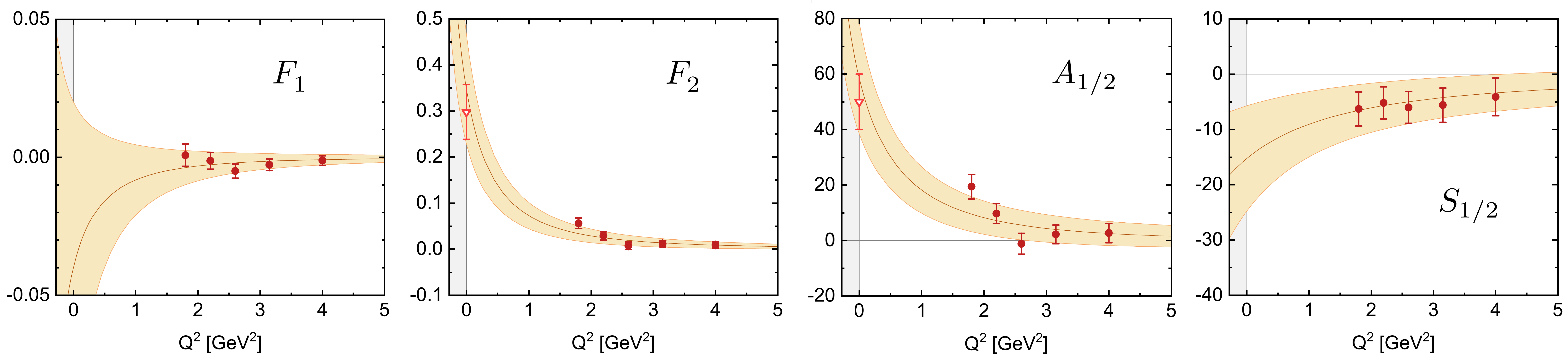}}
        \caption{Same as Fig.~\ref{fig:res-N(1440)} but for $\gamma^\ast N \to N(1710)$.
                 The data are from PDG~\cite{PDG2016} and CLAS/JLab~\cite{Park15a}.}
        \label{fig:res-N(1710)}

     \center{
     \includegraphics[width=1\textwidth]{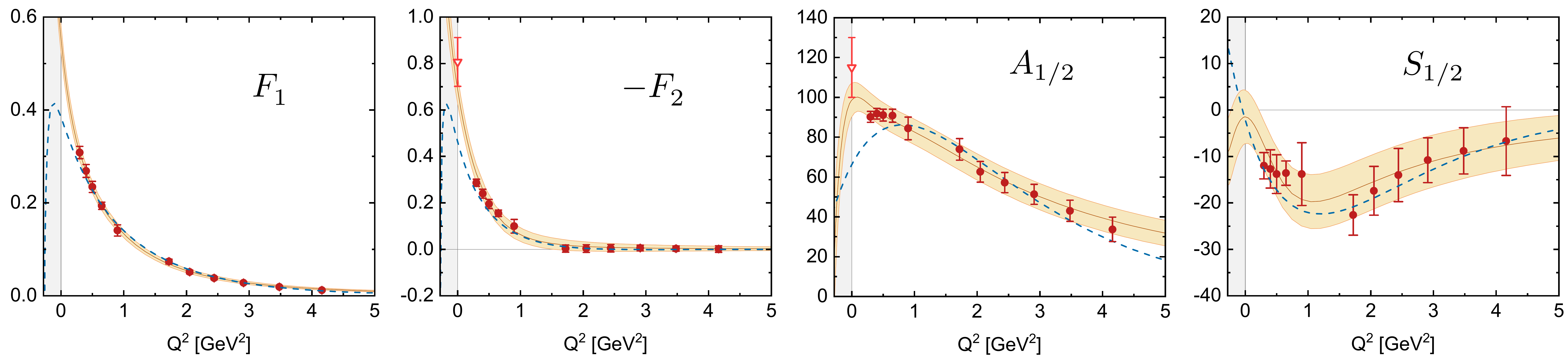}}
        \caption{Same as Fig.~\ref{fig:res-N(1440)} but for $\gamma^\ast N \to N(1535)$.
                 The data are from PDG~\cite{PDG2016} and CLAS/JLab~\cite{CLAS09}.}
        \label{fig:res-N(1535)}

     \center{
     \includegraphics[width=1\textwidth]{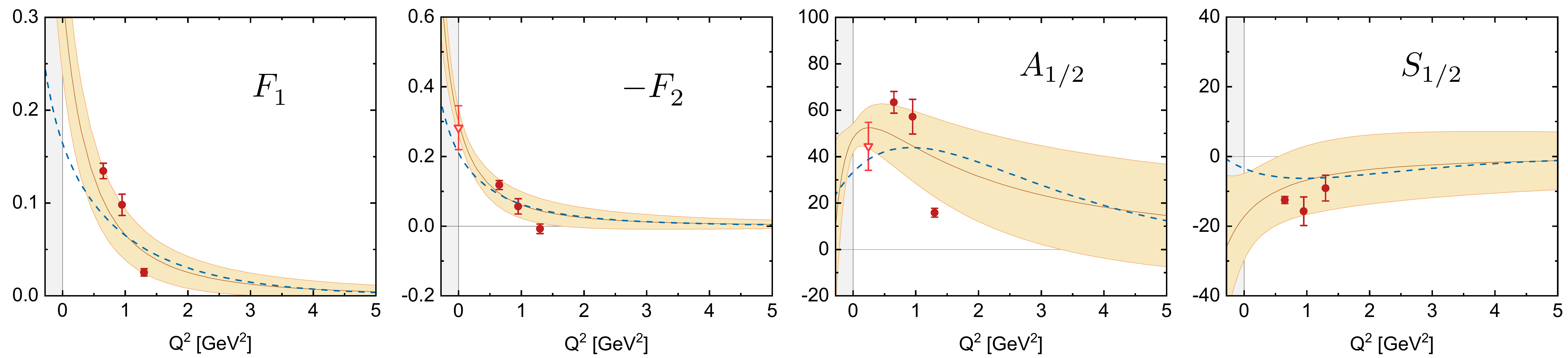}}
        \caption{Same as Fig.~\ref{fig:res-N(1440)} but for $\gamma^\ast N \to N(1650)$.
                 The data are from PDG~\cite{PDG2016} and CLAS/JLab~\cite{Mokeev-2pion}.}
        \label{fig:res-N(1650)}

     \end{figure*}

     \begin{figure*}[t]

     \center{
     \includegraphics[width=1\textwidth]{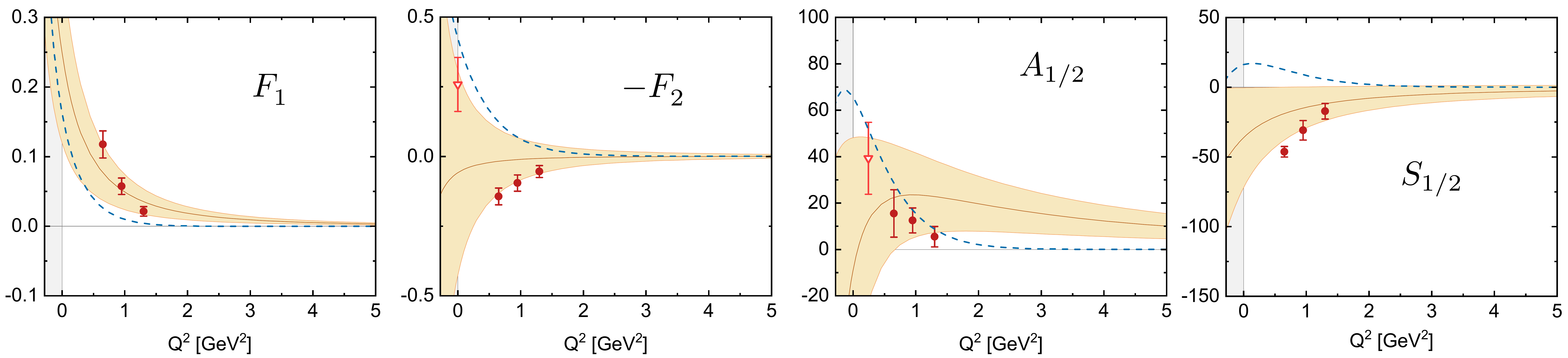}}
        \caption{Same as Fig.~\ref{fig:res-N(1440)} but for $\gamma^\ast N \to \Delta(1620)$.
                 The data are from PDG~\cite{PDG2016} and CLAS/JLab~\cite{Mokeev-2pion}.}
        \label{fig:res-D(1620)}

     \end{figure*}

        \pagebreak

        \vspace{.1cm}
        \smallskip
        \textbf{N(1440):}
        The Roper resonance is the first excitation in the $(I)J^P=(1/2)\,1/2^+$ channel.
        As such, $F_2$ has a zero crossing at intermediate $Q^2$, which is
        visible in Fig.~\ref{fig:res-N(1440)} and also found in theoretical calculations~\cite{Aznauryan07a,Lin:2008qv,Obukhovsky:2011sc,Gutsche:2012wb,Bauer:2014cqa,Segovia15,Roper-AdS}.
        By contrast, the data for $F_1$ agree with a monotonous rise.
        The MAID parametrizations implement a vanishing $F_1^\ast(0)=0$, however with a negative derivative;
        this implies a small negative value for $F_1(0)$ which produces the turnover at low $Q^2$ in the leftmost panel of Fig.~\ref{fig:res-N(1440)}.
        In the helicity amplitudes the difference is visible in $S_{1/2}$,
        where MAID is compatible with the recent
        A1/MAMI measurement for $S_{1/2}$ at very low $Q^2$~\cite{Stajner17}
        but does not reproduce the behavior~\eqref{pseudothreshold} at the pseudothreshold. 
        These relations follow automatically when we parametrize
        the form factors directly, as can be seen in the plots.

        \vspace{.1cm}
        \smallskip
        \textbf{N(1710):}
         Since this is the second excited state in the $(I)J^P=(1/2)\,1/2^+$ channel,
         one might expect two zero crossings in $F_2$.
         The five points in Fig.~\ref{fig:res-N(1710)} are recent data from JLab;
         they may indicate a slight trend in that direction but
         are too sparse to draw conclusions.
         $F_1$ is very small.
         In this case we simply fit the $H_x$ in Eq.~\eqref{ff-fit} to constants by setting $a_2=b_2=0$.
         Also here the resulting helicity amplitudes have sharp turnovers at the respective pseudothreshold
         $Q^2=-(m_R-m)^2$, which lies outside of the displayed region.

        \vspace{.1cm}
        \smallskip
        \textbf{N(1535):}
        The parity partner of the nucleon is the ground state
        in the $(I)J^P=(1/2)\,1/2^-$ channel and so we expect a monotonous behavior for both form factors,
        which is indeed visible in Fig.~\ref{fig:res-N(1535)}.
        As noted in Refs.~\cite{N1535,N1535-scaling,N1535-MesonCloud},
        the magnitude of $F_2$ quickly falls off with $Q^2$
        and is compatible with zero above $Q^2 \approx 1.5$ GeV$^2$. 
        In Table~\ref{tabCoeff-Nstar1} this amounts to the coefficient $a_2$,
        which dominates at large $Q^2$, being small compared to $a_0$.
        Model calculations typically  yield values of $F_2$
        with a different sign compared to the data~\cite{N1535,SRapproach}
        but they also do not include the $\rho$ pole;
        this may suggest cancellation effects between
        the vector-meson pole contributions and the remainder, or
        also large meson-cloud contributions at low $Q^2$~\cite{N1535-MesonCloud}.
        The oscillatory behavior of $S_{1/2}$ near the pseudothreshold
        is again a consequence of Eq.~\eqref{pseudothreshold}.

             \begin{table*}[t] \renewcommand{\arraystretch}{1.2}
             \begin{center}
             \begin{tabular}{ c @{\quad} | @{\quad} c @{\quad} | @{\quad} r @{\quad} | @{\quad} r @{\quad} r @{\quad} r @{\quad} c @{\quad} | @{\quad} c @{\quad} c @{\quad} | @{\quad} c @{\quad}}
                       &       &  $n$ & $a_0$ & $\;\;\;a_1$ & $a_2$ & $b_2$ & $e_0$ & $e_1$ & $\chi^2$   \\[1mm] \hline\hline \rule{-0.0mm}{0.4cm}

              $N(1440)$ & $F_1$ & 3 & 0.28 & $\;\;\;$ & 0.71 & 0.25 & 0.06 & 0.14 & 0.55  \\
                        & $F_2$ &  3 & $-0.45$ &  & 1.92 &  0.22 &  0.05 & 8.36 &  1.16             \\[1mm] \hline \rule{-0.0mm}{0.4cm}

             $N(1710)$ & $F_1$ & 3 &  $-0.04$  &  &    &   & 0.06 & 0.02 & 0.80 \\
                       & $F_2$ & 3 &  0.35  &  &    &   &  0.12 & 0.12 & 1.34                \\[1mm] \hline\hline \rule{-0.0mm}{0.4cm}

             $N(1535)$ & $F_1$ & 3 & 0.56 &  & 0.85 & 0.46  & 0.03 & 0.42 & 0.29  \\
                       & $F_2$ & 3 & $-0.69$ &  & $-0.07$ & 0.47  & 0.06 & 2.18 &  1.32                  \\[1mm] \hline \rule{-0.0mm}{0.4cm}

             $N(1650)$ & $F_1$ & 3 &  0.33  &  &  &  & 0.09 & 0.47  & 25.7   \\
                       & $F_2$ & 3 &  $-0.30$   &  &  & & 0.04 & 7.86 & 5.8                               \\[1mm] \hline\hline \rule{-0.0mm}{0.4cm}

             $\Delta(1620)$ & $F_1$ & 3 & 0.25  &     &    &  & 0.13 & 0.00     & 4.32  \\
                            & $F_2$ & 3 & $-0.06$  &    &    &   & 0.37 &  0.02 & 18.0            
             \end{tabular}
             \end{center}
             \caption{Fit parameters for the $J^P = 1/2^\pm$ resonance transition form factors.}
             \label{tabCoeff-Nstar1}
             \end{table*}

        \vspace{.1cm}
        \smallskip
        \textbf{N(1650):}
        The first excited state in the $(I)J^P=(1/2)\,1/2^-$ channel is shown in Fig.~\ref{fig:res-N(1650)}.
        So far there are only three data points from JLab.
        $F_2$ may be compatible with a zero crossing but in the absence of data
        we fit both $H_x$ to constants.

        \vspace{.1cm}
        \smallskip
        \textbf{$\Delta$(1620):}
        Also for the $(I)J^P=(3/2)\,1/2^-$ ground state  the data are sparse. In addition,
        Fig.~\ref{fig:res-D(1620)} displays some tension between the two data sets for $F_2(Q^2)$: the three JLab points rise towards
        a negative value at $Q^2=0$ whereas the PDG estimate is positive.
        Studies of  negative-parity states suggest a falloff $F_1(Q^2)\propto 1/Q^{8}$
        at large $Q^2$ due to the suppression of valence-quark contributions,
        which corresponds to $A_{1/2}(Q^2) \propto 1/Q^5$~\cite{SQTM}.
        Due to the lack of data we take a neutral point of view and fit the $H_x$ again by constants, so that the resulting parametrizations
        implement the usual $\propto 1/Q^6$ falloff.

\subsection{States with  $J^P=3/2^\pm$} \label{sec:ff-spin-3/2}

        \smallskip
        The $J^P=3/2^\pm$ resonances are determined by three transition form factors $F_i(Q^2)$
        or, equivalently, the helicity amplitudes $A_{3/2}(Q^2)$, $A_{1/2}(Q^2)$ and $S_{1/2}(Q^2)$.
        Their relations with the form factors are given in Eqs.~(\ref{ffs-ha32-1}--\ref{ffs-ha32-2}).
        It is also common to express them in terms of the Jones-Scadron form factors~\cite{Jones:1972ky,Devenish:1975jd}:
        magnetic dipole $G_M^\ast$, electric quadrupole $G_E^\ast$, and Coulomb quadrupole $G_C^\ast$;
        see Eq.~\eqref{delta-JS-Fi} for their relations with the $F_i$.
        Their electromagnetic ratios are defined as
        \begin{equation}\label{eqRatios}
            R_{EM} = - \frac{G_E^\ast}{G_M^\ast}, \quad
            R_{SM} = - \frac{|{\bf k}|}{2 m_R} \frac{G_C^\ast}{G_M^\ast }.
        \end{equation}

        At the pseudothreshold $|{\bf k}|=0$, the helicity amplitudes satisfy the constraints~\cite{Drechsel92,Tiator-Trento,Delta-Siegert}
        \begin{equation}\label{pseudothreshold-2}
        \begin{split}
            J^P = 3/2^+: \quad &  A_{3/2}, \, A_{1/2} \propto |{\bf k}|\,, \quad   S_{1/2} \propto |{\bf k}|^2\,,  \\
            J^P = 3/2^-: \quad &  S_{1/2} \propto |{\bf k}|
        \end{split}
        \end{equation}
        which are a direct consequence of Eqs.~\eqref{h1h2h3} and~\eqref{ffs-ha32-2}. Likewise,
        the kinematic relations between the Jones-Scadron form factors at the pseudothreshold
        follow from the definition~\eqref{eqRatios} and Eq.~\eqref{delta-JS-Fi}:
        \begin{equation}\label{pseudothreshold-3}
        \begin{split}
            J^P = 3/2^+: \quad &  2m\,G_E^\ast - (m_R-m)\,G_C^\ast \propto |{\bf k}|^2 \,, \\
                               &  R_{SM} \propto |{\bf k}| \,, \\[2mm]
            J^P = 3/2^-: \quad &  G_M^\ast \propto |{\bf k}|^2 \,, \\
                               &  m\,G_E^\ast + (m_R-m)\,G_C^\ast \propto |{\bf k}|^2 \,, \\
                               &  R_{EM} \propto 1/|{\bf k}|^2 \,, \\
                               &  R_{SM} \propto |{\bf k}| \,.
        \end{split}
        \end{equation}
        We emphasize again that the $F_i(Q^2)$ are free of kinematic constraints.

         At asymptotically large $Q^2$, the structure of the transition currents
         implies the relation $G_M^\ast \simeq  - G_E^\ast$, which amounts to $R_{EM}\to 1$~\cite{Carlson98}.
         From Eq.~\eqref{delta-JS-Fi} one obtains
         \begin{equation*}
            R_{EM} \stackrel{Q^2\to\infty}{\longlongrightarrow} \left\{ \begin{array}{ll}
               -F_2/(2F_1+F_2) & \dots J^P = 3/2^+ \\[2mm]
               -(F_1+2F_2)/F_1 & \dots J^P = 3/2^- \end{array}\right.
         \end{equation*}
         and thus $F_1 \simeq - F_2$ in both cases.
         In terms of the helicity amplitudes this entails a dominance of $A_{1/2}$ over  $A_{3/2}$~\cite{Jones:1972ky,Devenish:1975jd,Carlson86,CLAS2012}.
         We chose not to enforce this property in our fits (i.e., by constraining the respective coefficients $a_2$)
         because it is effectively absorbed in our error bands which grow with $Q^2$.
         The onset of such behavior may very well happen only at very large $Q^2$ and
         logarithmic corrections may spoil it; and
         with the exception of the $\Delta(1232)$
         the available data are in certain conflict with the constraint.

        \smallskip
        \textbf{$\Delta$(1232):}
        The $\Delta$ resonance with $(I)J^P=(3/2)\,3/2^+$ is the lowest-lying and best known nucleon resonance,
        both in terms of precision and $Q^2$ range.
        A significant amount of data for its helicity amplitudes  have been collected
        in several experiments~\cite{Sparveris05,CLAS09,Frolov99,Villano09,Sparveris13,Blomberg16}.
        For our fits we used the comprehensive database of Ref.~\cite{Mokeev-Database}
        but replaced the older data for $Q^2 < 0.2$ GeV$^2$
        by the most recent analysis from Ref.~\cite{Blomberg16}.
        At $Q^2=0$ we use the PDG estimate~\cite{PDG2016}.

\begin{figure*}[p]

     \center{
     \includegraphics[width=1\textwidth]{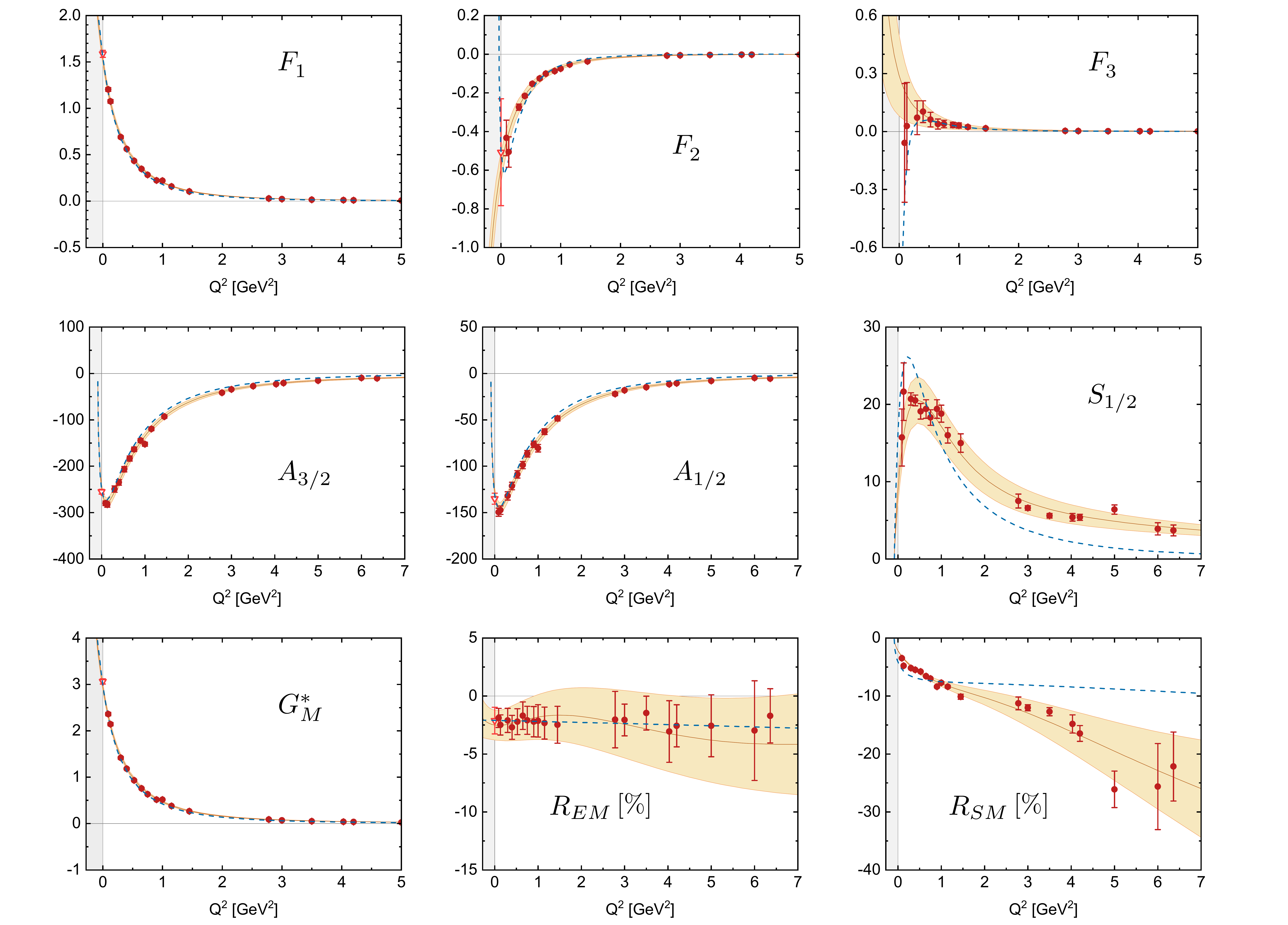}}
        \caption{$\gamma^\ast N \to \Delta(1232)$ transition form factors and helicity amplitudes.
                 The data are from Refs.~\cite{PDG2016,Frolov99,CLAS09,Villano09,Sparveris13,Blomberg16}.
                 The form factors are dimensionless and the helicity amplitudes carry units of $10^{-3}$ GeV$^{-1/2}$.}
        \label{fig:res-D(1232)}

     \center{
     \includegraphics[width=1\textwidth]{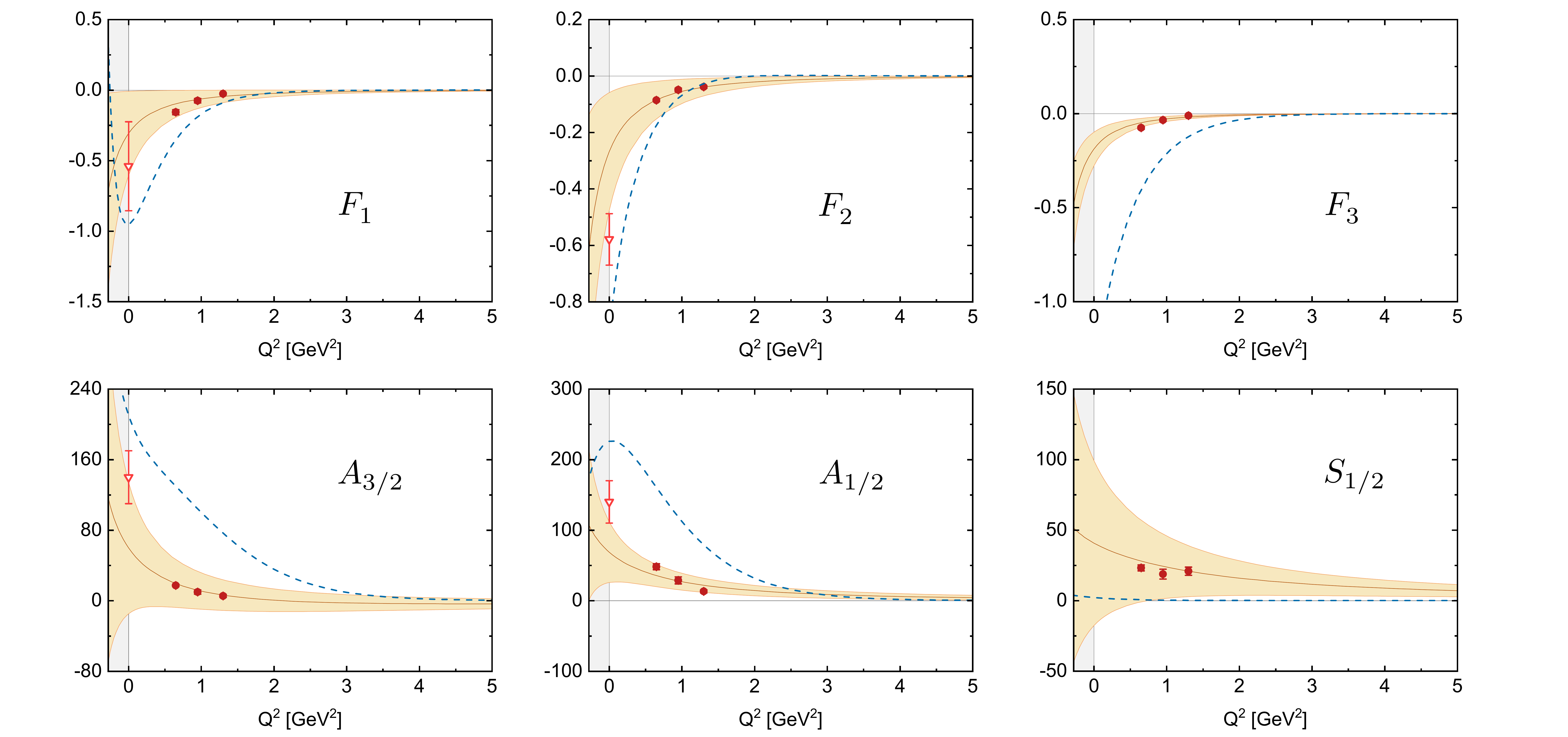}}
        \caption{Same as Fig.~\ref{fig:res-D(1232)} but for $\gamma^\ast N \to \Delta(1700)$.
                 The data are from PDG~\cite{PDG2016} and CLAS/JLab~\cite{Mokeev-2pion}.}
        \label{fig:res-D(1700)}

     \end{figure*}

        Fig.~\ref{fig:res-D(1232)} shows that $F_1$ and $F_2$ are well described
        by simple monotonous multipole functions.
        For $F_3$ the situation is less clear due to the low-$Q^2$ data, but
        since they come with large error bars
        our fit still returns a positive value for $a_0$ and thus a monotonous function.
        The resulting helicity amplitudes are plotted in the second row and
        they all vanish at the pseudothreshold.

      The $\gamma^\ast N \to \Delta(1232)$ transition is usually discussed in terms of the Jones-Scadron form factors
      displayed in the third row of Fig.~\ref{fig:res-D(1232)}.
      At $Q^2=0$, Eq.~\eqref{delta-JS-Fi} entails that
           \begin{equation}\label{JS-F-at-0}\renewcommand{\arraystretch}{1.2}
           \begin{split}
              G_M^\ast &= \sqrt{\frac{2}{3}}\,(2\delta_+ F_1 - \delta_- F_2)\,, \\
              \left[ \begin{array}{c} R_{EM} \\ R_{SM} \end{array}\right] &= \frac{\delta}{8\delta_+^2\,F_1 - \delta F_2}
              \left[ \begin{array}{c} F_2 \\ F_2 + \frac{\delta}{2r}\,F_3 \end{array}\right],
           \end{split}
           \end{equation}
      where $\delta$, $\delta_\pm$ and $r$ are defined in Eq.~\eqref{nr-var-2}.
      With $\delta_+ \gg \delta_-$,
      the dominance of the magnetic dipole form factor $G_M^\ast$
      then translates into the dominance of $F_1$,
      whereas $F_2$ and $F_3$ enter in the small ratios.
      Note that $R_{SM}$ must vanish at the pseudothreshold due to Eq.~\eqref{pseudothreshold-3}.

      Quark models can explain the dominance of $G_M^\ast$ but
      typically underestimate its magnitude by about 30--40\% at low $Q^2$~\cite{NDelta,NDelta2,Diaz08,Delta-EBAC,Burkert04,Pascalutsa:2006up}.
      In dynamical coupled-channel models
      that gap is usually attributed to meson-cloud effects~\cite{Delta-EBAC,Burkert04,Diaz08}.
      Model calculations and large-$N_c$ estimates also suggest a small valence-quark contribution
      to $R_{EM}$ and $R_{SM}$,
      indicating that these ratios may be dominated by  pion-cloud effects~\cite{NDelta-lattice,Pascalutsa:2006up,Pascalutsa:2007wz,Buchmann04,RSM-Siegert}.
      By contrast, in Dyson-Schwinger calculations
      the valence-quark components are significant due to relativistic effects~\cite{Eichmann:2011aa,Segovia14,Sanchis-Alepuz:2017mir,Eichmann:2016yit}.
      Eq.~\eqref{JS-F-at-0} shows that in the absence of $F_2$ and~$F_3$
      also $R_{EM}(0)$ and $R_{SM}(0)$ must vanish at $Q^2=0$,
      and if $R_{EM}$ were mainly a pion-cloud effect the same would be true for $F_2$.
      Concerning $F_3$,
      large-$N_c$ estimates predict $F_3(0) \simeq 0$ and thus
      $R_{SM}(0) \simeq R_{EM}(0)$~\cite{RSM-Siegert,Blomberg16,Pascalutsa:2007wz}.
      In large-$N_c$ based meson cloud models~\cite{Pascalutsa:2007wz,Buchmann04,Quadrupole-Siegert,RSM-Siegert}
      $F_3$  is small and negative below $Q^2=0$.

     \begin{figure*}[t]

     \center{
     \includegraphics[width=1\textwidth]{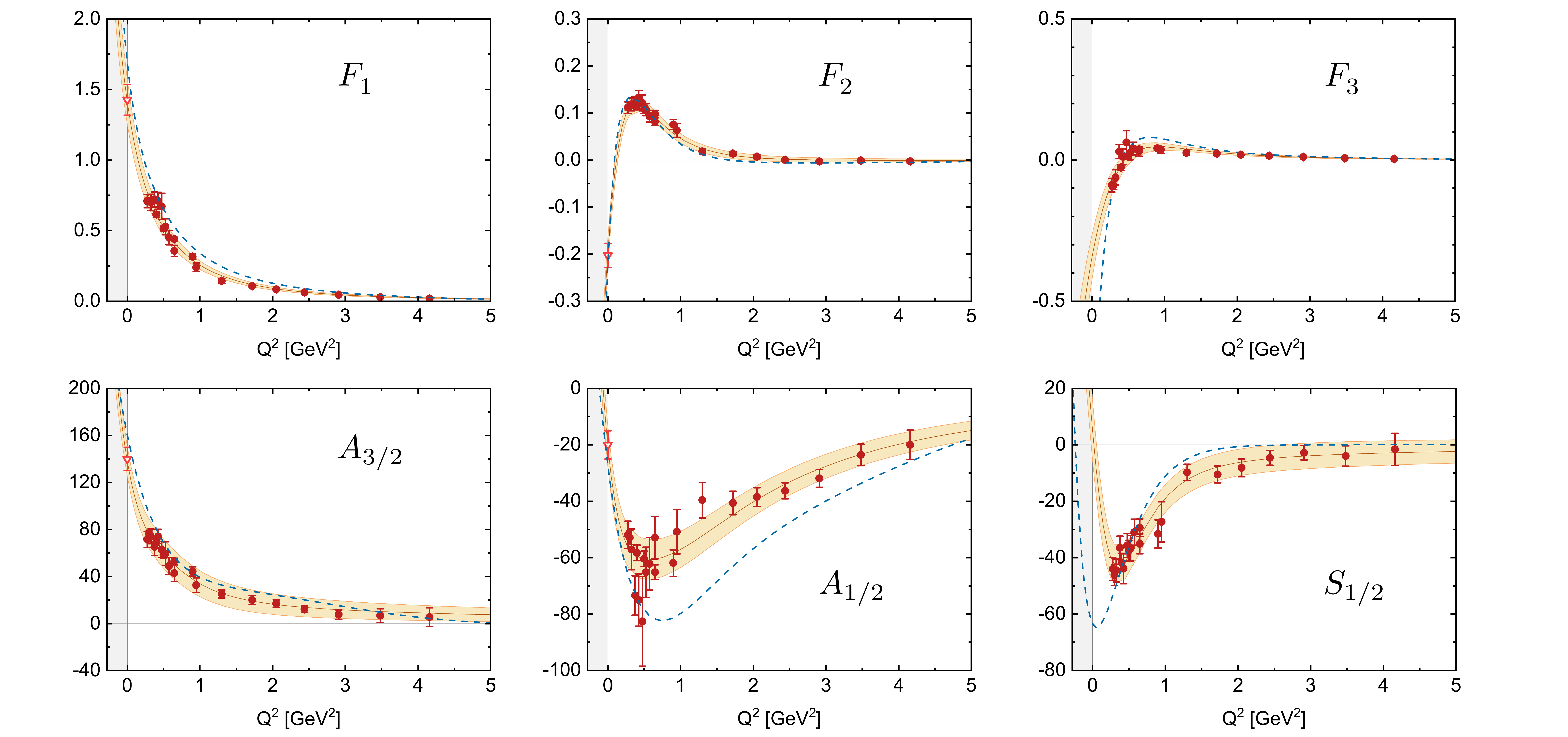}}
        \caption{Same as Fig.~\ref{fig:res-D(1232)} but for $\gamma^\ast N \to N(1520)$.
                 The data are from PDG~\cite{PDG2016} and CLAS/JLab~\cite{CLAS09,CLAS2012,CLAS2016}.}
        \label{fig:res-N(1520)}

     \center{
     \includegraphics[width=1\textwidth]{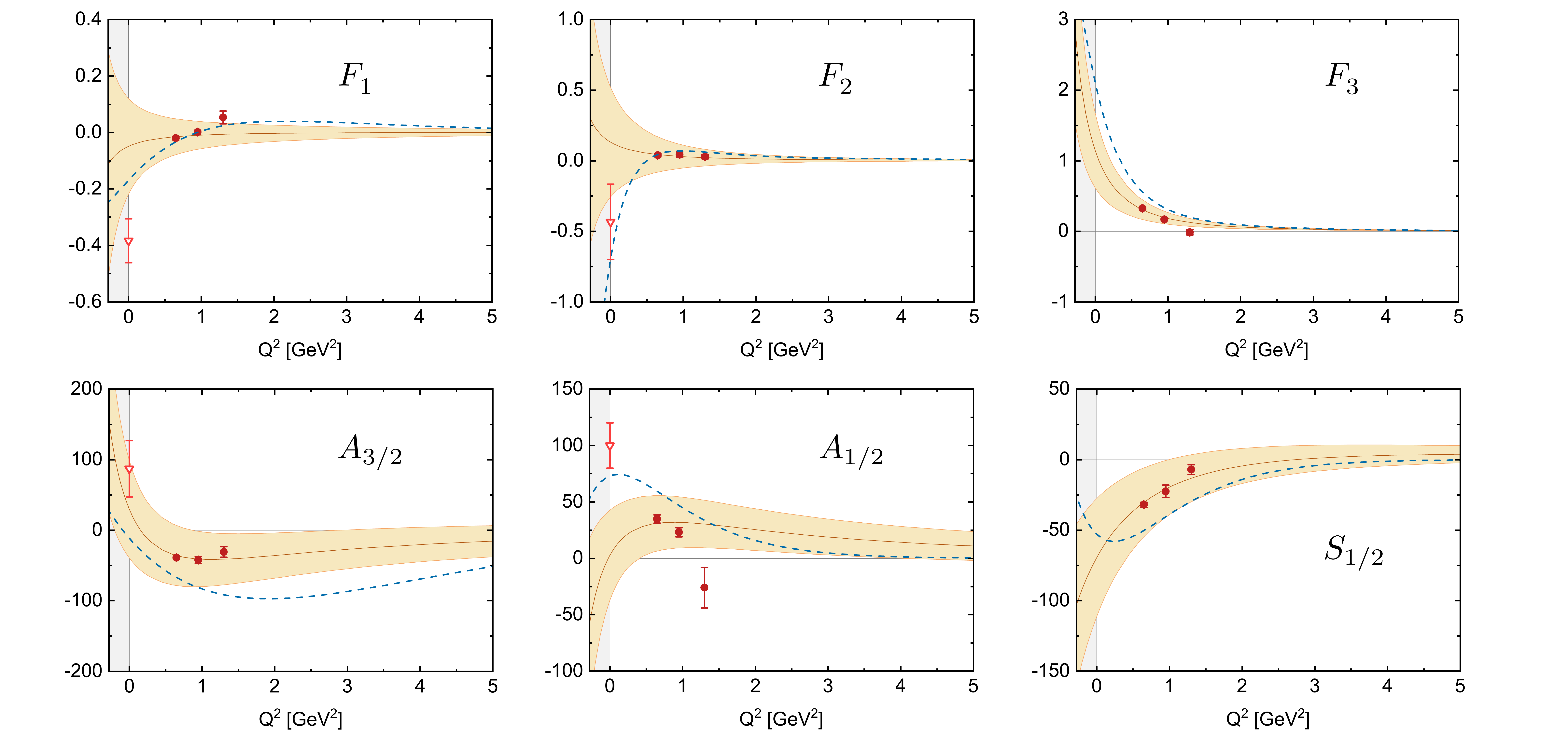}}
        \caption{Same as Fig.~\ref{fig:res-D(1232)} but for $\gamma^\ast N \to N(1720)$.
                 The data are from PDG~\cite{PDG2016} and CLAS/JLab~\cite{Mokeev-2pion}.}
        \label{fig:res-N(1720)}

     \end{figure*}

            \begin{table*}[t] \renewcommand{\arraystretch}{1.2}
            \begin{center}
            \begin{tabular}{c @{\quad} | @{\quad} c @{\quad} | @{\quad} r @{\quad} | @{\quad} r @{\quad} r @{\quad} r @{\quad} c @{\quad} | @{\quad} c @{\quad} c @{\quad} | @{\quad} c @{\quad}}
                      &       &  $n$ & $a_0$ &  $\;\;\;a_1$ & $a_2$ & $b_2$   & $e_0$ & $e_1$ & $\chi^2$ \\[1mm] \hline\hline \rule{-0.0mm}{0.4cm}

            $\Delta(1232)$ & $F_1$ & 3 & 1.53 & $\;\;$ &  0.87 & 0.04 &   0.06 & 0.02& 1.89\\
                           & $F_2$ & 3 & $-$0.59 &  &$-$0.25 & 0.11 &   0.08 & 0.00  &0.83\\
                           & $F_3$ & 4 & 0.29  &  &1.22 & 0.01 &    0.21    & 0.00 & 0.37           \\[1mm] \hline\hline \rule{-0.0mm}{0.4cm}

            $\Delta(1700)$   &  $F_1$  & 3 & $-0.31$ &   &      &       & 0.30  & 0.00  &  7.14    \\
                         &  $F_2$  & 3   & $-0.27$  &   &       &       & 0.21  & 0.00  &  4.55    \\
                         &  $F_3$  & 4   & $-0.19$  &   &       &       & 0.09  & 0.00  &  11.0      \\[1mm] \hline\hline \rule{-0.0mm}{0.4cm}

            $N(1520)$
                         &  $F_1$  & 3   & 1.42     &       & 1.09  & 0.03 & 0.16 & 0.03 &  1.43 \\
                         &  $F_2$  & 3   & $-0.20$  & 1.03  & $-0.23$  &  0.94  & 0.03 & 0.80 & 0.68  \\
                         &  $F_3$  & 4   & $-0.35$  & 0.21  & 0.61   &   0.50  & 0.08 & 0.16 &  1.11  \\[1mm] \hline\hline \rule{-0.0mm}{0.4cm}

            $N(1720)$    &  $F_1$  & 3   &  $-0.05$   &     &       &      & 0.17    & 0.30 &   9.78   \\
                         &  $F_2$  & 3   &  $0.13$    &     &       &      & 0.39  &  0.00 & 1.85  \\
                         &  $F_3$  & 4   &  1.14      &     &       &      & 0.53    & 0.00 & 8.26

            \end{tabular}
            \end{center}
            \caption{Fit parameters for the $J^P = 3/2^\pm$ resonance transition form factors.}
            \label{tabCoeff-Nstar2}
            \end{table*}

      Finally, given the asymptotic constraint $R_{EM} \to 1$ this ratio must cross zero at some value $Q^2>0$.
      From Eq.~\eqref{delta-JS-Fi} the location of the zero in $R_{EM}$ is
      \begin{equation}
         Q^2\left( 1 + \frac{2m_R}{m}\,\frac{F_3}{F_2}\right) = m_R^2-m^2\,.
      \end{equation}
      In the absence of $F_3$ the zero crossing would happen early on, but
      because $F_2$ is negative the presence of $F_3$ pushes it to larger $Q^2$.
      The existing data do not show a clear trend in any direction but stay essentially constant.
      Note that the ratios in Fig.~\ref{fig:res-D(1232)}
      are plotted in percent, so the constraint entails $R_{EM} \to +100 \%$.
      The central value of our fit crosses zero at $Q^2 \sim 20$ GeV$^2$
      but within the uncertainty band any other value above $Q^2 \sim 7$ GeV$^2$ is equally possible.
      Similarly, the large-$Q^2$ behavior for the ratio $R_{SM}$ also depends on $F_3$:
      \begin{equation}
         R_{SM} \xrightarrow[F_1 \to - F_2]{Q^2 \to \infty} - 1 + \frac{Q^2}{2m m_R} \frac{F_3}{F_1} \,.
      \end{equation}

     \pagebreak

        \smallskip
        \textbf{$\Delta$(1700):}
        The ground state in the $(I)J^P=(3/2)\,3/2^-$ channel is
        again an example where data are scarce. In this case the data points
        are compatible with all form factors being monotonous and negative,
        although this does not reproduce the large-$Q^2$ constraint $F_1 \simeq -F_2$.
        In accordance with our previous strategy we fit the $H_x$ form factor data by constants.

        \vspace{.1cm}
        \smallskip
        \textbf{N(1520):} 
        The transition form factors of the $(I)J^P=(1/2)\,3/2^-$ ground state
        in Fig.~\ref{fig:res-N(1520)} display rather peculiar features.
        $F_1$ is clearly monotonous but $F_2$ and $F_3$ are not.
        $F_3$ crosses zero at low $Q^2$, although the situation is somewhat reminiscent of the $\Delta(1232)$.
        $F_2$, on the other hand, appears to have \textit{two} zero crossings: one at very low $Q^2$ between the PDG value and the CLAS data,
        and another one around $Q^2 \sim 3$ GeV$^2$ (although within the error bars the data are still compatible
        with zero).
        A negative value at large $Q^2$ would indeed be consistent with the constraint $F_2 \simeq -F_1$.
        Still, this hints towards an interesting structure in the timelike region:
        $F_2$ is small compared to $F_1$, so potential meson-cloud effects induced by the cut structure could be magnified.
        Given the amount and precision of the data for this resonance, it is also the only example among all states
        considered where such features are clearly visible in a form factor.
        For these reasons we also include the parameter $a_1$ in our fit
        to achieve good parametrizations for $F_2$ and $F_3$.
        Significant meson-cloud contributions  for the transverse amplitude $A_{3/2}$,
        which is generally underestimated by quark models~\cite{SRapproach,N1520-1,N1520-2,Koniuk80,Santopinto12a},
        have also been suggested by dynamical coupled-channel calculations~\cite{Diaz08,CLAS2016}.

        \smallskip
        \textbf{N(1720):}
         The ground state in the $(I)J^P=(1/2)\,3/2^+$ channel is
         presently the highest-lying resonance where electroproduction data exist.
         In Fig.~\ref{fig:res-N(1720)} one can see that here it is not even possible to pin down
         the sign for any form factor: all three $F_i$ contain data with
         both positive and negative signs, even among the three CLAS points.
         This clearly calls for more measurements in the future.
         One should also note that another $N(1720)$ state with the same quantum numbers was recently proposed
         to explain the $\gamma^\ast N \to \pi \pi N$ data~\cite{Mokeev:2015moa,Mokeev-NSTAR}.
         We follow our previous strategy and fit the $H_x$ form factor data by constants;
         the resulting uncertainty bands provide at least a rough estimate for the magnitude
         of each form factor.

\subsection{Discussion}

We constructed parametrizations for the $\gamma^\ast N \to R$ transition form factors based on analytic properties.
Instead of fitting the data for the helicity amplitudes   
we directly fitted the constraint-free form factors.
In the majority of cases these show a monotonous behavior which is well described by simple parametrizations.
For the cases where data coverage is still poor we did not attempt to achieve a pointwise description
but rather employed qualitative fits.
In any case, the resulting helicity amplitudes automatically satisfy the kinematic constraints \textit{e.g.} at the pseudothreshold,
which can lead to significant deviations from the MAID parametrizations especially near the photon point.

The fits can be improved when more data become available.
The $Q^2$-dependence of several transition form factors is still poorly known, especially at low $Q^2$:
even the best known resonances such as the $N(1440)$, $N(1520)$ and $N(1535)$ do not have any data below $Q^2 \lesssim 0.3$ GeV$^2$.
This is particularly relevant for the form factors $F_1(Q^2)$ for $J=1/2$ states
and $F_3(Q^2)$ for $J=3/2$ states, which cannot be extracted at the photon point
because $S_{1/2}(0)$ cannot be measured directly.
It is then mandatory to expand the databases in this region to pin down the trend
of the transition form factors near $Q^2=0$, which is also the relevant region for Compton scattering.

          \begin{figure*}[t]
         \begin{center}

         \includegraphics[width=0.95\textwidth]{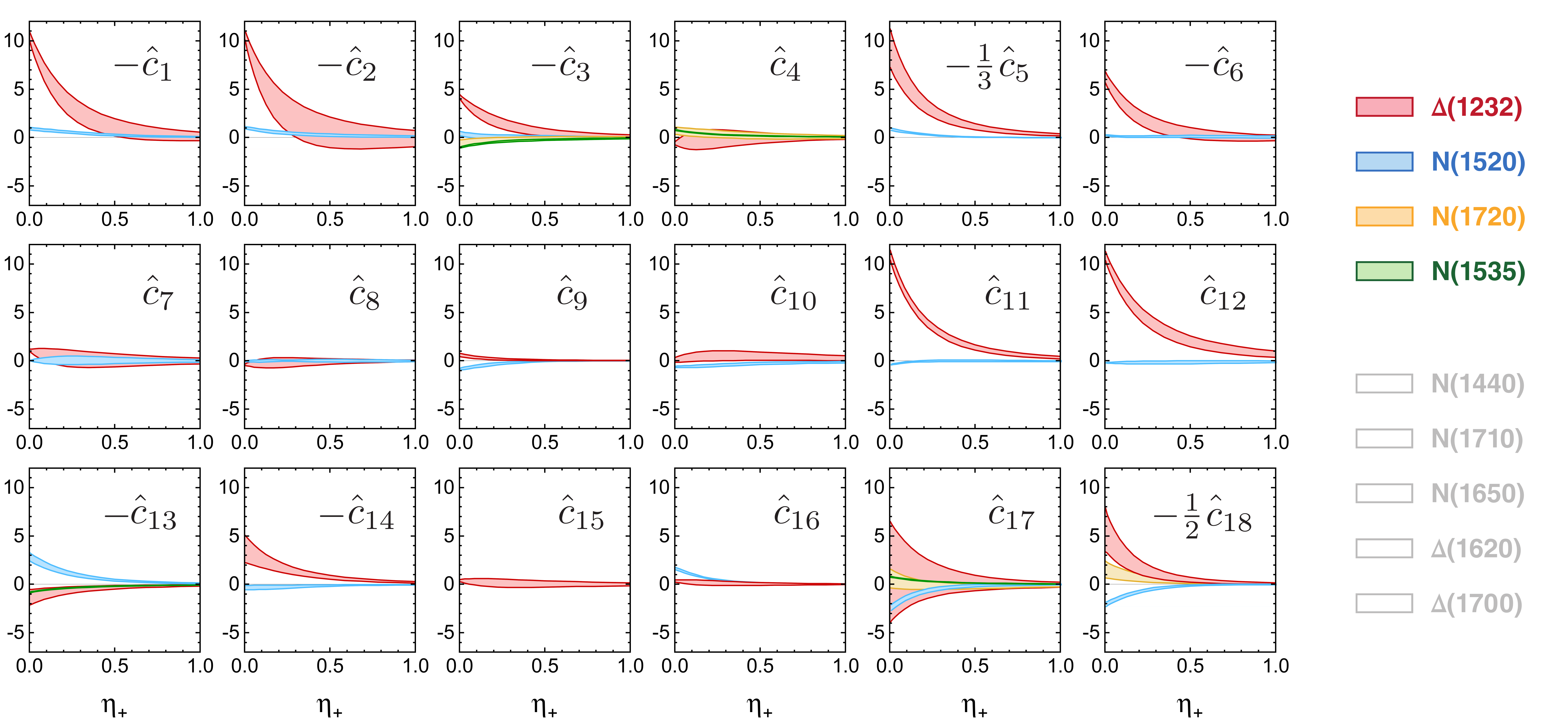}
        \caption{Residues of the Compton form factors inside the TPE cone, plotted as functions of $\eta_+$.
                 The bands contain the variation inside the cone
                 as well as the uncertainties from the form factor parametrizations.
                 The contributions from states other than the $\Delta(1232)$ and $N(1520)$
                 are only shown for $\hat c_{3,4,13,17,18}$ where they are visibly different from zero.}
        \label{fig:results-cffs-1}

        \end{center}
        \end{figure*}

   \newpage

        \section{Compton form factors} \label{sec:cffs}

        We now have everything in place to work out the nucleon resonance contributions to the CFFs. 
        In practice we set up the resonance terms in Eqs.~\eqref{qcv-born-res} and~\eqref{qcv-born-res-3/2}
        in a specific Lorentz frame, Eqs.~(\ref{simple-frame}--\ref{VCS-frame}),
        and extract the CFFs by matrix inversion.
        For the $J^P=1/2^\pm$ resonances
        we employ the offshell transition vertex~\eqref{J12-transition-current}
        and for the $J^P=3/2^\pm$ cases we use Eq.~\eqref{spin-3/2-offshell-1}, together with our parametrizations for the transition form factors.

        Fig.~\ref{fig:results-cffs-1} shows the resulting CFFs inside the TPE cone.
        The bands include the dependence on all four variables $\eta_+$, $\eta_-$, $\omega$ and $\lambda$
        as well as the uncertainty bands from our form factor parametrizations.
        In particular, we plot the residues $\hat c_i$ at the $s$ and $u$-channel poles defined by
        \begin{equation}
           c_i(\eta_+,\eta_-,\omega,\lambda) = \delta^2\,\frac{\hat c_i(\eta_+,\eta_-,\omega,\lambda)}{(\eta_-+\delta)^2-4\lambda^2}\,,
        \end{equation}
        so that the poles do not appear in the plots but the static values
        at $\eta_+=\eta_-=\lambda=\omega=0$ can be read off directly.

        One can see that the $\hat c_i$ typically fall into relatively thin bands.
        In the few cases where the bands are broader this is mainly due to the uncertainties from the form factors.
        Obviously this would not have been possible with a non-minimal basis: if some of the CFFs had
        kinematic singularities inside the cone or on its boundary, the spread of the bands would become infinite.
        Instead, the bands are narrow so that
        the dependence on four variables effectively reduces to a one-dimensional dependence on $\eta_+$.
        This is very helpful because
        instead of facing the need for studying many different kinematic slices the essential information is already encoded in
        a single variable.

        Concerning the individual resonances, the $\Delta(1232)$ clearly provides the largest contribution to most CFFs.
        The higher-lying states usually only have little impact.
        The biggest subleading contributions come from the $N(1520)$ and $N(1720)$,
        which all carry spin $3/2$ as well, whereas the $J=1/2$ resonances such as the Roper resonance or the $N(1535)$ are almost negligible.

            \begin{table*}[t] \renewcommand{\arraystretch}{1.2}
            \begin{center}
            \begin{tabular}{c @{\quad} | @{\;\;} r @{\quad} | @{\;\;} r @{\quad} r @{\quad} r @{\quad} r @{\quad} r @{\;\;} | @{\;\;} r  @{\quad} r @{\quad} r @{\quad} r @{\quad} r @{\quad} r @{\quad} r @{\quad} r @{\quad} r }
                                 &  Exp. $\;$ & $N(1440)$   &  $N(1710)$  & $N(1535)$   & $N(1650)$   & $\Delta(1620)$  & $\Delta(1232)$ & $\Delta(1700)$ & $N(1520)$   & $N(1720)$ \\[1mm] \hline\hline \rule{-0.8mm}{0.4cm}

            $-c_1$                & $ 20.2\,(4)$   & $ 0.2\,(0)$ & $ 0.1\,(0)$ & $-0.3\,(1)$ &             &  & $ 10.6\,(8)$   & $-0.1\,(2)$    & $ 0.9\,(2)$ & $0.0\,(1)$  \\
            $-c_2$                & $  3.7\,(6)$   & $ 0.2\,(0)$ & $ 0.1\,(1)$ & $ 0.1\,(0)$ &             &  & $ 10.8\,(8)$   & $ 0.0\,(1)$    & $ 1.0\,(2)$ &   \\[1mm] \hline \rule{-0.8mm}{0.4cm}
            $-c_6$                & $ 27.8\,(4.1)$ &             &             & $-0.4\,(1)$ & $-0.1\,(0)$ &  & $  6.3\,(9)$   & $ 0.0\,(1)$    & $ 0.3\,(1)$ & $0.0\,(1)$  \\
            $c_{10}$             & $  9.0\,(8.6)$ & $ 0.1\,(0)$ &             & $-0.1\,(0)$ &             &  & $  0.1\,(4)$   & $ 0.1\,(1)$    & $-0.6\,(1)$ &                \\
            $c_{11}$             & $  3.3\,(6.7)$ & $-0.2\,(0)$ &             & $ 0.2\,(0)$ &             &  & $ 11.1\,(8)$   &                & $-0.4\,(1)$ &                 \\
            $c_{12}$             & $  8.6\,(5.1)$ & $-0.2\,(0)$ &             & $-0.1\,(0)$ &             &  & $ 11.0\,(8)$   & $ 0.1\,(1)$    & $-0.2\,(0)$ &                 \\[1mm] \hline \rule{-0.8mm}{0.4cm}

            $\alpha$             & $ 11.2\,(4)$ &             &             & $-0.3\,(0)$ &             &  & $ -0.1\,(0)$   & $-0.1\,(1)$    & $-0.1\,(0)$ &          \\
            $\beta$              & $  2.5\,(4)$ & $ 0.1\,(0)$ & $ 0.1\,(0)$ & $ 0.1\,(0)$ &             &  & $  7.3\,(6)$   & $ 0.0\,(1)$    & $ 0.7\,(2)$ &          \\[1mm] \hline \rule{-0.8mm}{0.4cm}

            $\gamma_{E1E1}$      & $- 3.5\,(1.2)$ &             &             & $ 0.1\,(0)$ &             &  & $- 0.4\,(0)$   &                & $-0.1\,(0)$ &          \\
            $\gamma_{M1M1}$      & $  3.2\,(0.9)$ & $-0.1\,(0)$ &             &             &             &  & $  3.6\,(3)$   &                & $ 0.1\,(0)$ &          \\
            $\gamma_{E1M2}$      & $ -0.7\,(1.2)$ &             &             &             &             &  & $- 0.4\,(0)$   &                & $-0.1\,(0)$ &          \\
            $\gamma_{M1E2}$      & $  2.0\,(0.3)$ &             &             &             &             &  & $  0.4\,(1)$   &                &             &          \\
            $\gamma_{0}$         & $- 0.9\,(0.1)$ &             &             & $-0.1\,(0)$ &             &  & $- 3.2\,(2)$   &                & $ 0.1\,(0)$ &          \\
            $\gamma_{\pi}$       & $  8.0\,(1.8)$ & $-0.1\,(0)$ &             & $-0.2\,(0)$ &             &  & $  4.9\,(4)$   &                & $ 0.3\,(0)$ &          \\

            \end{tabular}
            \end{center}
            \caption{Resonance contributions to the nucleon's scalar and spin polarizabilities.
                     The experimental values of $\alpha$ and $\beta$ are from the PDG~\cite{PDG2016} and
                     those for the spin polarizabilities from Refs.~\cite{Martel:2014pba,Pasquini:2010zr,Camen:2001st}.
                     For the resonance masses that enter in Eqs.~(\ref{pol-12}--\ref{pol-32})
                     we use the PDG 2016 estimates for the real parts of the pole positions~\cite{PDG2016}.
                     The CFFs are dimensionless, $\alpha$ and $\beta$ carry units of $10^{-4}$ fm$^3$, and the spin polarizabilities
                     are given in $10^{-4}$ fm$^4$.
                     We do not display the numbers if both their absolute values and uncertainties are smaller than 0.05 in the respective units.}
            \label{tab:polarizabilities}
            \end{table*}

        Table~\ref{tab:polarizabilities} collects the CFFs $c_1$, $c_2$, $c_6$, $c_{10}$, $c_{11}$ and $c_{12}$
        in the static limit where all kinematic variables vanish.
        From the results in Tables~\ref{tab:j=1/2+born} and~\ref{tab:j=3/2+born} one extracts the following relations for $J^P=1/2^\pm$ states in that limit:
             \begin{equation} \label{pol-12}\renewcommand{\arraystretch}{1.2}
                 \left[ \begin{array}{c} c_1 \\ c_2 \\ c_6 \\ c_{10} \\ c_{11} \\ c_{12} \end{array}\right] = \frac{F_2^2}{\delta^2}
                 \left[ \begin{array}{c} \mp\delta\\
                                         -\delta\delta_\pm \\
                                         \delta\delta_\mp \\
                                         \pm\tfrac{1}{2}\delta \\
                                         \mp 1 \\
                                         -\delta_\pm
                        \end{array}\right],
             \end{equation}
        whereas for $J^P=3/2^\pm$ resonances one obtains
             \begin{equation} \label{pol-32} \renewcommand{\arraystretch}{1.2}
                 \left[ \begin{array}{c} c_1 \\ c_2 \\ c_6 \\ c_{10} \\ c_{11} \\ c_{12} \end{array}\right] = \frac{2}{3\delta^2}
                 \left[ \begin{array}{c} -4\delta\,G_2\\
                                         -4\delta\delta_\pm F_1^2 \\
                                         2\delta F_2\left( 3\delta_\pm F_1 - \delta_\mp F_2\right) \\
                                         -\delta\left( G_2 \pm 3F_1 F_2 \right) \\
                                         \pm 2\left( G_1 - \delta F_1 F_2 \right) \\
                                         2\delta_\pm\left( G_1 + (r\mp 2)\delta_\mp F_1 F_2\right)
                        \end{array}\right].
             \end{equation}
        We abbreviated $G_1 = \delta_\pm^2 F_1^2 + \delta_\mp^2 F_2^2$ and $G_2 = \delta_\pm F_1^2 - \delta_\mp F_2^2$, and
        the $F_i \equiv F_i(Q^2=0)$ denote the static values of the transition form factors.
        With our parametrizations they are fully specified by the parameters $a_0 \pm e_0$.
        For a spin-$1/2$ resonance only the form factor $F_2(0)$ contributes and for a spin-$3/2$ state only $F_1(0)$ and $F_2(0)$ survive.
        These CFFs are related to the scalar and spin polarizabilities through Eqs.~(\ref{polarizability-1}--\ref{polarizability-3});
        for example the scalar polarizabilities $\alpha$ and $\beta$ become 
             \begin{equation} \renewcommand{\arraystretch}{1.2}
                 \left[ \begin{array}{c} \alpha \\ \beta \end{array}\right] = \frac{\alpha_\text{em}}{m^3} \left\{
                 \begin{array}{ll}
                 \displaystyle \frac{F_2^2}{\delta}
                 \left[ \begin{array}{c} -\delta_\mp \\
                                         \delta_\pm
                        \end{array}\right]  & \dots \; J^P = \tfrac{1}{2}^\pm, \\[5mm]
                 \displaystyle \frac{8}{3\delta}
                 \left[ \begin{array}{c}  -\delta_\mp F_2^2 \\
                                         \delta_\pm F_1^2
                        \end{array}\right] &  \dots \; J^P = \tfrac{3}{2}^\pm.
                 \end{array} \right.
             \end{equation}

          \begin{figure}[b!]
         \begin{center}

         \includegraphics[width=0.5\textwidth]{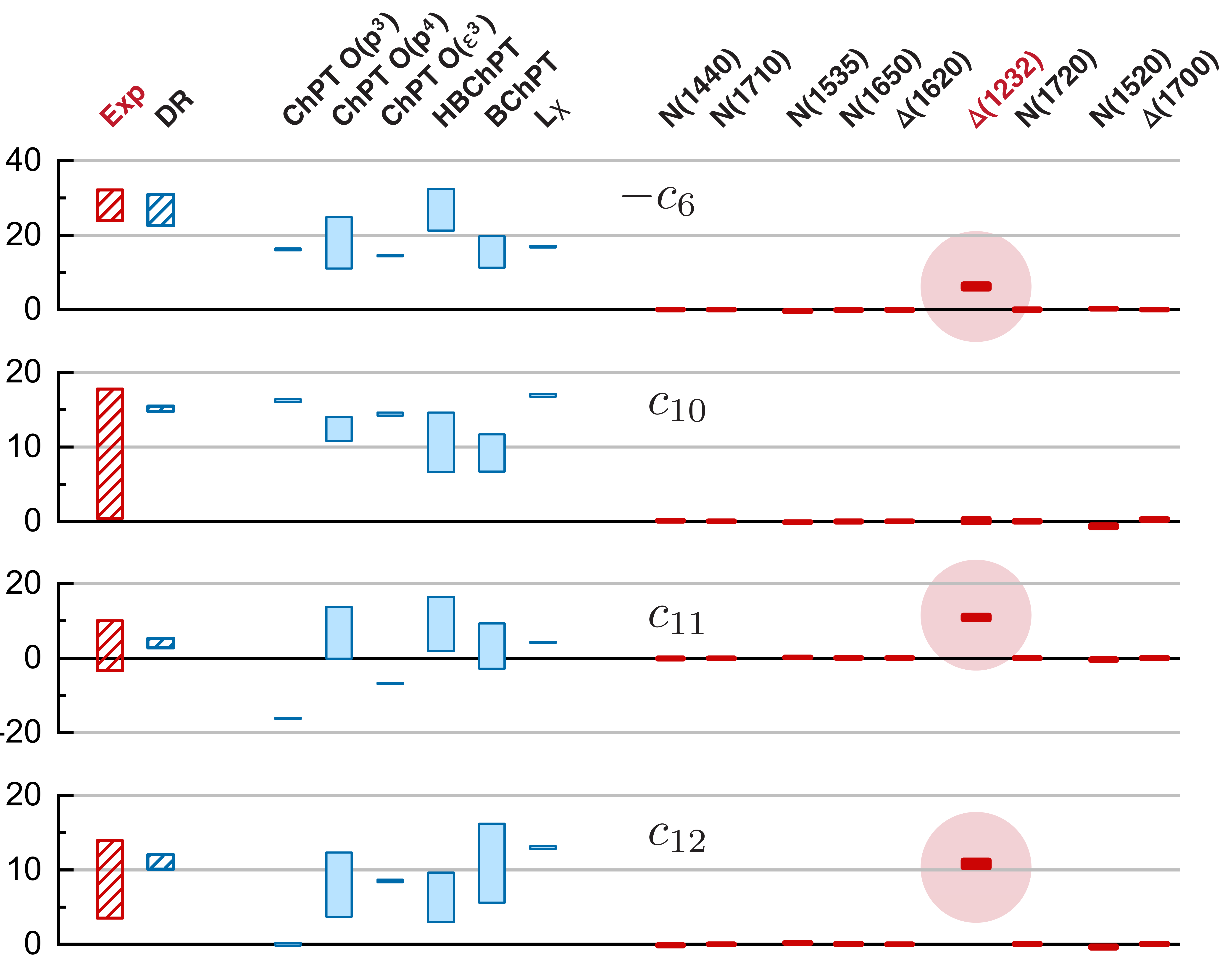}
        \caption{Resonance contributions to the spin polarizabilities encoded in the static values of $c_6$, $c_{10}$, $c_{11}$ and $c_{12}$.
                 We compare to experiment~\cite{Martel:2014pba},
                 dispersion theory~\cite{Holstein:1999uu,Babusci:1998ww,Drechsel:2002ar},
                 leading-order ChPT~\cite{Bernard:1995dp}, and higher-order chiral approaches~\cite{Gellas:2000mx,VijayaKumar:2000pv,Hemmert:1997tj,McGovern:2012ew,Lensky:2015awa,Gasparyan:2011yw};
                 see Refs.~\cite{Drechsel:2002ar,Hagelstein:2015egb} for compilations. For $c_6$ the pion pole contribution from Eq.~\eqref{spin-pol-chpt} has been excluded.}
        \label{fig:results-spin-polarizabilities}

        \end{center}
        \end{figure}

             Those CFFs in Fig.~\ref{fig:results-cffs-1} that are more sensitive to the higher-lying resonances do not contribute to
             the polarizabilities, so that also here mainly the $\Delta(1232)$ is relevant.
             The largest subleading effects come from the $N(1520)$ and $N(1535)$ but they are very small.
             That the $\Delta$ plays an important role is of course well known, and
             chiral effective field theory and dispersive approaches provide a more quantitative description
             than the simple tree-level expressions that we
             collect here~\cite{Drechsel:2002ar,Schumacher:2005an,Griesshammer:2012we,Holstein:2013kia,Hagelstein:2015egb}.
             For example, pion loops reduce the large $\Delta$ contribution 
             to the magnetic polarizability $\beta$ and,
             as a result of this cancellation, $\beta$ is small compared to the electric polarizability $\alpha$.
             The sum $\alpha+\beta$ is proportional to $c_1$ and constrained by the Baldin sum rule~\cite{Baldin1960310}.
             Indeed, Table~\ref{tab:polarizabilities} shows that
             none of the resonances contributes anything substantial to $\alpha$.
             What is noteworthy is the $N(1520)$ contribution to $\beta$,
             which is about a quarter of the size of its PDG value.

             \begin{figure*}[t]
                    \begin{center}
                    \includegraphics[width=1.0\textwidth]{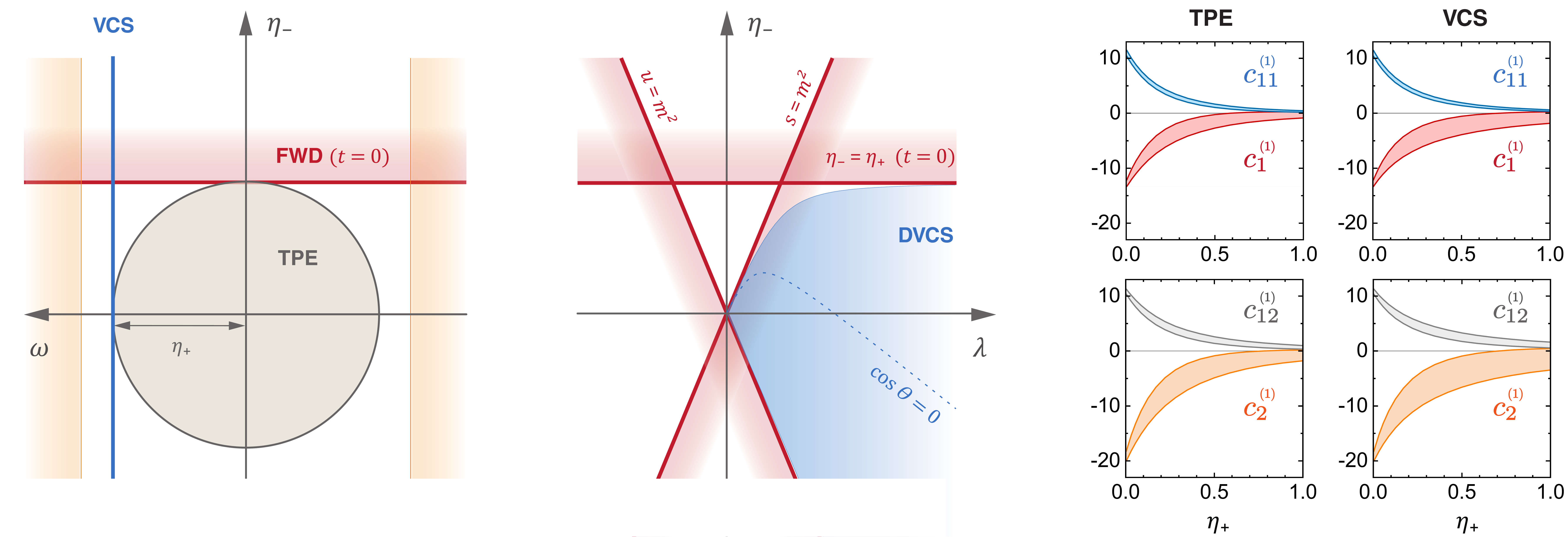}
                    \caption{\textit{Left:} Singularity structure of the Compton scattering amplitude at fixed $\eta_+$ in the $\{\omega,\eta_-\}$
                             and $\{\lambda,\eta_-\}$ planes; see text for a discussion.
                             \textit{Right:} Compton form factor residues for the $\Delta(1232)$ resonance
                             inside the TPE and VCS regions.
                             }\label{fig:vcs}
                    \end{center}
            \end{figure*}

            \pagebreak

             For the spin polarizabilities the situation is less well established. 
             Fig.~\ref{fig:results-spin-polarizabilities} contrasts experimental results from A2/MAMI~\cite{Martel:2014pba} with predictions from dispersion theory and
             chiral perturbation theory.
             Instead of the spin polarizabilities,
             we plot the CFFs directly because their leading ChPT values from Eq.~\eqref{spin-pol-chpt} are simple:
             $c_{12}$ vanishes  and after removing the pion pole in $c_6$ the remaining ones are identical up to signs.
             The various chiral approaches display sizeable uncertainties.
             The difficulty arises from the $\Delta$ resonance, which has a large effect
             and should be incorporated in the description.
             In Fig.~\ref{fig:results-spin-polarizabilities} one can see that even
             by simply adding together the $\Delta$ and leading-order ChPT results
             one obtains values that
             are roughly compatible with experiment and dispersion theory.
             On the other hand, it is reassuring that it is practically \textit{only} the $\Delta$ resonance that has an impact on spin polarizabilities
             because all other resonances are negligible.

           Returning to the CFFs in general kinematics,
           the practical result is that the same 18 CFFs describe the Compton scattering process in general.
           If one had complete knowledge of the Compton amplitude that information could be condensed
           in the 18 panels of Fig.~\ref{fig:results-cffs-1} as well.
           For Compton scattering on a pointlike scalar particle only $c_1$ and $c_2$ survive; for a non-pointlike scalar the first five $c_i$
           contribute; the scalar $t-$channel poles can only appear in $c_2$ and $c_3$; pseudoscalar poles can only appear in $c_6$;
           the nucleon Born term contributes to eight CFFs in Fig.~\ref{fig:nucleon-born}; in RCS the six CFFs discussed above remain;
           in VCS the twelve CFFs from Eq.~\eqref{CFFs-VCS}  and in the forward limit the four combinations in Eq.~\eqref{CFFs-FWD} survive.

           The question is then whether the observation from Fig.~\ref{fig:results-cffs-1} also holds
           in other kinematic limits, i.e., whether the momentum dependencies in the
           variables $\eta_-$, $\omega$ and $\lambda$ are generally weak.
           In general the answer depends on the singularity structure:
           viewed as analytic functions, the CFFs are determined by their physical singularities.
           Our present situation is of course rather special because we merely add up tree-level resonances.
           The formulas for the resonance terms contain $s$ and $u$-channel nucleon resonance poles
           together with timelike poles in $Q^2$ and ${Q'}^2$ from the transition form factors.

           Consider for example the situation in VCS, which is illustrated in Fig.~\ref{fig:vcs}.
           At fixed $\eta_+$, the TPE cone becomes a circle and the VCS and forward planes
           become the lines with $\omega=\eta_+$ and $\eta_-=\eta_+$, respectively.
           In the limit $\eta_+ \to 0$, the circle shrinks to a point and VCS collapses into RCS.
           The vector-meson poles in the transition form factors appear at timelike photon virtualities, which correspond to
           $\omega = \pm(\eta_+ + m_\rho^2/m^2)$ as indicated by the vertical (orange) bands.
           They are symmetric in $\omega$ and the CFFs depend on $\omega$ only quadratically, so
           it is clear that a multipole falloff in the form factors
           cannot induce an overly strong $\omega$ dependence in the interior $|\omega| \lesssim \eta_+$
           but mainly affects the momentum dependence in $\eta_+$.

             \begin{figure}[t]
                    \begin{center}
                    \includegraphics[width=0.73\columnwidth]{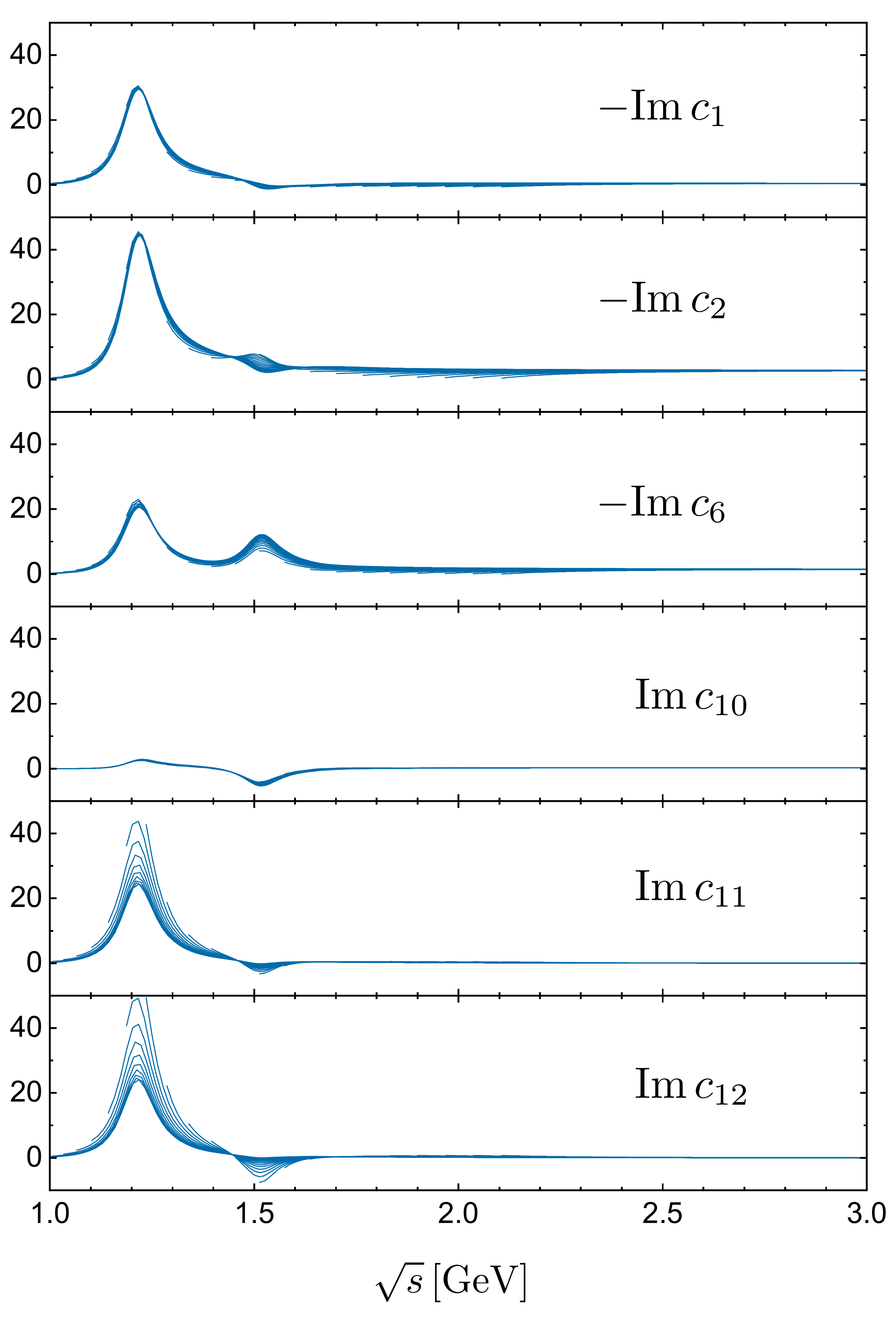}
                    \caption{Imaginary parts of the six Compton form factors in RCS, plotted in the physical $s$-channel region
                             for different values of $\eta_-$ between $-2 < \eta_- < 0$.}\label{fig:rcs}
                    \end{center}
            \end{figure}

           Including also the remaining variable $\lambda$ (again at fixed $\eta_+$) yields the Mandelstam plane in $\lambda$ and $\eta_-$.
           This is where the resonance poles at $\eta_-=\pm 2\lambda$ become visible, which are illustrated by the diagonal (red) bands in Fig.~\ref{fig:vcs}.
           The physical VCS region is the shaded (blue) area with $|\cos\theta| < 1$, where $\theta$ is the CM scattering angle.
           In RCS it would be the domain enclosed between $\eta_-=0$ and 
           $\eta_-=1-\sqrt{1+4\lambda^2}$;
           for increasing $\eta_+$ the line $t=0$ moves upwards and `drags' the physical region with it.
           DVCS is the domain of large $\lambda$ and moderate $t$.

           Clearly, the dependence in $\lambda$ cannot be weak because this is where the resonance bumps appear.
           Recalling the discussion around Eq.~\eqref{nucleon-born-split}, the $\lambda$ dependence of the resonance terms is trivial and can be removed
           by splitting the CFFs into non-resonant and resonant parts,
           \begin{equation*}
              c_i(\eta_+,\eta_-,\omega,\lambda) = c_i^{(0)}(\eta_+,\eta_-,\omega) + \frac{\delta^2\,c_i^{(1)}(\eta_+,\eta_-,\omega)}{(\eta_-+\delta)^2-4\lambda^2}\,,
           \end{equation*}
           where $c_i^{(0)}$ and $c_i^{(1)}$ no longer depend on $\lambda$.
           At fixed $\eta_+$ and with the $\omega$ dependence being weak, the remaining question is how strong their variation in $\eta_-$ is.

           The right panels in Fig.~\ref{fig:vcs} compare some of the  $c_i^{(1)}$ inside the TPE cone and in VCS.
           In the VCS case we limit the range of $\eta_-$ to $-\eta_+ < \eta_- < \eta_+$ with the same extent as the cone,
           whereas inside the cone the functions vary over the full $\eta_-$ and $\omega$ range.
           The bands do not change substantially, which generally also holds for the remaining CFFs and
           means that the VCS region is still sufficiently close to the cone. Thus,
           in principle one could predict the functional dependence of the CFFs in VCS
           from their knowledge, for example, near the symmetric limit where all variables except $\eta_+$ vanish.

           While these observations are particular to the case of resonances,
           they can be useful in more general situations.
           From the viewpoint of analyticity,
           the fact that the CFFs (and therefore structure functions)
           depend on $\lambda$ at all is tied to the $s$- and $u$-channel resonance structure,
           which points to the idea of quark-hadron duality~\cite{Bloom:1970xb,Melnitchouk:2005zr,Ent:2007zz}.
           On the other hand, in the microscopic decomposition of Compton scattering
           the nucleon resonance structure is produced by different quark-gluon topologies than the handbag and $t$-channel diagrams~\cite{Eichmann:2012mp}.
           Without an underlying mechanism to generate singularities in $\lambda$, the momentum dependence in that variable would disappear;
           unless it creates unphysical singularities, but those must ultimately cancel with other parts of the amplitude.
           Such calculations are typically easier to perform inside the TPE cone (the `Euclidean region')
           where one can avoid timelike singularities in the underlying correlation functions like the quark propagator;
           cf.~Refs.~\cite{Eichmann:2017wil,Weil:2017knt} for studies of the pion transition form factor.
           Thus it may be possible to analytically continue results inside the cone, where $\lambda$ is imaginary,
           also to the domain of large and real~$\lambda$ which is accessible in DVCS.

        Finally, the discussion in terms of Lorentz-invariant and constraint-free CFFs can be useful
        for amplitude analyses of Compton scattering~\cite{Krupina:2017pgr}.
        For illustration we plot in Fig.~\ref{fig:rcs} the reconstructed Compton amplitude in RCS,
        in particular the imaginary parts of the six associated CFFs, inside the physical RCS region and between $-2 < \eta_- < 0$.
        We equipped the resonances with widths according to Eq.~\eqref{delta-def},
        where $m_R$ and $-\Gamma/2$ are taken to be the real and imaginary parts of the respective pole positions from the PDG~\cite{PDG2016}.
        Analogous plots can be drawn in the forward limit, where the imaginary parts are proportional to the nucleon's structure functions, or in VCS.
        The $\Delta(1232)$ clearly dominates, whereas other resonances such as the $N(1520)$ are enhanced in particular CFFs; and
        with Tables~\ref{tab:j=1/2+born} and~\ref{tab:j=3/2+born} one can form CFF combinations where resonances with specific $J^P$ drop out.

        \section{Summary and outlook} \label{sec:outlook}

        In this work we have detailed the tensor basis construction for Compton scattering on the nucleon,
        which implements the constraints of electromagnetic gauge invariance, crossing symmetry, and the absence of kinematic singularities.
        The resulting 18 Lorentz-invariant Compton form factors are free of kinematic constraints and describe the process in general kinematics.

        As a practical application we worked out the Compton form factor contributions coming
        from intermediate $J^P=1/2^\pm$ and $3/2^\pm$ nucleon resonances.
        We derived the general forms for their offshell nucleon-to-resonance transition vertices
        according to electromagnetic and spin-$3/2$ gauge invariance. This automatically defines constraint-free
        onshell transition form factors, for which we constructed fits using the available experimental data.
        We find that apart from the $\Delta(1232)$ the resonance contributions to the scalar and spin polarizabilites are very small,
        although the $N(1520)$ could play a role for the proton's magnetic polarizability.

        Our study can be extended to calculate resonance contributions to two-photon exchange processes or baryons with higher spin.
        Moreover, since the tensor basis construction follows a general recipe it provides a template for other systems, such as
        the hadronic light-by-light scattering amplitude which enters in the Standard Model prediction
        for the muon anomalous magnetic moment.

         \subsection*{Acknowledgments}

         We acknowledge interactions with B.~Bakker, C.~S.~Fischer, R.~Gothe and V.~Mokeev.
         This work was supported by the FCT Investigator Grant IF/00898/2015
         and the Funda\c{c}\~ao de Amparo \`a Pesquisa do Estado de S\~ao Paulo (FAPESP) under
         project no.~2017/02684-5 and grant no.~2017/17020-BCO-JP.


\begin{appendix}

   \section{Conventions and formulas} \label{sec:euclidean}

   \subsection{Euclidean vs. Minkowski conventions}

   We use Euclidean conventions throughout this paper, but
   we took care in constructing the notation
   such that many identities are the same in Euclidean and Minkowski conventions.
   In the following we summarize the transcription rules from a Minkowski metric with signature $(+,-,-,-)$
   to a Euclidean metric $(+,+,+,+)$;
   more relations can be found in App.~A of Ref.~\cite{Eichmann:2016yit}.

         The replacement rules for vectors $a^\mu$ and tensors $T^{\mu\nu}$ are given by
         \begin{equation}\label{rules1}\renewcommand{\arraystretch}{1.0}
             a^\mu_E = \left[\begin{array}{c}\vect{a}\\ia_0\end{array}\right], \quad
             T^{\mu\nu}_E = \left[\begin{array}{cc} T^{ij} & iT^{i0} \\ iT^{0i} & -T^{00}\end{array}\right],
         \end{equation}
         where `$E$' stands for Euclidean and no subscript refers to the Minkowski quantity.
         As a consequence, the Lorentz-invariant scalar product of any two four-vectors differs by a minus sign from its Minkowski counterpart:
         \begin{equation}\label{consequences1}
             a_E \cdot b_E = \sum_{k=1}^4 a_E^k \, b_E^k = -a\cdot b\,.
         \end{equation}
         Therefore, a vector is spacelike if $a^2>0$ and timelike if $a^2<0$.
         Because the metric is positive,  the distinction between upper and lower indices disappears.
         To preserve the meaning of the slash $\slashed{a} = a^0 \gamma^0 - \vect{a}\cdot\vect{\gamma}$
         we must also redefine the $\gamma-$matrices:
         \begin{equation}\label{rules2}\renewcommand{\arraystretch}{1.0}
             i\gamma^\mu_E = \left[\begin{array}{c} \vect{\gamma}\\ i\gamma_0\end{array}\right], \qquad
             \gamma^5_E = \gamma^5\,,
         \end{equation}
         so that
         \begin{equation}\label{consequences2}
             \slashed{a}_E = a_E \cdot \gamma_E = i\slashed{a}, \qquad
             \{ \gamma^\mu_E, \gamma^\nu_E \} = 2\delta^{\mu\nu}\,.
         \end{equation}

         Our sign convention for the Euclidean $\gamma-$matrices changes all signs in the Clifford algebra relation~\eqref{consequences2} to be positive,
         and since this implies $(\gamma_E^i)^2=1$ for $i=1 \dots 4$ we can choose them to be hermitian: $\gamma_E^\mu = \left( \gamma^\mu_E\right)^\dag$.
         For example, in the standard representation they read
            \begin{equation*}\renewcommand{\arraystretch}{1.0}
                \gamma^k_E  =  \left[ \begin{array}{cc} 0 & -i \tau_k \\ i \tau_k & 0 \end{array} \right] , \;\;
                \gamma^4_E  =  \left[ \begin{array}{c@{\quad}c} \mathds{1} & 0 \\ 0 & \!\!-\mathds{1} \end{array} \right] , \;\;
                \gamma^5  =  \left[ \begin{array}{c@{\quad}c} 0 & \mathds{1} \\ \mathds{1} & 0 \end{array} \right] ,
            \end{equation*}
            where the $\tau_k$ are the usual Pauli matrices.
         Also the generators of the Clifford algebra are then hermitian:
         \begin{equation}
             \sigma^{\mu\nu} = \frac{i}{2}\,[\gamma^\mu,\gamma^\nu] \quad \Rightarrow \quad
             \sigma_E^{\mu\nu} = -\frac{i}{2}\,[\gamma^\mu_E, \gamma^\nu_E]
         \end{equation}
         with $(\sigma_E^{\mu\nu})^\dag = \sigma_E^{\mu\nu}$.

         The resulting replacement rules for some typical quantities are collected in Table~\ref{tab:minkowski-euclidean}.
         For expressions involving the $\varepsilon-$tensor the situation is slightly more complicated but follows from the same principles:
         the spatial parts of Lorentz tensors are identical in Minkowski and Euclidean conventions, so this must also hold for
         $\varepsilon^{\mu\nu\alpha\beta} a_\alpha  b_\beta$.
         In Euclidean space the $\varepsilon-$tensor is defined by $\varepsilon_{1234} = \varepsilon^{1234} = 1$, whereas
         in Minkowski conventions one has $\varepsilon_{0123} = -\varepsilon^{0123} = 1$, i.e., the spatial components switch sign when lowering or raising indices.
         Denoting spatial indices by $i,j,k$ and summing over $k$, one has
         \begin{equation}
         \begin{split}
             \varepsilon^{ij\alpha\beta} a_\alpha b_\beta
              &= \varepsilon^{ijk0}\,(a_k  b_0 - a_0   b_k) \\
              &= -\varepsilon^{ijk0}\,(a^k b^0 - a^0  b^k) \\
              &= i\varepsilon^{ijk4}\,(a^k b^4 - a^4 b^k)_E  \\
              &= \big( i\varepsilon^{ij\alpha\beta}  a^\alpha  b^\beta \big)_E\,,
         \end{split}
         \end{equation}
         because $\varepsilon^{1234} = 1 = \varepsilon^{1230}$ and $a^0 = -ia^4_E$.
         Repeating this for rank-1 and rank-3 tensors results in the analogous identities in Table~\ref{tab:minkowski-euclidean}
         (which would also follow from Eq.~\eqref{eps-01} below).

        \begin{table}[t]

             \begin{equation*}
             \renewcommand{\arraystretch}{1.2}
             \begin{array}{ @{\quad} l @{\quad} | @{\quad} l  @{\quad}  }

                 \text{Minkowski} &  \text{Euclidean} \\[1mm] \hline  \rule{-0.0mm}{0.5cm}

                 a\cdot b      & -a\cdot b  \\
                 a^\mu      & a^\mu  \\
                 \gamma^\mu & i\gamma^\mu \\
                 \gamma_5   & \gamma_5 \\
                 \slashed{a}  & -i\slashed{a} \\
                 g^{\mu\nu}   & -\delta^{\mu\nu} \\
                 a^\mu b^\nu  & a^\mu b^\nu \\
                 {[} \gamma^\mu,\gamma^\nu {]} & - {[} \gamma^\mu,\gamma^\nu {]} \\
                 {[} \gamma^\mu,\slashed{a} {]} &  {[} \gamma^\mu,\slashed{a} {]} \\
                 {[} \gamma^\mu,\gamma^\nu, \slashed{a} {]} & i{[}\gamma^\mu,\gamma^\nu, \slashed{a} {]} \\
                 {[} \slashed{a},\slashed{b} {]} & -{[} \slashed{a},\slashed{b} {]} \\[2mm]
                 \varepsilon^{\mu\nu\rho\alpha} a_\alpha & i\varepsilon^{\mu\nu\rho\alpha} a^\alpha \\
                 \varepsilon^{\mu\nu\alpha\beta} a_\alpha \,b_\beta & i\varepsilon^{\mu\nu\alpha\beta} a^\alpha  b^\beta \\
                 \varepsilon^{\mu\alpha\beta\gamma} a_\alpha \,b_\beta \,c_\gamma & i\varepsilon^{\mu\alpha\beta\gamma} a^\alpha  b^\beta  c^\gamma\\
                 \varepsilon^{\mu\nu\alpha\beta} a_\alpha  \gamma_\beta  & -\varepsilon^{\mu\nu\alpha\beta} a^\alpha  \gamma^\beta

             \end{array}
             \end{equation*}

             \caption{Replacement rules for some typical quantities.
             For expressions with Lorentz indices, the right column defines their Euclidean version in the sense of Eqs.~\eqref{rules1} and \eqref{rules2}.
             Each additional Minkowski summation over Lorentz indices leads to a minus sign in Euclidean conventions.}

             \label{tab:minkowski-euclidean}

          \end{table}

         With these rules it is straightforward to transform Euclidean tensors, such as for example those in Table~\ref{basis-type-I},
         to Minkowski space. To further facilitate the transcription, we have introduced the variables $\eta_+$, $\eta_-$, $\lambda$ and $\omega$
         in the main text, Eq.~\eqref{li-1}, because they allow for a common definition in Euclidean and Minkowski space. For example, with $q_E^\mu = Q^\mu$:
         \begin{equation}
            \eta_+ = \frac{Q^2+{Q'}^2}{2m^2} = -\frac{q^2+{q'}^2}{2m^2} \,,
         \end{equation}
         and so on for the remaining variables. Once the momentum variables and Lorentz tensors are given appropriate names,
         all Lorentz-covariant and Lorentz-invariant relations between them
         are the same in Euclidean and Minkowski conventions.

            The advantage of the Euclidean metric is that one can 
            perform numerical calculations directly in a given frame, with explicit $\gamma$ matrices and without the need for inserting the metric tensor in each summation.
            A practical Lorentz frame for the momenta $p$, $\Sigma$ and $\Delta$ defined in Eq.~\eqref{kinematics-1} is~\cite{Eichmann:2012mp}
             \begin{equation}\label{simple-frame}
             \begin{split}
                 \frac{\Delta^\mu}{m} &= 2\sqrt{t}\left[\begin{array}{c} 0 \\ 0 \\ 0 \\ 1 \end{array}\right], \quad
                 \frac{\Sigma^\mu}{m}=\sqrt{\sigma}\left[\begin{array}{c} 0 \\ 0 \\ \sqrt{1-Z^2} \\ Z \end{array}\right],  \\
                 &  \frac{p^\mu}{m}=i \sqrt{1+t}\left[\begin{array}{c} 0 \\ \sqrt{1-Y^2} \\ Y \\ 0 \end{array}\right],
             \end{split}
             \end{equation}
             where $t$, $\sigma$, $Z$ and $Y$ are defined in Eqs.~(\ref{t-def}--\ref{li-3}).
             Inside the TPE cone, the angular variables $Z$ and $Y$ fill the interval $[-1,1]$ whereas in its exterior they can become complex.
             In the VCS limit it is more convenient to use the frame where $p^\mu = im\sqrt{1+t}\,[0,0,0,1]$ and
             \begin{equation}\label{VCS-frame}
                 \frac{{Q'}^\mu}{m} = \alpha\left[\begin{array}{c} 0 \\ 0 \\ 1 \\ i \end{array}\right], \quad
                 \frac{Q^\mu}{m}=\left[\begin{array}{c} 0 \\ \sqrt{4t-\beta^2} \\ \alpha + \beta \\ i\alpha \end{array}\right],
             \end{equation}
             with $\alpha = \lambda/\sqrt{1+t}$ and $\beta=\eta_-/\alpha$.
             In any case, this does not affect the CFFs because they are frame-independent.

   \subsection{Formulas} \label{sec:formulas}

            Dropping the index `$E$', we collect some useful Euclidean formulas. The $\gamma_5$ matrix is defined by
            \begin{equation}
                \gamma^5 = -\gamma^1 \gamma^2 \gamma^3 \gamma^4 = -\frac{1}{24} \,\varepsilon^{\mu\nu\rho\sigma} \gamma^\mu \gamma^\nu \gamma^\rho \gamma^\sigma
            \end{equation}
            with $\varepsilon^{1234} = 1$.
            It is convenient to define the fully antisymmetric combinations of Dirac matrices via the commutators
            \begin{align}
            [A,B] &= AB-BA \,, \\[1mm]
            [A,B,C] &= [A,B]\,C + [B,C]\,A + [C,A]\,B\,, \label{three-commutator} \\[1mm]
            [A,B,C,D] &= [A,B,C]\,D + [B,C,D]\,A  \nonumber \\
                      &+ [C,D,A]\,B + [D,A,B]\,C\,.
            \end{align}
            Inserting $\gamma-$matrices, this yields the antisymmetric combinations
            \begin{align}
            [\gamma^\mu,\gamma^\nu] &= \gamma_5\,\varepsilon^{\mu\nu\alpha\beta}\,\gamma^\alpha \gamma^\beta \,, \\
            \tfrac{1}{6}\,[\gamma^\mu,\gamma^\nu,\gamma^\rho] &= \tfrac{1}{2}\,(\gamma^\mu \gamma^\nu \gamma^\rho - \gamma^\rho \gamma^\nu \gamma^\mu) \nonumber \\
                                                              &= \tfrac{1}{4}\left\{ [\gamma^\mu,\gamma^\nu], \gamma^\rho \right\} \label{triple-commutator}\\
                                                              &= -\gamma_5\,\varepsilon^{\mu\nu\rho\sigma}  \gamma^\sigma\,, \nonumber \\
            \tfrac{1}{24}\,[\gamma^\mu,\gamma^\nu,\gamma^\alpha,\gamma^\beta] &= -\gamma_5\,\varepsilon^{\mu\nu\alpha\beta}\,. \label{eps-01}
            \end{align}
            The various contractions of $\varepsilon-$tensors are given by
            \begin{equation}
            \begin{split}
               \varepsilon^{\mu\nu\rho\lambda}\,\varepsilon^{\alpha\beta\gamma\lambda} &= \delta^{\mu\alpha}\,( \delta^{\nu\beta}\,\delta^{\rho\gamma}
                                                                                         - \delta^{\nu\gamma}\,\delta^{\rho\beta}) \\
                                                                                       & + \delta^{\mu\beta}\,( \delta^{\nu\gamma}\,\delta^{\rho\alpha} - \delta^{\nu\alpha}\,\delta^{\rho\gamma}) \\
                                                                                       & + \delta^{\mu\gamma}\,(\delta^{\rho\beta}\,\delta^{\nu\alpha} - \delta^{\rho\alpha}\,\delta^{\nu\beta})\,, \\
               \tfrac{1}{2}\,\varepsilon^{\mu\nu\lambda\sigma}\,\varepsilon^{\alpha\beta\lambda\sigma} &= \delta^{\mu\alpha}\,\delta^{\nu\beta} - \delta^{\mu\beta}\,\delta^{\nu\alpha}\,, \\
               \tfrac{1}{6}\,\varepsilon^{\mu\lambda\sigma\tau}\,\varepsilon^{\alpha\lambda\sigma\tau} &= \delta^{\mu\alpha}\,, \\
               \tfrac{1}{24}\,\varepsilon^{\lambda\sigma\tau\omega}\,\varepsilon^{\lambda\sigma\tau\omega} &= 1
            \end{split}
            \end{equation}
            and the $\varepsilon-$tensor satisfies
            \begin{equation}
               a^{ \{ \mu }   \varepsilon^{ \alpha\beta\gamma\delta \} }  = 0\,,
            \end{equation}
            where $a^\mu$ is an arbitrary four-vector and $\{ \dots \}$ denotes a symmetrization of indices.

            Four-momenta are conveniently expressed through hyperspherical coordinates:
            \begin{equation}\label{APP:momentum-coordinates}\renewcommand{\arraystretch}{1.0}
                p^\mu = \sqrt{p^2} \left[ \begin{array}{l} \sqrt{1-z^2}\,\sqrt{1-y^2}\,\sin{\phi} \\
                                                           \sqrt{1-z^2}\,\sqrt{1-y^2}\,\cos{\phi} \\
                                                           \sqrt{1-z^2}\;\;y \\
                                                           \;\; z
                                         \end{array}\right],
            \end{equation}
            and a four-momentum integration reads:
            \begin{equation} \label{hypersphericalintegral}
            \begin{split}
                \int \!\!\frac{d^4 p}{(2\pi)^4}
                       &=  \frac{1}{(2\pi)^4}\,\frac{1}{2} \int_0^{\infty} dp^2 \,p^2 \int_{-1}^1 dz\,\sqrt{1-z^2}  \\
                       & \qquad \qquad \times \int_{-1}^1 dy \int_0^{2\pi} d\phi \,.
            \end{split}
            \end{equation}

        The positive- and negative-energy onshell spinors for spin-1/2 particles satisfy the Dirac equations
        \begin{equation}
        \begin{split}
            (i\slashed{p}+m)\,u(\vect{p}) &= 0 = \conjg{u}(\vect{p})\,(i\slashed{p}+m)\,, \\
            (i\slashed{p}-m)\,v(\vect{p}) &= 0 = \conjg{v}(\vect{p})\,(i\slashed{p}-m)\,,
        \end{split}
        \end{equation}
        where the conjugate spinor is $\conjg{u}(\vect{p}) = u(\vect{p})^\dag \gamma^4$.
        Since the onshell spinors only depend on $\vect{p}$ they are the same as in Minkowski space;
        for example in the standard representation:
        \begin{equation}\renewcommand{\arraystretch}{1.2}
            u_s(\vect{p}) =   \sqrt{\frac{E_p+m}{2m}}\left(\begin{array}{c} \xi_s \\ \frac{\vect{p}\cdot\vect{\tau}}{E_p+m}\,\xi_s \end{array}\right)
        \end{equation}
        with
        \begin{equation*}
            \xi_+ = \left(\begin{array}{c} 1 \\ 0 \end{array}\right),  \quad
            \xi_- = \left(\begin{array}{c} 0 \\ 1 \end{array}\right),  \qquad
            E_p = \sqrt{\vect{p}^2 + m^2}\,.
        \end{equation*}
        We have normalized them to unity,
        \begin{equation}
        \begin{split}
           \conjg{u}_s(\vect{p})\,u_{s'}(\vect{p}) &=  -\conjg{v}_s(\vect{p})\,v_{s'}(\vect{p}) = \delta_{ss'}\,, \\
           \conjg{u}_s(\vect{p})\,v_{s'}(\vect{p}) &= \conjg{v}_s(\vect{p})\,u_{s'}(\vect{p}) = 0\,,
        \end{split}
        \end{equation}
        and their completeness relations define the positive- and negative-energy projectors:
        \begin{equation}\label{positive-energy-projector}
        \begin{split}
           \sum_s u_s(\vect{p})\,\conjg{u}_s(\vect{p}) &= \frac{-i\slashed{p}+m}{2m} = \Lambda_+(p)\,, \\
           \sum_s v_s(\vect{p})\,\conjg{v}_s(\vect{p}) &= \frac{-i\slashed{p}-m}{2m} = -\Lambda_-(p)\,.
        \end{split}
        \end{equation}
        Therefore, $\Lambda_+(p)\,u(\vect{p}) = u(\vect{p})$ and $\Lambda_-(p)\,u(\vect{p}) = 0$.

    \renewcommand{\arraystretch}{1.3}

             \begin{table*}[t]

             \begin{equation*}\label{tarrach-K}
             \begin{array}{@{\quad} c @{\quad} | @{\quad} rr @{\!\;\,} l @{\!\;\,}c r @{\!\;\,} l @{\quad}   }

                 n & \multicolumn{4}{l}{\text{Basis element}} \\[1mm] \hline\hline \rule{-0.0mm}{0.5cm}
                 0 &       K_1    = & -               & T_1    &=&               & \; \delta^{\mu\nu} \\
                 2 & m^2 \,K_2    = &                 & T_2    &=&               & \; {Q'}^\mu Q^\nu \\
                 2 & m^2 \,K_3    = &                 & T_3    &=&               & \; Q^\mu {Q'}^\nu \\
                 2 & m^2 \,K_4    = &                 & T_4    &=&               & \; Q^\mu Q^\nu + {Q'}^\mu {Q'}^\nu \\
                 4 & m^2 \,K_5    = &  \omega         & T_5    &=& \omega        & (Q^\mu Q^\nu - {Q'}^\mu {Q'}^\nu) \\[2mm]

                 0 & m^2 \,K_6    = &                 & T_6    &=&               & \; p^\mu p^\nu \\
                 2 & m^2 \,K_7    = &  \lambda        & T_7    &=& \lambda       & (p^\mu {Q'}^\nu + Q^\mu p^\nu) \\
                 4 & m^2 \,K_8    = & -\lambda\omega  & T_8    &=& \lambda\omega & (p^\mu {Q'}^\nu - Q^\mu p^\nu) \\
                 2 & m^2 \,K_{9}  = &  \lambda        & T_9    &=& \lambda       & (p^\mu {Q}^\nu + {Q'}^\mu p^\nu) \\
                 4 & m^2 \,K_{10} = & -\lambda\omega  & T_{10} &=& \lambda\omega & (p^\mu {Q}^\nu - {Q'}^\mu p^\nu)  \\[2mm]

                 3 & m^2 \,K_{27} = &  \omega         & T_{27} &=& \omega        & \big[ \, p^\mu \gamma^\nu + \gamma^\mu p^\nu, \,\slashed{\Sigma}\,\big] \\
                 1 & m^2 \,K_{28} = & -               & T_{28} &=&               & \big[ \, p^\mu \gamma^\nu - \gamma^\mu p^\nu, \,\slashed{\Sigma}\,\big] \\
                 5 & m^2 \,K_{29} = &  \lambda\omega  & T_{29} &=& \lambda\omega & \big[ \, {Q'}^\mu \gamma^\nu + \gamma^\mu {Q}^\nu, \,\slashed{\Sigma}\,\big] \\
                 3 & m^2 \,K_{30} = & -\lambda        & T_{30} &=& \lambda       & \big[ \,{Q'}^\mu \gamma^\nu - \gamma^\mu {Q}^\nu, \,\slashed{\Sigma}\,\big] \\
                 5 & m^2 \,K_{31} = &  \lambda\omega  & T_{31} &=& \lambda\omega & \big[ \,{Q}^\mu \gamma^\nu + \gamma^\mu {Q'}^\nu, \,\slashed{\Sigma}\,\big] \\
                 3 & m^2 \,K_{32} = & -\lambda        & T_{32} &=& \lambda       & \big[ \,{Q}^\mu \gamma^\nu - \gamma^\mu {Q'}^\nu, \,\slashed{\Sigma}\,\big] \\[2mm]

                 1 & K_{33}       = &  \lambda        & T_{33} &=& \lambda       & [\gamma^\mu, \gamma^\nu]

             \end{array} \qquad\quad
             \begin{array}{@{\quad} c @{\quad} | @{\quad} rr @{\!\;\,} l @{\!\;\,}c r @{\!\;\,} l @{\quad}   }

                 n & \multicolumn{4}{l}{\text{Basis element}} \\[1mm] \hline\hline \rule{-0.0mm}{0.5cm}
                 2 & im   \,K_{11} = & -i\lambda        & T_{11} &=& \lambda       & \; \delta^{\mu\nu} \,\slashed{\Sigma} \\
                 4 & im^3 \,K_{12} = &  i\lambda        & T_{12} &=& \lambda       & \; {Q'}^\mu Q^\nu  \,\slashed{\Sigma} \\
                 4 & im^3 \,K_{13} = &  i\lambda        & T_{13} &=& \lambda       & \; Q^\mu {Q'}^\nu  \,\slashed{\Sigma} \\
                 4 & im^3 \,K_{14} = &  i\lambda        & T_{14} &=& \lambda       & (Q^\mu Q^\nu + {Q'}^\mu {Q'}^\nu)\, \slashed{\Sigma} \\
                 6 & im^3 \,K_{15} = &  i\lambda\omega  & T_{15} &=& \lambda\omega & (Q^\mu Q^\nu - {Q'}^\mu {Q'}^\nu)\, \slashed{\Sigma} \\[2mm]

                 2 & im^3 \,K_{16} = &  i\lambda        & T_{16} &=& \lambda       & \; p^\mu p^\nu \,\slashed{\Sigma}\\
                 2 & im^3 \,K_{17} = &  i               & T_{17} &=&               & (p^\mu {Q'}^\nu + Q^\mu p^\nu)\,\slashed{\Sigma} \\
                 4 & im^3 \,K_{18} = & -i\omega         & T_{18} &=& \omega        & (p^\mu {Q'}^\nu - Q^\mu p^\nu)\,\slashed{\Sigma} \\
                 2 & im^3 \,K_{19} = &  i               & T_{19} &=&               & (p^\mu {Q}^\nu + {Q'}^\mu p^\nu)\,\slashed{\Sigma} \\
                 4 & im^3 \,K_{20} = & -i\omega         & T_{20} &=& \omega        & (p^\mu {Q}^\nu - {Q'}^\mu p^\nu)\,\slashed{\Sigma}  \\[2mm]

                 0 & im   \,K_{21} = & -i               & T_{21} &=&               & \; p^\mu \gamma^\nu + \gamma^\mu p^\nu \\
                 2 & im   \,K_{22} = &  i\omega         & T_{22} &=& \omega        & (p^\mu \gamma^\nu - \gamma^\mu p^\nu) \\
                 2 & im   \,K_{23} = & -i\lambda        & T_{23} &=& \lambda       & ({Q'}^\mu \gamma^\nu + \gamma^\mu {Q}^\nu) \\
                 4 & im   \,K_{24} = &  i\lambda\omega  & T_{24} &=& \lambda\omega & ({Q'}^\mu \gamma^\nu - \gamma^\mu {Q}^\nu) \\
                 2 & im   \,K_{25} = & -i\lambda        & T_{25} &=& \lambda       & ({Q}^\mu \gamma^\nu + \gamma^\mu {Q'}^\nu) \\
                 4 & im   \,K_{26} = &  i\lambda\omega  & T_{26} &=& \lambda\omega & ({Q}^\mu \gamma^\nu - \gamma^\mu {Q'}^\nu) \\[2mm]

                 1 & im   \,K_{34} = &  i               & T_{34} &=&               & \big\{ \, [\gamma^\mu, \gamma^\nu], \,\slashed{\Sigma}\,\big\}

             \end{array}
             \end{equation*}

               \caption{Elementary tensors for the nucleon Compton scattering amplitude. We suppressed the Lorentz indices for $K_i^{\mu\nu}$ and $T_i^{\mu\nu}$ to avoid clutter.
               The $K_i$ are dimensionless and invariant under photon crossing and charge conjugation.
               The $T_i$ are the same as in Eq.~(8) of Ref.~\cite{Tarrach:1975tu}.
               $n$ counts the powers of photon momenta; $\lambda$ contributes one power and $\omega$ two.
               }
               \label{basis-type-I}

             \end{table*}

    \renewcommand{\arraystretch}{1.0}

   \newpage

   \section{Tensor basis}\label{sec:tensor-basis}

        In the following we derive the tensor basis of the nucleon Compton amplitude given in Table~\ref{transverse-basis-0}
        and Eq.~\eqref{basis-explicit}. To begin with, we follow the construction by Tarrach~\cite{Tarrach:1975tu}
        and define the 34 auxiliary tensors $K_i^{\mu\nu}$ in Table~\ref{basis-type-I}.
        The $T_i^{\mu\nu}$ are the Euclidean versions of Tarrach's Eq.~(8) according to the replacement rules in App.~\ref{sec:euclidean}.
        We construct the $K_i^{\mu\nu}$ by attaching prefactors of $\lambda$, $\omega$ or $\lambda\omega$,
        which makes them even under Bose symmetry and charge conjugation, cf.~Eq.~\eqref{bose+cc}, and powers of the nucleon mass $m$
        to make them dimensionless. Thus, the initial nucleon Compton scattering amplitude has the form
        \begin{equation}\label{ca-01}
            \Gamma^{\mu\nu}(p,Q',Q) = \Lambda_+^f \left[ \sum_{i=1}^{34} g_i\,K_i^{\mu\nu} \right] \Lambda_+^i\,.
        \end{equation}
        Here we abbreviated the positive-energy projectors from Eq.~\eqref{pos-energy-proj} by $\Lambda_+^{f,i}$ and
        the dressing functions $g_i$ depend on the variables $\eta_+$, $\eta_-$, $\omega$ and $\lambda$
        defined in Eq.~\eqref{li-1}.

        That the $K_i$ are fully symmetric will be important in what follows, because it implies that the
        $g_i$ are even in both $\omega$ and $\lambda$.
        The analysis in terms of the $T_i$ would complicate the discussion of kinematic singularities;
        take for example the contribution from $K_{33}$:
        \begin{equation}\label{kin-sing-disc-1}
           g_{33}\,K_{33} = \lambda\,g_{33}\,T_{33} = g_{33}'\,T_{33}\,.
        \end{equation}
        Clearly, $g_{33}=g_{33}'/\lambda$ does not have a kinematic singularity at $\lambda=0$ because $g_{33}'$ is odd under
        photon crossing and therefore must be proportional to $\lambda$.

        There are two non-trivial linear dependencies between the $K_i^{\mu\nu}$, namely~\cite{Tarrach:1975tu}
        \begin{equation}\label{linear-relations-1}
        \begin{split}
             & K_{17}-K_{19}+ K_{22} - K_{23} + K_{25} + K_{28} - K_{33} \\
            &+ \left( 1 + \frac{\eta_+-\eta_-}{2}\right)\frac{K_{34}}{2} = 0
        \end{split}
        \end{equation}
        and
        \begin{equation}\label{linear-relations-2}
        \begin{split}
             &\frac{K_8+K_{10}}{2} + K_{12}-K_{13} \\
             &- \frac{K_{24}+K_{26}}{2} + \frac{K_{29}-K_{31}}{4} \\
            & -\lambda^2\left( K_2-K_3 + K_{28} + \frac{K_{34}}{2}\right) \\
            &+ \frac{\eta_++\eta_-}{2}\,(K_7-K_9-K_{23}+K_{25}) \\
            &+ \frac{\eta_+-\eta_-}{4}\,(K_{30}+K_{32})  \\
            &+ \left( \lambda^2 + \frac{\eta_+^2-\eta_-^2 - \omega^2}{4}\right) K_{33} = 0\,.
        \end{split}
        \end{equation}
        These relations hold inside the positive-energy projectors of Eq.~\eqref{ca-01}.
        Therefore, only 32 tensors are linearly independent.
        This is analogous to the discussion of the light-by-light scattering amplitude in Ref.~\cite{Eichmann:2015nra}:
        an $n$-point function depends on $n-1$ momenta, but with increasing $n$ one can only construct a limited
        number of orthogonal momenta due to the fixed dimension of spacetime.
        In practice this leads to relations between the basis elements and thus
        to a smaller number of independent tensors than what can be naively written down.

        One must therefore eliminate two tensors in such a way that the resulting 32 coefficients $g_i$ do not pick up kinematic singularities.
        To do so, it is sufficient to eliminate one tensor from the first row in Eq.~\eqref{linear-relations-1}
        and another from the first two rows in Eq.~\eqref{linear-relations-2}.
        Within this constraint, any choice is equivalent
         and we choose to
        eliminate $K_{22}$ from the first equation and $K_{31}$ from the second one.
        After crossing off
        these two tensors from Eq.~\eqref{ca-01} the sum goes over 32 linearly independent tensors.

   \subsection{Transverse part}
   \label{appendix-TransversePart}

        To derive the transverse part $\Gamma^{\mu\nu}_{\perp\perp}$ in Eq.~\eqref{generic-decomposition}, we work out the transversality conditions
        \begin{equation}\label{derive-transverse-1}
        \begin{split}
            {Q'}^\mu\,\Gamma^{\mu\nu}(p,Q',Q) &\stackrel{!}{=} 0, \\
            {Q}^\nu\,\Gamma^{\mu\nu}(p,Q',Q) &\stackrel{!}{=} 0.
        \end{split}
        \end{equation}
        Either one of these is sufficient as long as we respect photon-crossing and charge-conjugation invariance.
        For example, the contraction of~\eqref{ca-01} with ${Q'}^\mu$,
        \begin{equation}
             \sum_{ i=1 \atop \backslash \{ 22,31\}}^{34} g_i\,{Q'}^\mu K_i^{\mu\nu} = \sum_{j=1}^8 A_j\,K_j^\nu \stackrel{!}{=} 0\,,
        \end{equation}
         produces eight linearly independent tensors $K_j^\nu$:
        \begin{equation*}
            p^\nu, \;\; {Q'}^\nu, \;\;  Q^\nu, \;\;
                            p^\nu \slashed{\Sigma}, \;\;   {Q'}^\nu\slashed{\Sigma}, \;\;    Q^\nu\slashed{\Sigma}, \;\;
                            \gamma^\nu, \;\; [\gamma^\nu, \slashed{\Sigma} ]\,.
        \end{equation*}
        This leads to eight conditions $A_j=0$ for their Lorentz-invariant coefficients
        and thus eight relations between the dressing functions $g_i$.
        Whereas the $A_j$ are either even or odd in $\lambda$, they are superpositions of even and odd pieces in $\omega$:
        \begin{equation}\label{derive-transverse-2}
           A_j = A_j^{(1)} + \omega\,A_j^{(2)} \stackrel{!}{=} 0\,.
        \end{equation}
        Thus we arrive at 16 conditions $A_j^{(1)} = 0$ and $A_j^{(2)} = 0$,
        where it turns out that only 14 are independent.

    \renewcommand{\arraystretch}{1.4}

             \begin{table}[t]
             \begin{equation*}
             \begin{array}{rl}

                 X_1    &=   \lambda^2 \,K_1 + \eta_- K_6 + K_7 \\
                 X_2    &=  \eta_- K_1 - K_3 \\
                 X_3    &=  (\eta_+^2 - \omega^2)\,K_1 + \eta_- K_2 - \eta_+ K_4 + K_5 \\
                 X_4    &=  \lambda^2 K_2 + (\eta_+^2-\omega^2)\,K_6 + \eta_+ K_9 - K_{10}\\
                 X_5    &=  \lambda^2\, (-2\eta_+ K_1 + K_4) - \eta_+ K_7 - K_8 + \eta_- K_9 \\[2mm]

                 X_6    &=  -K_{28} + K_{33} - \tfrac{1}{2} K_{34}  \\
                 X_7    &=  2\,(-K_{19} + K_{22} - K_{23} + \tfrac{1}{4} \eta_+ K_{34} ) \\
                 X_8    &=  2\,(K_{20} - \eta_+ K_{22} - K_{24} - \tfrac{1}{4} \omega^2 K_{34}) \\
                 X_9    &=  4\,(K_{18} - \eta_- K_{22} - K_{26}) + X_8 \\[2mm]

                 X_{10} &=  -2K_{11} + \eta_- K_{21} + 2K_{25} - \tfrac{1}{4}\eta_- K_{34} + 2X_1\\
                 X_{11} &=  4K_{16} + 2\lambda^2 (K_{21} + \tfrac{1}{4} K_{34}) +\lambda^2 X_6 + \tfrac{1}{4} X_{13}   \\
                 X_{12} &=  -4\lambda^2 K_1 - 2K_7 + 4K_{11} -2 K_{25}  + K_{32} + \eta_- K_{33}    \\
                 X_{13} &=  -4\lambda^2 K_2  + 2K_{10} + 4K_{12} -2K_{24} + K_{29} \\
                          & \quad -2\eta_ +( K_9 + K_{23} -\tfrac{1}{2} K_{30})  + (\eta_+^2-\omega^2)\,K_{33} \\
                 X_{14} &=  2\, (-K_8  + K_{26}) +\eta_- (2K_{22} + K_{27}) + K_{31}  \\[2mm]

                 X_{15} &=  2\,(-K_9+K_{19}) + K_{27} - K_{30}  - \eta_+ \widetilde{K}-X_7  \\
                 X_{16} &=  2\,(-K_{10}+K_{20}) + \eta_+K_{27} + K_{29}  - \omega^2 \widetilde{K} + X_8  \\
                 X_{17} &=  -K_{14} + \eta_+(2K_{11}-K_{25}) + \eta_- K_{23} + K_{26} \\
                 X_{18} &=  K_{15} - \omega^2(2K_{11}-K_{25}) + \eta_- K_{24} - \eta_+ K_{26}

             \end{array}
             \end{equation*}

               \caption{Basis of Table~\ref{transverse-basis-0} expressed through the elementary tensors $K_i$ of Table~\ref{basis-type-I}.
                        The equalities hold inside the nucleon spinors since we exploited the relations~(\ref{linear-relations-1}--\ref{linear-relations-2}).
                        For $X_{15}$ and $X_{16}$ we abbreviated $\widetilde{K} = X_6+4K_6+2K_{21}+K_{33}$.}
               \label{X-basis-in-terms-of-K}
             \end{table}

        The resulting relations are rather complicated but they can be solved without divisions:
        similarly to~(\ref{linear-relations-1}--\ref{linear-relations-2}) one can eliminate 14 dressing functions $g_j$ (for example, those for
        $j=$ 1, 2, 4, 6, 9, 11, 12, 14, 19, 21, 23, 27, 28, 33) without dividing by terms
        depending on the kinematic variables $\eta_+$, $\eta_-$, $\omega^2$ and $\lambda^2$.
        If we relabel the independent functions by $c_i$ with $i=1\dots 18$, the relations take the form
        $g_j = g_j(c_1, \dots c_{18})$ and
        reinserting them into~\eqref{ca-01} yields the transverse Compton amplitude
        \begin{equation}\label{ca-02}
            \Gamma^{\mu\nu}_{\perp\perp} = \Lambda_+^f  \left[ \sum_{i=1}^{18} c_i\,X_i^{\mu\nu} \right] \Lambda_+^i\,.
        \end{equation}
        The transverse tensors $X_i$ are given in Table~\ref{X-basis-in-terms-of-K} and identical to those in Table~\ref{transverse-basis-0}
        in the main text.

    \renewcommand{\arraystretch}{1.4}

             \begin{table*}[t]

             \begin{equation*}\label{tarrach-T}
             \begin{array}{ r l   }

                    \tau_1/m^2                  &= X_2 \\
                   \tau_2/m^4                   &= -X_3 \\
                   \tau_3/m^4                   &= -X_1 \\
                   \lambda \,\tau_4/m^4         &= -X_5 \\
                   \tau_{19}/m^6                &= 2X_4 \\[2mm]

                   \lambda\,\tau_6/m^4          &= -\tfrac{1}{4} X_{13} \\
                   \lambda\,\tau_7/m^3          &= -2\lambda^2 X_6 + 2X_{11} -\tfrac{1}{2} X_{13} \\
                   \tau_8/m^3                   &= -\tfrac{1}{2} X_7 \\
                   \omega\,\tau_9/m^3           &= -\tfrac{1}{2} X_8 \\
                    \tau_{10}/m^4               &= 4X_{10}+2X_{12}  \\
                    \omega \,\tau_{11}/m^3      &= \tfrac{1}{4} (X_8-X_9)

             \end{array}\qquad
             \begin{array}{ r l   }

                   \lambda\,\tau_{12}/m^4       &= X_5 - \tfrac{1}{2} \eta_+ X_{12} + X_{17} \\
                    \lambda\omega\,\tau_{13}/m^4 &= (\eta_+^2-\omega^2)  X_1 + \lambda^2 X_3 - \eta_- X_4 + \eta_+ X_5   - \tfrac{1}{2} \,\omega^2 X_{12} - X_{18} \\
                   \omega\,\tau_{14}/m^4        &= X_{14} \\
                   \omega\,\tau_{20}/m^4        &= -(\omega^2 X_6 - 2X_8 + X_{16})  \\
                   \tau_{21}/m^4                &= -(\eta_+ X_6 + 2X_7 + X_{15}) \\
                   \lambda\,\tau_{17}/m^2       &= -X_{12} \\
                   \tau_{18}/m^3                &= 4X_6-2X_7 \\[7mm]

                   \lambda\omega\, \tau_5/m^4   &= (\eta_+^2-\omega^2) X_1 + \lambda^2 X_3 - \eta_- X_4 + \eta_+ X_5 \\
                   \lambda\omega\,\tau_{15}/m^4 &= \omega^2\, ( \eta_- X_6 + 2X_{10} + X_{12})   - \eta_-( 2X_8 - X_{16}) - \eta_+ X_{14} \\
                   \lambda\,\tau_{16}/m^4       &= \eta_+\, ( \eta_- X_6 + 2X_{10} + X_{12})   + \eta_-(2X_7 + X_{15})  - X_{14}

             \end{array}
             \end{equation*}

               \caption{Tarrach's basis for the Compton amplitude~\cite{Tarrach:1975tu}.
                 The $\tau_i$ are the transverse tensors in Eqs.~(12--13) therein;
                 with appropriate prefactors $\lambda$, $\omega$ or $\lambda\omega$ they become symmetric under photon crossing and charge conjugation.
                 The $X_i$ are our Compton tensors from Table~\ref{X-basis-in-terms-of-K}.}

               \label{tab:tarrach-1}

             \end{table*}

        We did check
        other choices of eliminating two tensors from~(\ref{linear-relations-1}--\ref{linear-relations-2})
        within the aforementioned constraints. They all produced equivalent bases in the sense that the determinants of the transformation matrices
        between the bases are constant and not momentum-dependent (so they can never vanish or diverge).

        The procedure by Bardeen and Tung~\cite{Bardeen:1969aw} and Tarrach~\cite{Tarrach:1975tu} for deriving the transverse basis is slightly different from ours.
        In that case one enforces transversality by acting with projectors on the initial amplitude:
        \begin{equation}
           \frac{t^{\mu\alpha}_{Q'Q}}{Q\cdot Q'}\,\Gamma^{\alpha\beta}\,\frac{t^{\beta\nu}_{Q'Q}}{Q\cdot Q'} \stackrel{!}{=} \Gamma^{\mu\nu}\,,
        \end{equation}
        which gives 18 tensors with single and double poles in the variable $Q\cdot Q'$.
        By forming appropriate linear combinations one then eliminates as many poles as possible, multiplies the remaining double-pole structures
        with $Q\cdot Q'$, repeats, and finally multiplies the single-pole tensors by $Q\cdot Q'$.
        In contrast to Eqs.~(\ref{derive-transverse-1}--\ref{derive-transverse-2}), however, this does not \textit{automatically} lead to a minimal basis.
        Tarrach derives the tensors $\tau_{1 \dots 18}$, given in Table~\ref{tab:tarrach-1}, but notes that the resulting basis is not minimal
        due to $\tau_5$, $\tau_{15}$ and $\tau_{16}$, which are subsequently exchanged with new tensors $\tau_{19}$, $\tau_{20}$ and $\tau_{21}$
        to form a minimal basis. Written in terms of the $X_i$, the problem with these tensors is evident
        as one can see in the table: for example,
        all coefficients of the $X_i$ in the equation for $\tau_5$ are momentum-dependent and thus the determinant of the basis transformation
        from $\tau_{1\dots 18}$ to $X_{1\dots 18}$ would depend on the kinematics. Phrased differently, the crossing- and charge-conjugation symmetric combination
        $\lambda\omega\,\tau_5$ has a higher photon momentum power (namely $n=6$) than its replacement $\tau_{19} \sim X_4$ with $n=4$;
        cf. the discussion in Sec.~\ref{sec:cs-tensor-basis}.

        For these reasons we prefer the more direct method of Eqs.~(\ref{derive-transverse-1}--\ref{derive-transverse-2}) for deriving the basis,
        because it is failsafe and also provides a safety check: if it were \textit{not} possible to solve~\eqref{derive-transverse-2} without divisions,
        this would point to a problem with minimality. Fortunately, in the case of Compton scattering there is no such problem.
        Our main reasons for working with the $X_i$ instead of Tarrach's (modified) basis
        are a cleaner physical interpretation (see the remarks at the end of Sec.~\ref{sec:tensor-basis}),
        a simpler form of Table~\ref{transverse-basis-0}, and simpler expressions for Tables~\ref{tab:nucleon-born}, \ref{tab:j=1/2+born} and~\ref{tab:j=3/2+born}.

        As discussed in connection with Eq.~\eqref{vvcs}, it is still possible to identify kinematic limits
        where kinematic singularities cannot be avoided. In that case the 18 tensors $X_i$ collapse into a set of fewer tensors
        whose coefficients are linear combinations of the $c_i$ but with singular denominators. However, this does not change
        the fact that the $c_i$ themselves are still finite in those limits. Take for example our results in Fig.~\ref{fig:results-cffs-1}:
        the $\hat c_i$ obtained from the intermediate nucleon resonances are non-singular everywhere inside the TPE cone, although the cone
        contains the VVCS limit $\omega=0$. This means we can go arbitrarily close to that limit and extract the $\hat c_i$,
        which also remain finite in the limit. On the other hand, had we calculated \textit{directly} in the VVCS limit,
        the $\hat c_i$ would have collapsed into fewer functions with kinematic singularities.

        In any case, this situation does not affect
        the kinematic limits of RCS, VCS and the forward limit where direct measurements are possible.
        This is evident from the discussion below Eq.~\eqref{RCS-1}, Eqs.~(\ref{VCS-1}--\ref{CFFs-VCS}), and Eq.~\eqref{CFFs-FWD},
        as well as the following subsection~\ref{sec:kinematic-limits-2}:
        in all those cases the $X_i$ collapse into fewer tensors but the respective coefficients do not pick up kinematic singularities.
        Hence, in principle the CFFs (or their combinations) can be measured  directly in these limits.

   \subsection{Kinematic limits} \label{sec:kinematic-limits-2}

        In the following we collect the relations between the $K_i$ in Table~\ref{basis-type-I} in the various kinematic limits,
        which leads to the reduced transverse bases discussed in Sec.~\ref{sec:kinematic-limits}.
        We further relate the CFFs in those limits to some common amplitude choices employed in the literature.

        \smallskip
        \textbf{RCS:}
        Here the condition $\omega=0$ eliminates the tensors
             \begin{equation} \label{omega=0}    \renewcommand{\arraystretch}{1.2}
                \begin{array}{l}
                   K_5  \\
                   K_8  \\
                   K_{10}
                \end{array}\qquad
                \begin{array}{l}
                   K_{15}  \\
                   K_{18}  \\
                   K_{20}
                \end{array}\qquad
                \begin{array}{l}
                   K_{22}  \\
                   K_{24}  \\
                   K_{26}
                \end{array}\qquad
                \begin{array}{l}
                   K_{27}  \\
                   K_{29}  \\
                   K_{31}
                \end{array}
             \end{equation}
        from the basis. In addition, applying polarization vectors for the onshell photon momenta has the same effect
        as crossing off tensors which contain instances of either ${Q'}^\mu$ or $Q^\nu$, so that also
             \begin{equation*}     \renewcommand{\arraystretch}{1.2}
                \begin{array}{l}
                   K_2  \\
                   K_4  \\
                   K_9
                \end{array}\qquad
                \begin{array}{l}
                   K_{12}  \\
                   K_{14}  \\
                   K_{19}
                \end{array}\qquad
                \begin{array}{l}
                   K_{23} \\
                   K_{30}
                \end{array}
             \end{equation*}
        vanish in RCS. From Table~\ref{X-basis-in-terms-of-K} one then infers that only the transverse tensors
        $X_1$, $X_2$, $X_6$, $X_{10}$, $X_{11}$ and $X_{12}$ survive in RCS.
        The relations between our CFFs and the RCS amplitudes $A_i(\eta_-,\lambda)$
        defined by L'vov \textit{et al.}~\cite{Lvov:1980wp,Lvov:1996rmi} are given in
        Table~\ref{tab:lvov}.

             \begin{table}[t]  \renewcommand{\arraystretch}{1.3}
             \begin{equation*}
                 \left[ \begin{array}{c} A_1 \\ A_2 \\ A_3 \\ A_4 \\ A_5 \\ A_6 \end{array}\right] = -\frac{e^2}{m^3}
                 \left[ \begin{array}{c} \tfrac{1}{4} (\eta_--2)\,c_1 + c_2 + \tfrac{1}{2} \eta_- c_{10} + \lambda^2 c_{11} \\
                                         -(c_6 + c_{10}) \\
                                         -(\tfrac{1}{2} c_1 + c_{10} - c_{11}) \\
                                         c_{11} \\
                                         \tfrac{1}{2}(\eta_--2)\,c_{11} + 2c_{12} \\
                                         c_{10} + \tfrac{1}{2} (\eta_--2)\,c_{11}
                        \end{array}\right]
             \end{equation*}

               \caption{Relations between the RCS amplitudes $A_i$ of Refs.~\cite{Lvov:1980wp,Lvov:1996rmi} and our CFFs in RCS ($\eta_+=\omega=0$).}
               \label{tab:lvov}

             \end{table}

        \smallskip
        \textbf{VCS:}
        The same strategy applied to VCS amounts to dropping instances of ${Q'}^\mu$ only. With $\eta_+=\omega$ this implies
             \begin{equation*}     \renewcommand{\arraystretch}{1.2}
                \begin{array}{rl}
                   K_2 &= 0\,, \\
                   K_5 &= \eta_+ K_4\,, \\
                   K_{10} &= \eta_+ K_9\,, \\
                   K_{24} &= -\eta_+ K_{23} \,,
                \end{array}\qquad
                \begin{array}{rl}
                   K_{12} &= 0\,, \\
                   K_{15} &= \eta_+ K_{14}\,, \\
                   K_{20} &= \eta_+ K_{19}\,, \\
                   K_{29} &= -\eta_+ K_{30}
                \end{array}
             \end{equation*}
        which induce the linear relations~\eqref{VCS-1} between the $X_i$. The resulting CFFs are those in Eq.~\eqref{CFFs-VCS}.

        The relations between the Compton tensors $T_i$ employed
        by Drechsel \textit{et al.},
        defined in App. A of Ref.~\cite{Drechsel:1997xv},
        with Tarrach's $\tau_i$ are given by:
        \begin{equation}
           \begin{array}{rl}
              T_1 &= -\tau_1\,, \\
              T_2 &= -4\tau_3\,, \\
              T_3 &= \tau_2\,, \\
              T_4 &= 2\tau_4\,, \\
              T_5 &= -2\tau_5\,,
           \end{array}\qquad
           \begin{array}{rl}
              T_{6\dots 18} &= \tau_{6 \dots 18}\,, \\[2mm]
              T_{19} &= 2\tau_{19}\,, \\
              T_{20} &= \tau_{20}\,, \\
              T_{21} &= \tau_{21}\,.
           \end{array}
        \end{equation}
        These are subsequently used to define the VCS tensors $\rho_i$ and corresponding dressing functions $f_i$, cf.~Eq.~(A10) in~\cite{Drechsel:1997xv}.
        Their relations with our CFFs are given in Table~\ref{tab:drechsel}.
        The nucleon's generalized polarizabilities can then be reconstructed using Eqs.~(3--8)  in Ref.~\cite{Drechsel:1998zm}
        or in a manifestly covariant form via Eq.~(A1) in Ref.~\cite{Lensky:2016nui}.
        Note that in the conventions of Drechsel \textit{et al.} the average nucleon and photon momenta differ by a factor 2 and
        one has to interchange the Lorentz indices $\mu \leftrightarrow \nu$.

             \begin{table}[t]  \renewcommand{\arraystretch}{1.4}
             \begin{equation*}
             \begin{array}{rl}

                 m^3 f_1 &= -c_2 \\
                 m^5 f_2 &= \tfrac{1}{4} \,c_1\\
                 m^5 f_3 &= -\tfrac{1}{2} \lambda\,(c_5-\overline c_{17}) \\
                 m^4 f_4 &= \tfrac{1}{2} \lambda\,c_{11} \\

                 m^4 f_5 &= -(c_6 + 2\,\overline c_7 - 2\eta_+ c_9  + \lambda^2 c_{11} - (\eta_+ + 4)\,\overline c_{15})  \\
                 m^5 f_6 &= \tfrac{1}{4}  c_{10} \\
                 m^4 f_7 &= -4\eta_+ c_9 \\
                 m^5 f_8 &=  \lambda\,\overline c_{17} \\

                 m^5 f_9 &= \eta_+ c_{14} \\
                 m^3 f_{10} &= \tfrac{1}{2} \lambda \,( c_{10} - 2c_{12} - \eta_+ \overline c_{17}) \\
                 m^4 f_{11} &= \tfrac{1}{4} ( c_6 + \lambda^2 c_{11} - \eta_+ \overline c_{15}) \\
                 m^5 f_{12} &= -\overline c_{15}

             \end{array}
             \end{equation*}

               \caption{Relations between the $f_i$ defined in Ref.~\cite{Drechsel:1997xv} and our CFFs in VCS ($\eta_+=\omega$).
                        The relations between the nucleon's generalized polarizabilities and the $f_i$ can be found in Refs.~\cite{Drechsel:1998zm,Lensky:2016nui};
                        the $f_i$ in~\cite{Drechsel:1997xv,Drechsel:1998zm} are identical to the $A_i$ in Ref.~\cite{Lensky:2016nui}.
                        We abbreviated $\overline c_7 = c_7-\eta_+ c_8$, $\overline c_{15} = c_{15} + \eta_+ c_{16}$ and $\overline c_{17} = c_{17} - \eta_+ c_{18}$.}
               \label{tab:drechsel}

             \end{table}

        \smallskip
        \textbf{FWD:}
        In the doubly-virtual forward limit the condition $\omega=0$ disposes again  of the tensors in Eq.~\eqref{omega=0}.
        In addition one has ${Q'}^\mu=Q^\mu$, but without
        any polarization vectors because the photons are still virtual.
        Because the incoming and outgoing nucleon momenta are the same, one exploits the identities
        \begin{equation}
           \Lambda_+(p)\,\gamma^\mu\,\Lambda_+(p) = \frac{p^\mu}{im}\,\Lambda_+(p)
        \end{equation}
        and
        \begin{equation}
        \begin{split}
           &\Lambda_+(p)\,[p^\mu \gamma^\nu - \gamma^\mu p^\nu, \gamma^\rho ]\,\Lambda_+(p) \\
                               &= \Lambda_+(p) \left( \frac{im}{3}\,[\gamma^\mu,\gamma^\nu,\gamma^\rho]  - [\gamma^\mu,\gamma^\nu]\,p^\rho \right)\Lambda_+(p)
        \end{split}
        \end{equation}
        to obtain further relations between the $K_i$:
             \begin{equation*}
             \begin{split}
                 K_3 &= \tfrac{1}{2} K_4 = K_2\,, \\
                 K_9 &= K_{17} = K_{19} = -K_{23} = -K_{25} = K_7\,, \\
                 K_{11} &= \lambda^2 K_{1}\,, \\
                 K_{12} &= K_{13} = \tfrac{1}{2} K_{14} = \lambda^2 K_2\,, \\
                 K_{16} &= \lambda^2 K_6\,, \\
                 K_{21} &= -2K_6\,, \\
                 K_{28} &= K_{33} - \tfrac{1}{2} K_{34}\,, \\
                 K_{32} &= K_{30}
             \end{split}
             \end{equation*}
        In total only seven tensors $K_1$, $K_2$, $K_6$, $K_7$, $K_{30}$, $K_{33}$ and $K_{34}$ remain independent.
        The resulting four transverse tensors in the forward limit are given in Eq.~\eqref{fwd-tensors-1}
        and the corresponding CFFs in Eq.~\eqref{CFFs-FWD}.

   \subsection{Non-transverse part}\label{sec:non-transverse-basis}

        Next, we work out the remaining non-transverse tensors of the basis,
        i.e., the $\Gamma^{\mu\nu}_\perp$ and $\Gamma^{\mu\nu}_\text{G}$
        from Eq.~\eqref{generic-decomposition}. 
        For the physical amplitude they are irrelevant because their coefficients vanish due to gauge invariance,
        but projecting onto the full 32-dimensional basis including all terms serves as a useful test of gauge invariance.

        The `partially transverse' piece $\Gamma^{\mu\nu}_\perp$ is subject to the weaker constraint
        \begin{equation}\label{transv-constraint-2}
           {Q'}^\mu \,\Gamma^{\mu\nu}(p,Q',Q)\,Q^\nu \stackrel{!}{=} 0\,.
        \end{equation}
        To derive it, we lift the requirement of gauge invariance and add the 14 tensors
        belonging to the $g_j$ (given above Eq.~\eqref{ca-02}) that we previously eliminated:
        \begin{equation*}\label{ca-03}
            \Gamma^{\mu\nu} = \Lambda_+^f  \left[ \sum_{j} g_j \,K_j^{\mu\nu} + \sum_{i=1}^{18} c_i\,X_i^{\mu\nu} \right] \Lambda_+^i\,.
        \end{equation*}
        Since the transverse tensors $X_i$ already satisfy Eq.~\eqref{transv-constraint-2} the condition only affects the first sum.
        Its contraction with ${Q'}^\mu$ and $Q^\nu$ generates two independent structures proportional to $\mathds{1}$ and $\slashed{\Sigma}$,
        \begin{equation}\label{pt-conditions}
             \sum_{ j} g_j \,{Q'}^\mu K_j^{\mu\nu} Q^\nu = A'_1 + A'_2\,i\lambda\,\slashed{\Sigma} \stackrel{!}{=} 0\,,
        \end{equation}
        with two resulting conditions $A'_1=A'_2=0$. This leaves 12 independent functions;
        if we relabel them by $c_i$ with $i=19 \dots 30$, the result is
        \begin{equation}\label{ca-04}
            \Gamma^{\mu\nu}_\perp = \Lambda_+^f  \left[ \sum_{i=19}^{30} c_i\,X_i^{\mu\nu} \right] \Lambda_+^i\,,
        \end{equation}
        where the tensors $X_{19\dots 30}$ are collected in Table~\ref{partially-transverse-basis-in-terms-of-K}.

        Unfortunately, here it is no longer possible to solve the system~\eqref{pt-conditions} without divisions,
        which means that some of the resulting tensors acquire kinematic singularities.
        Those are the ones proportional to $X_{19}$ and $X_{20}$, whose original form as a result of the equations is $X_{19}/\eta_-$ and $X_{20}/\eta_-$.
        Therefore, they must be multiplied by $\eta_-$ which removes the kinematic singularities at $\eta_- =0$
        and in the process raises their photon momentum powers.

        That the $X_{19\dots 30}$ satisfy Eq.~\eqref{transv-constraint-2} is a simple check; for example, Table~\ref{basis-type-I} entails
        \begin{equation*}
           {Q'}^\mu K_{22}^{\mu\nu} \,Q^\nu = \frac{\omega}{im}\,(p\cdot Q'\,\slashed{Q} - \slashed{Q}'\,p\cdot Q) =  im\lambda\omega\,\slashed{\Delta} \,,
        \end{equation*}
        which vanishes in the contraction with the positive-energy projectors. In analogy to Eq.~\eqref{basis-explicit},
        it is possible to recast the $X_{19\dots 30}$ in a form where the (partial) transversality is manifest.
        For example, using the definition~\eqref{new-transverse-projectors-2}:
        \begin{equation}
        \begin{split}
            X_{21}^{\mu\nu} &= \frac{1}{m^2}\left( t^{\mu\nu}_{Q'Q'} + t^{\mu\nu}_{QQ}\right),\\
            X_{23}^{\mu\nu} &= \frac{i\lambda}{m}\left( t^{\mu\nu}_{Q'\gamma} + t^{\mu\nu}_{\gamma Q}\right), \\
            X_{28}^{\mu\nu} &= \frac{i\lambda\omega}{m}\left( t^{\mu\nu}_{Q'\gamma} - t^{\mu\nu}_{\gamma Q}\right).
        \end{split}
        \end{equation}

        Finally, the remaining non-transverse part
        \begin{equation}\label{ca-05}
          \Gamma^{\mu\nu}_\text{G} = \Lambda_+^f  \,\big[ c_{31}\,X_{31}^{\mu\nu} + c_{32}\,X_{32}^{\mu\nu} \big]\, \Lambda_+^i
        \end{equation}
        depends on the two tensors $X_{31} = K_1$ and $X_{32} = K_{21}$ corresponding to the coefficients
        that we eliminated in the solution of Eq.~\eqref{pt-conditions}. In total, the sum of Eqs.~\eqref{ca-02}, \eqref{ca-04} and \eqref{ca-05}
        constitutes a complete 32-dimensional tensor basis according to Eq.~\eqref{generic-decomposition}.

    \renewcommand{\arraystretch}{1.4}

             \begin{table}[t]
             \begin{equation*}
             \begin{array}{rl}

                 X_{19} &= \eta_- K_2 - (\eta_+^2-\omega^2) K_1 \\
                 X_{20} &= \eta_- K_6 - \lambda^2 K_1\\
                 X_{21} &= 2\eta_+ K_1 - K_4 \\
                 X_{22} &= 2\eta_+ K_6 - K_9 \\[2mm]

                 X_{23} &= 2K_{11} - K_{25} \\
                 X_{24} &= 2K_{12} + (\eta_+^2-\omega^2) K_{21} \\
                 X_{25} &= K_{19} - K_{23} - 2\eta_+ K_{21} \\
                 X_{26} &= K_{28} + K_{33}+4K_{21}+8K_6

             \end{array} \qquad
             \begin{array}{rl}

                 X_{27} &= K_{22} \\
                 X_{28} &= K_{26} \\
                 X_{29} &= K_{27} \\
                 X_{30} &= K_{34} \\[2mm]

                        & \\
                 X_{31} &= K_{1} \\
                 X_{32} &= K_{21} \\
                        &

             \end{array}
             \end{equation*}

               \caption{Non-transverse basis tensors for the Compton amplitude.
               Taken together with those in Table~\ref{X-basis-in-terms-of-K}, they form a complete basis.}
               \label{partially-transverse-basis-in-terms-of-K}

             \end{table}

   \subsection{Effects of breaking gauge invariance}\label{sec:gi-broken}

    Finally we return to the question posed in Sec.~\ref{sec:nucleon-born}:
    what are the consequences of breaking gauge invariance on purpose?
    To investigate this, we consider the nucleon Born term from Eq.~\eqref{qcv-born} but
    implement a nucleon-photon vertex that differs from the Dirac form~\eqref{nucleon-off-curr-1}.
    For example:
        \begin{equation}\label{nucleon-off-curr-3}
            \Gamma^\mu(k,Q) = i\left[ F_1 \,G_1^\mu + F_2 \,\frac{T_3^\mu}{2}  + \alpha\,T_4^\mu\right]
        \end{equation}
    with the usual Dirac and Pauli Form factors $F_{1,2}(Q^2)$, but including
    the tensor $T_4^\mu$ from Table~\ref{n-offshell} with a constant coefficient $\alpha$.
    For simplicity, let us set $F_1=1$ and $F_2=0$. Then for $\alpha=0$ one obtains
    the CFFs for a pointlike Dirac particle: $\widetilde{c}_1=-4$ and $\widetilde{c}_{10}=2$.

    After switching on $\alpha$, the Compton amplitude is no longer gauge invariant. However, by projecting it onto
    its full 32-dimensional tensor basis,
        \begin{equation} \label{ca-07}
            \Gamma^{\mu\nu} =  \Gamma^{\mu\nu}_\text{G} + \Gamma^{\mu\nu}_\perp + \Gamma^{\mu\nu}_{\perp\perp} \,,
        \end{equation}
    with the transverse part $\Gamma^{\mu\nu}_{\perp\perp}$ from Eq.~\eqref{ca-02} and the
    remainders $\Gamma^{\mu\nu}_\perp$ and $\Gamma^{\mu\nu}_\text{G}$ from Eqs.~\eqref{ca-04} and ~\eqref{ca-05},
    we can isolate the terms that violate gauge invariance and work out the effects on the transverse CFFs.

    The result for the simple example above is given in Table~\ref{tab:nucleon-born-GI-violation}.
    The transverse CFFs pick up extra terms which depend on $\alpha$,
    but in addition we have also generated a gauge part $\propto X_{30}$.
    As in Eq.~\eqref{nucleon-born-split} we quote the residues $\widetilde{c}_i$ for the transverse parts
    but the CFF $c_{30}$ itself for the gauge part:
    only the transverse terms contain the nucleon poles whereas they drop out in the gauge part.

    Note also that no additional kinematic singularity has been generated in any CFF.
    Had we simply performed a transverse projection of the full amplitude, both gauge and transverse parts would have collapsed into 18 transverse functions;
    however, because the gauge part has a lower photon momentum power it will produce kinematic singularities in those functions.

        \begin{table}[t]
             \begin{equation*} \renewcommand{\arraystretch}{1.5}
             \begin{split}
             \begin{array}{rl   }
                 \widetilde c_1 &= - 4 + A-B+D  \\
                 \widetilde c_2 &= A + \eta_-(4\alpha-C)  \\
                 \widetilde c_6 &= -A - 2\alpha \,\eta_- \\
                 \widetilde c_{10} &=  2-4\alpha - \tfrac{1}{2}(A-B+D)
             \end{array}\quad
             \begin{array}{rl   }
                 \widetilde c_{11} &=  -4\alpha^2 \\
                 \widetilde c_{12} &=  2\alpha - C  \\
                 \widetilde c_{15} &= -\alpha^2 \,\eta_- \\
                 c_{30} &= -\tfrac{1}{2}\alpha
             \end{array} \\[2mm]
             \begin{array}{rl   }
                 A &= \alpha^2\, (\eta_-^2-4\lambda^2)  \\
                 B &= \alpha^2\, (\eta_+^2-\omega^2)
             \end{array} \quad
             \begin{array}{rl   }
                 C &= \alpha^2 \,(2\eta_+ - \eta_- + 4) \\
                 D &= 4\alpha \,(\eta_+-\alpha\eta_-)
             \end{array}
             \end{split}
             \end{equation*}
             \caption{Compton form factors for a pointlike nucleon
             but with an extra term $\propto T_4^\mu$ that violates gauge invariance,
             see~Eq.~\eqref{nucleon-off-curr-3}.}
             \label{tab:nucleon-born-GI-violation}
        \end{table}

\begin{figure}[t]
                    \begin{center}
     \includegraphics[width=1.0\columnwidth]{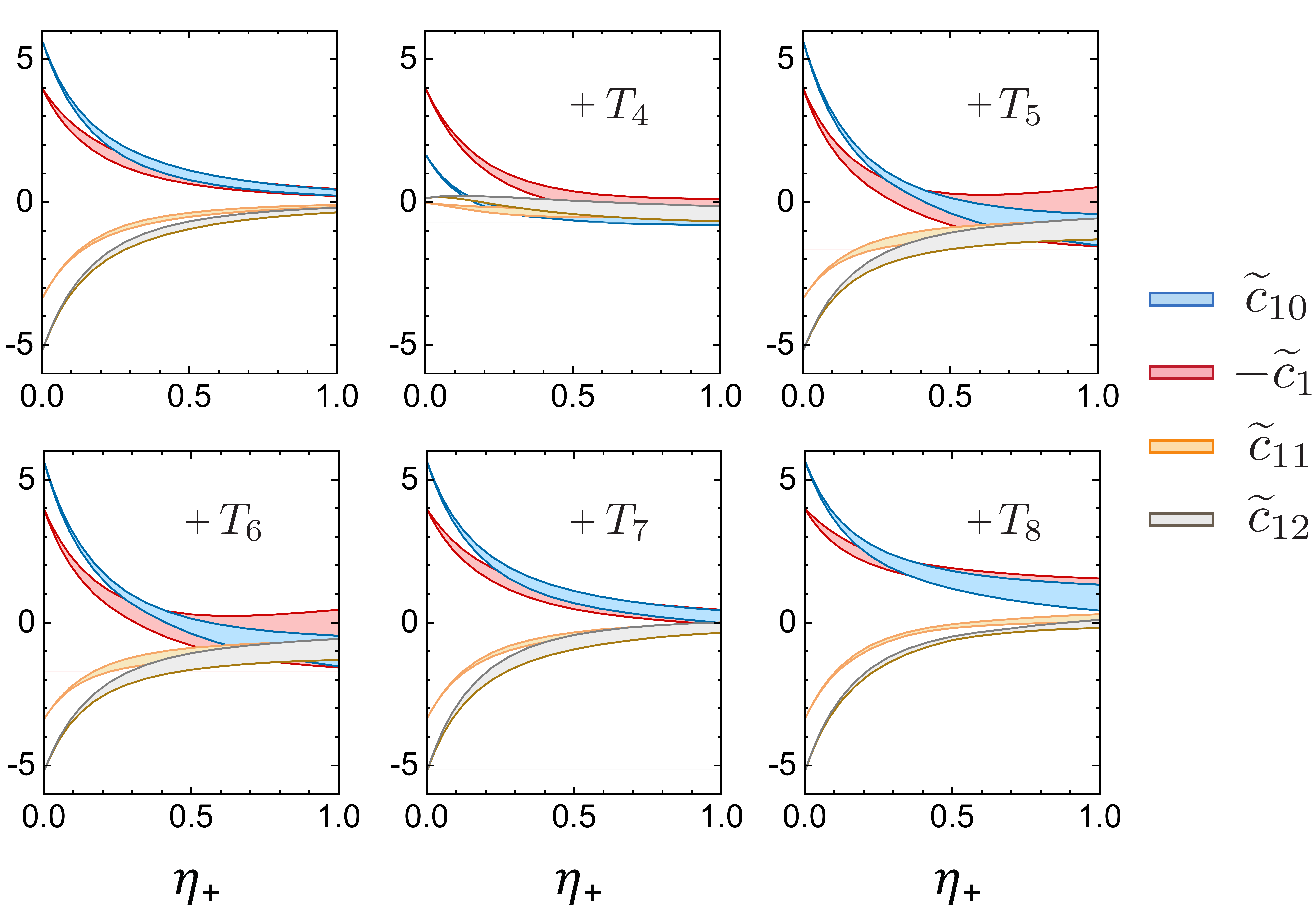}
        \caption{Compton form factor residues of the nucleon Born term from an offshell nucleon-photon vertex that violates electromagnetic gauge invariance.
                 The upper left panel shows the original result from Fig.~\ref{fig:nucleon-born} where gauge invariance is preserved.
                 In the remaining panels we switch on $T_{4\dots 8}^\mu$
                 as in Eq.~\eqref{gi-violating-2}, with $\alpha=1$ in all cases.}
        \label{fig:gi-breaking}
        \end{center}
        \end{figure}

    This principle can be taken further to test offshell effects in the nucleon Born term
    within a hadronic description~\cite{Miller:2011yw}.
    To do so, we restore the proton's Dirac and Pauli form factors in Eq.~\eqref{nucleon-off-curr-3}
    but add other tensors with a simple momentum dependence of the form
    \begin{equation}\label{gi-violating-2}
       \alpha\,T_4^\mu \; \to \; \frac{\alpha}{(1+Q^2/m^2)}\,T_i^\mu\,,
    \end{equation}
    with $i \neq 1$ and $i\neq 3$. 
    After the projection we drop the unphysical gauge parts and consider the transverse CFFs only.
    In Fig.~\ref{fig:gi-breaking} we show a few selected results for the leading transverse CFFs obtained with $\alpha=1$.
    Without the additional tensors (upper left panel) they are identical to those in Fig.~\ref{fig:nucleon-born}.
    As one can see, breaking gauge invariance has rather modest effects on the transverse part of the
    Compton amplitude because  the CFFs do not change their form dramatically.
    Thus, even if offshell effects played a role (as far as that can be judged within an effective hadronic theory)
    their correct implementation
    leads back to results which resemble the onshell forms.

    This observation is useful also in a different context, namely in
    microscopic calculations of Compton scattering.
    In that case the complete expression for the Compton amplitude in terms of quark and gluon degrees of freedom
    has been derived based on electromagnetic gauge invariance~\cite{Eichmann:2011ec,Eichmann:2012mp}.
    As usual only the sum of all diagrams is gauge invariant but not the individual terms.
    Unfortunately, some of those diagrams are numerically hard to calculate.
    Keeping only parts of the results, even if they provide the dominant contributions,
    would indeed be useless if one cannot quantify the effects of breaking gauge invariance.
    For example, with a naive transverse projection the resulting CFFs would be contaminated
    by unphysical kinematic singularities.
    The separation~\eqref{ca-07} resolves the problem:
     one can project the dominant diagrams onto the complete basis, which ensures both transversality and the absence of kinematic singularities,
     and subsequently retain the transverse CFFs.
     This leads to well-defined expressions, which can be systematically improved upon, where
     the subleading diagrams would mainly serve to cancel the gauge parts
     because the sum of all diagrams is known to be gauge invariant.

  \pagebreak

   \section{Spin-3/2 Lagrangians and point transformations}\label{sec:point-tfs}

              In this appendix we collect some further properties of spin-3/2 Lagrangians.
              We drop the label `R' that was used in the main text and denote the mass of the spin-3/2 particle by $m$,
              its tree-level propagator by $S^{\alpha\beta}$, and the nucleon-to-resonance transition vertex by $\Gamma^{\alpha\mu}$.

              The free spin-3/2 Lagrangian
              \begin{equation}\label{Lagrangian-3/2-2}
                  \mathcal{L} = \conjg{\psi}^\alpha \Lambda^{\alpha\beta}  \psi^\beta\,, \quad \Lambda^{\alpha\beta} = -\frac{i}{2} \left\{ \sigma^{\alpha\beta},\,i\slashed{k}+m \right\}
              \end{equation}
              with $i\slashed{k}=\slashed{\partial}$
               is a special case of more general possible forms that are related to each other by point transformations~\cite{Nath:1971wp,Benmerrouche:1989uc,Pascalutsa:1999zz}.
              Let us define the transverse and longitudinal projectors onto $\gamma-$matrices,
              \begin{equation}      \renewcommand{\arraystretch}{1.2}
                  \mathds{P}_\perp^{\alpha\beta} = \delta^{\alpha\beta} - \tfrac{1}{4} \gamma^\alpha \gamma^\beta\,, \qquad
                  \mathds{P}_\parallel^{\alpha\beta} = \tfrac{1}{4} \gamma^\alpha \gamma^\beta
              \end{equation}
              with the properties
              \begin{equation}
                  \mathds{P}_\perp^{\alpha\beta}\gamma^\beta =0, \quad
                  \mathds{P}_\parallel^{\alpha\beta}\gamma^\beta = \gamma^\alpha, \quad
                  \mathds{P}_\perp^{\alpha\gamma}\, \mathds{P}_\parallel^{\gamma\beta}=0\,.
              \end{equation}
              The so-called point transformation tensors
              \begin{equation}
                  \Theta^{\alpha\beta}(\lambda) = \mathds{P}_\perp^{\alpha\beta} + \lambda\,\mathds{P}_\parallel^{\alpha\beta}
              \end{equation}
              form a group:
              \begin{equation}\label{RS-point-tf-group}
              \begin{split}
                  &\Theta^{\alpha\gamma}(\lambda)\,\Theta^{\gamma\beta}(\lambda') = \Theta^{\alpha\beta}(\lambda\lambda')\,, \\
                  &\Theta^{\alpha\gamma}(\lambda)\,\Theta^{\gamma\beta}(\lambda^{-1}) = \delta^{\alpha\beta}\,,
              \end{split}
              \end{equation}
              where the group parameter $\lambda$ is the coefficient of the longitudinal part in $\Theta^{\alpha\beta}(\lambda)$.
              The general form of the inverse propagator $\Lambda^{\alpha\beta}$ can then be written as~\cite{Pascalutsa:1999zz}
              \begin{equation}\label{RS-gauge-general}
                  \Lambda^{\alpha\beta}(\xi) = \Theta^{\alpha\gamma}\big(\tfrac{1}{\xi}\big)\,\Lambda^{\gamma\delta}(\xi=1)\,\Theta^{\delta\beta}\big(\tfrac{1}{\xi}\big)\,.
              \end{equation}
              It depends on a gauge parameter $\xi$, where
              $\xi=1$ corresponds to the `Rarita-Schwinger gauge'
              and $\Lambda^{\alpha\beta}(\xi=1)$ is the expression in Eq.~\eqref{Lagrangian-3/2-2}.\footnote{To compare with the notation in the literature, e.g.~\cite{Pascalutsa:1999zz,Haberzettl:1998rw}, use $\xi=-1/(1+2A)$ and $\lambda=1+4a$.}
              From Eqs.~(\ref{RS-point-tf-group}--\ref{RS-gauge-general}) one has the general relation
              \begin{equation}\label{RS-gauge-TF}
                  \Lambda^{\alpha\beta}(\xi) = \Theta^{\alpha\gamma}\big(\tfrac{\xi'}{\xi}\big)\,\Lambda^{\gamma\delta}(\xi')\,\Theta^{\delta\beta}\big(\tfrac{\xi'}{\xi}\big)\,,
              \end{equation}
              which entails that the Lagrangian $\mathcal{L} = \conjg{\psi}^\alpha\,\Lambda^{\alpha\beta}(\xi)  \,\psi^\beta$ is invariant under the point transformation
              \begin{equation}\label{point-tf}
                  \xi\rightarrow\xi', \qquad \psi^\alpha \rightarrow {\psi'}^\alpha = \Theta^{\alpha\beta}\big(\tfrac{\xi'}{\xi}\big)\,\psi^\beta\,.
              \end{equation}
              Defining $\Omega=i\slashed{k}+m$, one can further show that Eq.~\eqref{RS-gauge-general} is identical to
              \begin{equation}\label{RS-Lambda-2}
              \begin{split}
                  \Lambda^{\alpha\beta}(\xi) &= \mathds{P}_\perp^{\alpha\gamma} \,\Omega \, \mathds{P}_\perp^{\gamma\beta} -\frac{3}{\xi^2}\,\mathds{P}_\parallel^{\alpha\gamma} \,\Omega \, \mathds{P}_\parallel^{\gamma\beta} \\
                                   &- \frac{1}{\xi}\,\big(\mathds{P}_\perp^{\alpha\gamma} \,\Omega \, \mathds{P}_\parallel^{\gamma\beta} +\mathds{P}_\parallel^{\alpha\gamma} \,\Omega \, \mathds{P}_\perp^{\gamma\beta}\big)\,.
              \end{split}
              \end{equation}

              The spin-3/2 propagator is the inverse of $\Lambda^{\alpha\beta}(\xi)$ in momentum space and therefore it satisfies
              \begin{equation}\label{RS-gauge-TF-propagator}
                  S^{\alpha\beta}(\xi) = \Theta^{\alpha\gamma}(\xi)\,S^{\gamma\delta}(\xi=1)\,\Theta^{\delta\beta}(\xi)\,,
              \end{equation}
              where $S^{\alpha\beta}(\xi=1)$ is the Rarita-Schwinger propagator in Eq.~\eqref{RS-prop-1}.
              The explicit form of the general tree-level propagator is
              \begin{equation}\label{rs-prop-11}
              \begin{split}
                   S^{\alpha\beta}(\xi) &= \frac{-i\slashed{k}+m}{k^2+m^2}\,\Delta^{\alpha\beta} + \frac{(1-\xi)^2}{24im^2}\, \gamma^\alpha \slashed{k}\,\gamma^\beta  \\
                                                & -\frac{ 1-\xi }{6im^2}\left(k^\alpha\gamma^\beta+\gamma^\alpha k^\beta \right) + \frac{1-\xi^2}{12m}\, \gamma^\alpha \gamma^\beta\,,
              \end{split}
              \end{equation}
              where only the first term survives for $\xi=1$.
              Using the spin-3/2 and spin-1/2 projectors defined in Eqs.\,(\ref{spin-3/2-proj}--\ref{spin-1/2-proj}) and~\eqref{spin-1/2-proj-mixed},
              the propagator can also be written as~\cite{Shklyar:2008kt}
              \begin{equation}\label{rs-prop-12}
              \begin{split}
                   S^{\alpha\beta}(\xi) &= \frac{-i\slashed{k}+m}{k^2+m^2}\,\mathds{P}_{3/2}^{\alpha\beta} \\
                   & + \frac{1-\xi}{4m}\left[(1+\xi)+(1-\xi)\,\frac{i\slashed{k}}{2m}\right] \mathds{P}_{11}^{\alpha\beta} \\
                  & + \frac{3+\xi}{12m}\left[ (3-\xi) - (3+\xi)\,\frac{i\slashed{k}}{2m}\right] \mathds{P}_{22}^{\alpha\beta} \\
                  & + \frac{3+\xi^2}{4\sqrt{3}\,m}\,(\mathds{P}_{12} + \mathds{P}_{21})^{\alpha\beta} \\
                   & + \frac{(1-\xi)(3+\xi)}{8\sqrt{3}\,m^2}\,i\slashed{k}\,(\mathds{P}_{12}-\mathds{P}_{21})^{\alpha\beta} \,,
              \end{split}
              \end{equation}
              which reduces to Eq.~\eqref{RS-prop-2} if $\xi=1$.

              The invariance of matrix elements under point transformations can be discussed along the same lines.
              A generic interaction term for the electromagnetic coupling of the nucleon to a spin-3/2 resonance has the form
              \begin{equation}
                 \mathcal{L}_{N\Delta\gamma} = \conjg{\psi}^\alpha \,\Gamma^{\alpha\mu} A^\mu \,\psi\,,
              \end{equation}
              where $\psi$ denotes the nucleon and $A^\mu$ the photon field.
              $\Gamma^{\alpha\mu}$ is the tree-level interaction vertex that satisfies $Q^\mu \Gamma^{\alpha\mu} =0$ and $k^\alpha \Gamma^{\alpha\mu}=0$ in momentum space,
              with $Q^\mu$ the photon momentum and $k^\alpha$ the spin-3/2 momentum. The first condition follows from electromagnetic gauge invariance
              and the second from spin-3/2 gauge symmetry.

              Following Ref.~\cite{Shklyar:2008kt}, one can interpret the tree-level vertex as the special case $\Gamma^{\alpha\mu}(\xi=1)$, so that its general form becomes
              \begin{equation}\label{RS-gauge-TF-vertex}
                  \Gamma^{\alpha\mu}(\xi) = \Theta^{\alpha\beta}\big(\tfrac{1}{\xi}\big)\,\Gamma^{\beta\mu}(\xi=1)\,,
              \end{equation}
              and the invariance of the Lagrangian under the point transformation~\eqref{point-tf} follows from
              \begin{equation}
                  \Gamma^{\alpha\mu}(\xi) = \Theta^{\alpha\beta}\big(\tfrac{\xi'}{\xi}\big)\,\Gamma^{\beta\mu}(\xi')\,.
              \end{equation}
              In combination with Eq.~\eqref{RS-gauge-TF-propagator}, under a change $\xi\rightarrow \xi'$
              these transformation matrices will trivially cancel each other in any Lagrangian
              \begin{equation}
                  \mathcal{L} = \conjg{\psi}^\alpha\,\Lambda^{\alpha\beta}(\xi)  \,\psi^\beta + \conjg{\psi}^\alpha \,\Gamma^{\alpha\mu}(\xi) \,A^\mu \, \psi
              \end{equation}
              as well as any matrix element with internal spin-3/2 legs.
              Since this always leads back to $\xi=1$, it is sufficient to restrict the discussion to the Rarita-Schwinger gauge.

              For the correct counting of degrees of freedom, the invariance under point transformations
              is also satisfied if either the propagator or the vertex is transverse to $\gamma^\alpha$.
              Imposing this condition on the propagator by setting $\xi=0$ in Eq.~\eqref{RS-gauge-TF-propagator} has the undesired
              consequence that it is no longer invertible. On the other hand, it is legitimate to set $\xi\rightarrow\infty$ in Eq.~\eqref{RS-gauge-TF-vertex}
              because we never need to invert vertices; this is equivalent to imposing $\gamma^\alpha \Gamma^{\alpha\mu}=0$.
              In that case the interaction term is already invariant by itself and any point transformation leads to the same result:
              \begin{equation}
                 \Theta^{\alpha\beta}(\lambda)\,\Gamma^{\beta\mu} = \Gamma^{\alpha\mu}\,.
              \end{equation}
              Matrix elements where the internal spin-3/2 propagator is connected with two vertices are also invariant,
              because the relation
              \begin{equation}
                  S^{\alpha\beta}(\xi) = \Theta^{\alpha\gamma}\big(\tfrac{\xi}{\xi'}\big)\,S^{\gamma\delta}(\xi')\,\Theta^{\delta\beta}\big(\tfrac{\xi}{\xi'}\big)
              \end{equation}
              entails $\overline{\Gamma}^{\mu\alpha} S^{\alpha\beta}(\xi)\,\Gamma^{\beta\nu} = \overline{\Gamma}^{\mu\alpha} S^{\alpha\beta}(\xi')\,\Gamma^{\beta\nu}$.
              The resulting dressed vertex admits 12 independent tensor structures, which translates to at most 12 possible independent electrocouplings
              in an effective Lagrangian.
              As discussed below Eq.~\eqref{Lagrangian-3/2},
              any vertex that is transverse in $k^\alpha$ and $\gamma^\alpha$ will also remove the
              spin-1/2 contributions from the propagators that appear in such matrix elements.
              In the Rarita-Schwinger gauge the condition $k^\alpha \Gamma^{\alpha\mu}=0$ is sufficient for this purpose
              because the projector $\mathds{P}_{11}^{\alpha\beta}$ decouples from the propagator.

        \section{Form factor relations} \label{sec:ffs-onshell}

        Here we give the relations between the constraint-free form factors $F_i(Q^2)$, which
        we use to parametrize the nucleon-to-resonance transition currents (with $i=1,2$ for $J^P= 1/2^\pm$ resonances and
        $i=1,2,3$ for $J^P = 3/2^\pm$ resonances), and the experimental helicity amplitudes.
        We also compare with the standard form factor conventions in the literature~\cite{Jones:1972ky,Devenish:1975jd};
        see also the reviews~\cite{Pascalutsa:2006up,Aznauryan:2011qj}.

        We first collect the necessary definitions to arrive at compact expressions.
        These are the relations between the nucleon and resonance masses,
        \begin{equation}
        \begin{split}
            r &= \frac{m_R}{m} \,, \\
            \delta &= \frac{m_R^2-m^2}{m^2} = r^2-1\,, \\
           \delta_\pm &= \frac{m_R \pm m}{2m} = \frac{r\pm 1}{2},
        \end{split}
        \end{equation}
        abbreviations for the photon momentum transfer,
        \begin{equation}
            \tau = \frac{Q^2}{4m^2}, \qquad
             \tau' = \tau - \frac{\delta}{4}\,,
        \end{equation}
        and
        \begin{equation}
        \begin{split}
            \lambda_\pm &= \frac{(m_R\pm m)^2+Q^2}{4m^2} =\tau +  \delta_\pm^2 \,, \\
            R_\pm &= e\sqrt{\frac{2\lambda_\pm}{m\delta}}\,, \\
            \kappa &=\frac{\sqrt{\lambda_+\lambda_-}}{\sqrt{2}\, r}\,,
        \end{split}
        \end{equation}
        where $e^2=4\pi\alpha_\text{em}$ is the electric charge.
        Note also that $\lambda_+ \lambda_- = r^2 \tau  + {\tau'}^2$.
        In the CMS frame of the pion electroproduction process $\gamma^\ast N \to \pi N$, the Lorentz-invariant quantity
        \begin{equation}\label{|k|}
           2\sqrt{2} m \kappa = \frac{2m}{r}\,\sqrt{\lambda_+\lambda_-} =  |\vect{k}|
        \end{equation}
        becomes the three-momentum of the virtual photon evaluated at the resonance position.
        Likewise, the three-momentum of the pion at the resonance position and for a vanishing pion mass is the so-called
        photon-equivalent energy: $|\vect{k}'|_{m_\pi=0} = m\delta/(2r)$.

        \subsection{\texorpdfstring{$J^P=\tfrac{1}{2}^\pm$}{JP = 1/2+-} transition form factors}

        We express the onshell $N\to 1/2^\pm$ transition matrix element in terms of Eq.~\eqref{J12-onshell-current},
        \begin{equation}
           J^\mu_\text{R}(k,Q) = \Lambda_+(k_+)\,\Gamma_\text{R}^\mu(k,Q)\,\Lambda_+(k_-)\,, 
        \end{equation}
        with the onshell kinematics as in Sec.~\ref{sec:spin-1/2-transition}:
        $k_-^\mu$ is the nucleon momentum, $k_+^\mu$ the resonance momentum,
        $Q^\mu$ the incoming photon momentum,
        and $k^\mu=(k_+^\mu + k_-^\mu)/2$ the average momentum of the nucleon and the resonance.
        On the mass shell: $k_-^2=-m^2$, $k_+^2=-m_R^2$ and therefore $k^2$ and $w=k\cdot Q$ are given by Eq.~\eqref{12-k2-w}.
        The transition form factors $F_1(Q^2)$ and $F_2(Q^2)$ are defined in~\eqref{current-onshell-resonance},
        \begin{equation}\label{current-onshell-resonance-2}\renewcommand{\arraystretch}{1.0}
           \Gamma_\text{R}^\mu(k,Q) =  i\left[ \begin{array}{c}  \mathds{1} \\ \gamma_5  \end{array}\right]\left( F_1 \,T_1^\mu + F_2\,\frac{T_3^\mu}{2}  \right),
        \end{equation}
        where the $T_i^\mu$ are given in Table~\ref{n-offshell} and
        the upper (lower) entry corresponds to positive (negative) parity.

        The definition in Ref.~\cite{Devenish:1975jd} is analogous
        but expressed in terms of two dimensionful transition form factors $G_1(Q^2)$ and $G_2(Q^2)$:
        \begin{equation}\renewcommand{\arraystretch}{1.0}
           \Gamma_\text{R}^\mu(k,Q) =  i m^2 \left[ \begin{array}{c}  \mathds{1} \\ \gamma_5  \end{array}\right]\left( G_1 \,T_1^\mu + G_2\,\frac{m^2}{w}\,T_7^\mu  \right).
        \end{equation}
        With $T_7^\mu$ being proportional to $T_3^\mu$ on the mass shell, cf.~Table~\ref{n-res-onshell-gordon},
        one can read off the onshell relations between the form factors:
        \begin{equation}
            G_1 = \frac{F_1}{m^2}\,, \qquad
            G_2 = \mp \frac{F_2}{2m^2 \delta_\mp}\,,
        \end{equation}
        where upper (lower) signs correspond to resonances with positive (negative) parity.

        The helicity amplitudes $A_{1/2}(Q^2)$ and $S_{1/2}(Q^2)$ are related with
        the form factors through~\cite{Aznauryan:2011qj}
        \begin{equation}\label{ffs-ha1}
        \begin{split}
            A_{1/2} &= R_\mp\left( 4\tau F_1 \pm \delta_\pm F_2\right), \\
            S_{1/2} &= \kappa R_\mp\left(\pm 4\delta_\pm \,F_1 - F_2\right),
        \end{split}
        \end{equation}
        with the inverse relations
        \begin{equation}\label{ffs-ha2}
        \begin{split}
           F_1 &= \frac{1}{4R_\mp \lambda_\pm}\left( A_{1/2} \pm \frac{\delta_\pm}{\kappa}\,S_{1/2}\right), \\
           F_2 &= \frac{1}{R_\mp \lambda_\pm}\left(\pm\delta_\pm A_{1/2} - \frac{\tau}{\kappa}\,S_{1/2}\right).
        \end{split}
        \end{equation}
        Note that because of the factors $R_\mp$ and $\kappa$
        the helicity amplitudes vanish either at $\lambda_+=0$, the pseudothreshold $\lambda_-=0$, or both.

        \subsection{\texorpdfstring{$J^P=\tfrac{3}{2}^\pm$}{JP = 3/2+-} transition form factors} \label{sec:32-onshell-ff-relations}

        The onshell $N\to 3/2^\pm$ transition matrix element is given by Eq.~\eqref{onshell-3/2-current-0},
        \begin{equation}\label{32-onshell-current-10}
           J_\text{R}^{\alpha\mu} = \Lambda_+(k)\,\mathds{P}_{3/2}^{\alpha\beta}(k)\,\Gamma^{\beta\mu}_\text{R}(k,Q)\,\Lambda_+(k-Q)\,, 
        \end{equation}
        where $k$ is the outgoing momentum of the resonance and $Q$ is the incoming photon momentum.
        On the mass shell: $(k-Q)^2=-m^2$ and $k^2=-m_R^2$,
        which entails $k\cdot Q = 2m^2\,\tau'$.
        The constraint-free transition form factors $F_i(Q^2)$ are defined via~\eqref{spin-3/2-offshell-1},
               \begin{equation} \renewcommand{\arraystretch}{1.0}
                   \Gamma_\text{R}^{\alpha\mu}(k,Q) =  \left[ \begin{array}{c} \gamma_5 \\ \mathds{1} \end{array}\right]
                   \left( F_1 \,T_1^{\alpha\mu} - F_2 \,T_2^{\alpha\mu}  - F_3 \,T_3^{\alpha\mu}  \right),
               \end{equation}
        where the upper (lower) entry corresponds to positive (negative) parity.
        The $T_i^{\alpha\mu}$ are defined in Table~\ref{n-delta-gamma-offshell}. 

        To write down the various different versions of the onshell currents used in the literature, we define
        the tensors $T_{21} \dots T_{25}$ in Table~\ref{32-additional-tensors}
        in addition to those in Table~\ref{n-delta-gamma-offshell}.
        On the mass shell and inside the positive-energy and Rarita-Schwinger projectors~\eqref{32-onshell-current-10} they are linearly related with $T_1$, $T_2$ and $T_3$, but
        with the exception of $T_{21}$ they do not satisfy the offshell constraint $k^\alpha \,T_i^{\alpha\mu}=0$.

        \begin{table}[t]

             \begin{equation*}  \renewcommand{\arraystretch}{1.6}
             \begin{array}{r@{\!\;\,}l   }

                 m^4\,T_{21}^{\alpha\mu} &= \varepsilon^{\alpha\beta}_{kQ}\,\varepsilon^{\beta\mu}_{kQ}  \\
                 m^4\,T_{22}^{\alpha\mu} &= Q^\alpha k^\beta t^{\beta\mu}_{QQ}  \\
                 m \,T_{23}^{\alpha\mu} &= -i t^{\alpha\mu}_{\gamma Q} \\
                  m^2\,T_{24}^{\alpha\mu} &= t^{\alpha\mu}_{QQ}  \\
                  m^4\,T_{25}^{\alpha\mu} &= \slashed{k}\,Q^\alpha \gamma^\beta\,\varepsilon^{\beta\mu}_{kQ}

             \end{array} \qquad
             \begin{array}{l   }

                 T_{21} -2\tau' T_2 - r T_3  \\
                 r T_{22} + 4r\tau T_2 - 2\tau' T_3  \\
                 r T_{23} - T_1 - T_2  \\
                   r T_{24} - T_3 \\
                  T_{25} + 2\lambda_- T_1 + 2\tau' T_2 + r T_3

             \end{array}
             \end{equation*}

               \caption{Additional tensors appearing in the $J^P=\tfrac{3}{2}^\pm$ currents~\eqref{3/2-current-helicity}, \eqref{spin-3/2-variant-1} and \eqref{3/2-current-js}.
               For positive parity,
               the right column gives their onshell relations
               which relate them to $T_1$, $T_2$ and $T_3$ defined in Table~\ref{n-delta-gamma-offshell}.
               For negative parity the same relations hold if one exchanges $r\to -r$ and $\lambda_- \to \lambda_+$.}
               \label{32-additional-tensors}

             \end{table}

        \renewcommand{\arraystretch}{1.1}

        Following~\cite{Aznauryan:2011qj}, 
        the experimentally extracted helicity amplitudes $A_{3/2}(Q^2)$, $A_{1/2}(Q^2)$ and $S_{1/2}(Q^2)$ are
        related to the helicity form factors $h_i(Q^2)$ via
        \begin{equation}\label{h1h2h3}
            \{ h_1, \,h_2, \,h_3\} = \sqrt{\frac{3}{2}}\,\frac{4}{R_\mp}\left\{ \frac{rS_{1/2}}{2\kappa }, \; \pm \frac{A_{3/2}}{\sqrt{3}}, \; A_{1/2} \right\},
        \end{equation}
        where upper (lower) signs denote positive (negative) parity.
        The corresponding form of the current is~\cite{Devenish:1975jd} 
        \begin{equation} \label{3/2-current-helicity}
        \begin{split}
            \Gamma_\text{R}^{\alpha\mu} &= \frac{1}{16 \lambda_+ \lambda_-}\left[ \begin{array}{c} \gamma_5 \\ \mathds{1} \end{array}\right]
                                           \Big(  -h_1\,T_{22}^{\alpha\mu} + 2h_2\,T_{21}^{\alpha\mu}  \\[-2mm]
                                           & \qquad\qquad\qquad\qquad\;\; + (h_2+h_3)\,T_{25}^{\alpha\mu}   \Big)\,,
        \end{split}
        \end{equation}
        which is neither free of kinematics nor satisfies the offshell constraints.
        Using the onshell relations in Table~\ref{32-additional-tensors},
        the helicity form factors (and thus helicity amplitudes) are related to the $F_i$ via
        \begin{equation}\label{ffs-ha32-1}
        \begin{split}
           F_1 &= -\frac{h_2+h_3}{8\lambda_\pm}\,, \\
           F_2 &= -\frac{1}{8\lambda_+ \lambda_-}\left[ 2\tau\,h_1 + \tau' (h_2-h_3) \right], \\
           F_3 &= \pm \frac{1}{8r\lambda_+\lambda_-}\left[ \tau' h_1 - \frac{r^2}{2}\,(h_2-h_3) \right]
        \end{split}
        \end{equation}
        and vice versa
        \begin{equation}\label{ffs-ha32-2}
        \begin{split}
           h_1 &= -4\left( r^2 F_2 \mp 2r\tau' F_3\right), \\
           h_2 &= -4\left( \lambda_\pm F_1 + \tau' F_2 \pm 2r\tau F_3\right), \\
           h_3 &= -4\left( \lambda_\pm F_1 - \tau' F_2 \mp 2r\tau F_3\right).
        \end{split}
        \end{equation}
        Also here the helicity amplitudes in Eq.~\eqref{h1h2h3} vanish either at $\lambda_+=0$ or $\lambda_-=0$
        due to the factors $R_\mp$.

        Another form of the current
        expressed in terms of three form factors $G_i(Q^2)$ is~\cite{Devenish:1975jd,Aznauryan:2011qj}: 
        \begin{equation}\label{spin-3/2-variant-1}
            \Gamma_\text{R}^{\alpha\mu} = \left[ \begin{array}{c} \gamma_5 \\ \mathds{1} \end{array}\right]
            \left( G_1\,T_{23}^{\alpha\mu} + G_2\,T_2^{\alpha\mu} + G_3\,T_{24}^{\alpha\mu}   \right).
        \end{equation}
        Here we defined the $G_i$ to be dimensionless
        (in the standard definition they carry dimensions: replace $G_1 \to m G_1$ and $G_{2,3} \to m^2 G_{2,3}$.)
        They are free of kinematics but again the current does not satisfy the offshell constraints.
        Their relation with the $F_i$ is
        \begin{equation*}
            G_1 = \pm r F_1, \quad
            G_2 = -\left(F_1+F_2\right), \quad
            G_3 = \mp r F_3\,.
        \end{equation*}

        Finally, the Jones-Scadron form of the current in terms of $G_E^\ast$, $G_M^\ast$ and $G_C^\ast$ is given by~\cite{Jones:1972ky,Devenish:1975jd}
        \begin{equation}\label{3/2-current-js}
        \begin{split}
             \Gamma_\text{R}^{\alpha\mu} & = \pm \sqrt{\frac{3}{2}}\,\frac{\delta_\pm}{2 \lambda_+ \lambda_-} \left[ \begin{array}{c} \gamma_5 \\ \mathds{1} \end{array}\right]
                              \bigg( \left[ \begin{array}{c} G_M^\ast-G_E^\ast \\ 2G_M^\ast \end{array}\right] \lambda_\mp \,T_1^{\alpha\mu}  \\
                              & \qquad + \left[ \begin{array}{c} -2G_E^\ast \\ G_M^\ast-G_E^\ast \end{array}\right]
                              \frac{T_{21}^{\alpha\mu}}{2}  \mp G_C^\ast\,\frac{T_{22}^{\alpha\mu}}{2}\,\bigg)\,.
        \end{split}
        \end{equation}
        As before, upper (lower) components and signs denote positive (negative) parity.
        The Jones-Scadron form factors are related to the helicity form factors via
        \begin{equation}\label{delta-JS-helicity}
        \begin{split}
            \left[\begin{array}{c} G_M^\ast \\ G_E^\ast \end{array}\right] &= -\sqrt{\frac{3}{2}}\,\frac{h_3 \pm 3h_2}{12\delta_\pm}\,, \\
            \left[\begin{array}{c} G_E^\ast \\ G_M^\ast \end{array}\right] &= \sqrt{\frac{3}{2}}\,\frac{h_3 \mp h_2}{12\delta_\pm}\,,\\
            G_C^\ast &= \sqrt{\frac{3}{2}}\,\frac{h_1}{6\delta_\pm}\,.
        \end{split}
        \end{equation}
        This coincides with the conventions in~\cite{Jones:1972ky,Aznauryan:2011qj}
        whereas in Ref.~\cite{Devenish:1975jd} $G_M^\ast$, $G_E^\ast$ and $G_C^\ast$ are defined without the factor $\sqrt{3/2}$ on the r.h.s.
        The relations between the Jones-Scadron form factors and the $F_i$ are given by
        \begin{equation}\label{delta-JS-Fi}
        \begin{split}
            G_M^\ast &= \sqrt{\frac{2}{3}}\,\frac{1}{\delta_\pm}
            \left[\begin{array}{c}
            2\lambda_+ F_1 + \tau' F_2 + 2r\tau F_3 \\
            -\lambda_- F_1
            \end{array}\right] , \\
            G_E^\ast &= \sqrt{\frac{2}{3}}\,\frac{1}{\delta_\pm}
            \left[\begin{array}{c}
            \tau' F_2 + 2r\tau F_3 \\
            -\lambda_- F_1 - 2\tau' F_2 + 4r\tau F_3
            \end{array}\right] , \\
            G_C^\ast &= \sqrt{\frac{2}{3}}\,\frac{1}{\delta_\pm}  \left( -r^2 F_2 \pm 2r \tau' F_3\right)
        \end{split}
        \end{equation}
        and vice versa
        \begin{equation*}
        \begin{split}
            F_1 &= \sqrt{\frac{3}{2}}\,\frac{\delta_\pm}{2\lambda_+\lambda_-}
            \left[\begin{array}{c}
            \lambda_- (G_M^\ast-G_E^\ast) \\
            -2\lambda_+ G_M^\ast
            \end{array}\right] , \\
            F_2 &= \sqrt{\frac{3}{2}}\,\frac{\delta_\pm}{\lambda_+\lambda_-}
            \left[\begin{array}{c}
            \tau' G_E^\ast - \tau G_C^\ast \\
            \tfrac{1}{2}\tau' (G_M^\ast-G_E^\ast) - \tau G_C^\ast
            \end{array}\right] , \\
            F_3 &= \sqrt{\frac{3}{2}}\,\frac{\delta_\pm}{2r\lambda_+\lambda_-}
            \left[\begin{array}{c}
            r^2 G_E^\ast + \tau' G_C^\ast \\
            -\tfrac{1}{2} r^2 (G_M^\ast-G_E^\ast) - \tau' G_C^\ast
            \end{array}\right] .
        \end{split}
        \end{equation*}

\end{appendix}

\newpage

\bibliographystyle{apsrev4-1-mod}

\bibliography{lit-resonances-2}

\end{document}